\documentclass[a4paper,fleqn,usenatbib]{mnras}

\usepackage[T1]{fontenc}
\usepackage{ae,aecompl}

\usepackage{natbib}
\usepackage{placeins}
\usepackage{multirow}
\usepackage{multicol}        
\usepackage{amsmath}
\usepackage{amssymb}
\usepackage{epsfig}
\usepackage{graphicx}
\usepackage{bm}		
\usepackage{pdflscape}	
\usepackage{color}

\newcommand{\kms}{\mbox{km s$^{-1}$}}

\newcommand{\Msun}{\mbox{$M_{\odot}$}}

\newcommand{\arcminspace}{\mbox{$\arcmin$ }}

\newcommand{\HI}{H\,{\sevensize I}}
\newcommand{\hii}{H\,{\sevensize II}}

\newcommand{\co}{\mbox{$^{12}$CO}}
\newcommand{\cother}{\mbox{$^{13}$CO}}

\newcommand{\degrees}{\degr}

\title[The Carina Nebula and Gum 31 molecular complex]{The Carina Nebula and Gum 31 molecular complex: I. Molecular gas distribution, column densities and dust temperatures.}

\author[D. Rebolledo et al.]{David Rebolledo,$^{1,2}$\thanks{E-mail: davidr@physics.usyd.edu.au}
Michael Burton,$^{2}$
Anne Green,$^{1}$
Catherine Braiding,$^{2}$
\newauthor Sergio Molinari,$^{3}$
Graeme Wong,$^{4,5}$
Rebecca Blackwell,$^{6}$
Davide Elia$^{3}$ 
\newauthor and Eugenio Schisano$^{3}$
\\
$^{1}$Sydney Institute for Astronomy, School of Physics, The University of Sydney, NSW 2006, Australia\\
$^{2}$School of Physics, The University of New South Wales, Sydney, NSW, 2052, Australia\\
$^{3}$INAF -- Istituto di Astrofisica e Planetologia Spaziali, via Fosso del Cavaliere 100, I-00133 Rome, Italy\\
$^{4}$School of Computing Engineering and Mathematics, Western Sydney University, Locked Bay 1797, Penrith, NSW 2751, Australia\\
$^{5}$CSIRO Astronomy and Space Science, PO Box 76, Epping, NSW 1710, Australia\\
$^{6}$Department of Physics, School of Physical Sciences, University of Adelaide, Adelaide, SA 5005, Australia
}

\date{{\bf Accepted for Publication in the Monthly Notices of the Royal Astronomical Society Journal.}}

\pubyear{2015}

\begin{document}
\label{firstpage}
\pagerange{\pageref{firstpage}--\pageref{lastpage}}
\maketitle

\begin{abstract}
We report high resolution observations of the $^{12}$CO$(1\rightarrow0)$ and $^{13}$CO$(1\rightarrow0)$ molecular lines in the Carina Nebula and the Gum 31 region obtained with the 22-m Mopra telescope as part of the The Mopra Southern Galactic Plane CO Survey.  We cover 8 deg$^2$ from $l = 285\degrees$ to 290$\degrees$, and from $b = -1.5\degrees$ to +0.5$\degrees$.  The molecular gas column density distributions from both tracers have a similar range of values.  By fitting a grey-body function to the observed infrared spectral energy distribution from Herschel maps, we derive gas column densities and dust temperatures.  The gas column density has values in the range from $6.3\times\ 10^{20}$ to $1.4\times 10^{23}$ cm$^{-2}$, while the dust temperature has values in the range from $17$ to $43$ K.  The gas column density derived from the dust emission is approximately described by a log-normal function for a limited range of column densities.  A high-column density tail is clearly evident for the gas column density distribution, which appears to be a common feature in regions with active star formation.  There are regional variations in the fraction of the mass recovered by the CO emission lines with respect to the total mass traced by the dust emission.  These variations may be related to changes in the radiation field strength, variation of the atomic to molecular gas fraction across the observed region, differences in the CO molecule abundance with respect to H$_{2}$, and evolutionary stage differences of the molecular clouds that compose the Carina Nebula-Gum 31 complex.
\end{abstract}




\section{Introduction}\label{intro}

The Carina Nebula Complex (NGC 3372, hereafter CNC) is the southern hemisphere's largest and brightest nebula.  Located at 2.3 kpc away (\citealt{2006MNRAS.367..763S}), it provides a laboratory to study ongoing star formation in the vicinity of some of the most massive stars known, including our Galaxy's most luminous star, Eta Carinae ($\eta$ Car). Across the entire CNC there are more than 65 O stars and several hundred protostars (\citealt{2010MNRAS.406..952S}), making it a rich environment to study the interplay between clustered star formation, massive star feedback and triggered star formation.  Two young stellar clusters, Trumpler 14 and Trumpler 16, dominate the central region and the CNC is home to three high mass-loss Wolf-Rayet stars.  For comparison, Orion Nebula is dominated by just a single O-type star (\citealt{1986ApJ...307..609H}).

Located at $\sim 1\degrees$ north-west of the CNC, the Gum 31 region is a bubble-shaped young \hii\ region containing the stellar cluster NGC 3324.  Due to its proximity to the more attractive star forming region CNC, the physical properties of the Gum 31 region have remained relatively unexplored compared to its southern neighbour (\citealt{2013A&A...552A..14O}).  Whether both regions are physical connected or not remains unclear, however; the distance of Gum 31 is $\sim$ 2.5 kpc (\citealt{2010MNRAS.402...73B}) and the distance of NGC 3324 is $\sim$ 2.3 kpc (\citealt{2005A&A...438.1163K}), placing both objects at the same distance as the CNC (\citealt{2006MNRAS.367..763S}).  

The CNC-Gum 31 complex is in an unobscured region, and is sufficiently close for us to resolve structures in detail.  Other similar star-forming complexes are more difficult to study because they are located within the dense ring of molecular gas surrounding the Galactic Centre, four times further away and heavily obscured by intervening dust. The CNC-Gum 31 region is not as extreme as 30 Doradus in the Large Magellanic Cloud (\citealt{1998ApJ...493..180M}), the Arches cluster near the Galactic Centre (\citealt{2004ApJ...611L.105N}), or the Galactic clusters NGC 3603 (\citealt{1983A&A...124..273M}) and W49 (\citealt{2005A&A...430..481H}).  However, the CNC-Gum31 complex location and properties make it an excellent analogue for understanding the star formation in the distant Universe.

Until the mid-1990s, the prevailing view was that the CNC was an evolved star-forming region, with a few remnant gas clouds that had been shredded by the energy input from the massive stars.  In recent years, and with the new wide-field, high resolution images at X-ray, optical and IR wavelengths, this view has changed dramatically.  We now recognise that the CNC is a hotbed of active star formation, with a reservoir of quiescent molecular gas where new bursts of star formation may one day occur (\citealt{2008hsf2.book..138S}).  The stunning images from HST and Spitzer (\citealt{2010MNRAS.405.1153S}) reveal numerous gigantic pillars of dust (sometimes called ``elephant trunks'' because of their morphology), situated around the periphery of the ionised nebula associated with $\eta$ Car and which point towards the central massive star clusters.  Star formation appears to have been triggered recently in these pillars, as there are hundreds of protostars of ages $\sim10^{5}$ years, possibly formed as a result of the impact of winds from the massive stars on the dust clouds (\citealt{2004A&A...418..563R}). The region also shows extended, diffuse X-ray emission (\citealt{2011ApJS..194...16T}), which may be from a previous supernova.  If so, the supernova's expanding blast wave could have triggered the new generation of star formation, but this also raises an intriguing question as to the location of the precursor star.  Was this star more massive than $\eta$ Car and why have not we seen direct evidence of its blast wave?  \citet{2008Natur.455..201S} has provided evidence to suggest a recent low-energy eruption occurred from $\eta$ Car itself, so a massive supernova explosion should be detectable.

There are several other unresolved issues concerning the CNC.  The total infrared luminosity detected from the region (re-radiated from ultraviolet heated dust) is only about half the energy input from the known stars, and the measured free-free radio continuum emission only accounts for about three-quarters of the ionising flux from these same stars (\citealt{2007MNRAS.379.1279S}).  While a few hundred thousand solar masses of cold gas have been observed from the whole nebula, it may be that much is in atomic phase in photon-dominated regions rather than as molecular gas.  This neutral gas is not currently participating in star formation and may be a reservoir for future cycles of star formation once the current hot young stars explode as supernovae (\citealt{2011ApJ...728..127D}). 

The CNC was originally mapped in CO as part of the Colombia Galactic plane survey (\citealt{1988ApJ...331..181G}) at 9$\arcminspace$ angular resolution.  The observations of the molecular emission covered a $\sim 2 \degrees \times 2\degrees$  region centred on $\eta$ Car.  The first $^{12}$CO$(1\rightarrow0)$ map of the CNC with the Mopra Telescope was obtained by \citet{2000PhDT.........2B}, achieving a 45\arcsec resolution.  Subsequently, maps of $3\arcminspace - 4\arcminspace$ resolution have been obtained for the $^{12}$CO$(1\rightarrow0)$ and $^{12}$CO$(4\rightarrow3)$ lines by the NANTEN and AST/RO telescopes, respectively (\citealt{2005ApJ...634..476Y}; \citealt{2001ApJ...553..274Z}).  The gas mass, estimated to be $\sim 10^{5}\ \Msun$ (\citealt{2005ApJ...634..476Y}), is typical of a giant molecular cloud (GMC).  The images show three distinct regions, each representing a different aspect of the star formation environment: (i) small globules near $\eta$ Car, the last shredded pieces of a GMC core; (ii) the northern cloud, surrounding the Tr 14 cluster, which is a relatively pristine cloud with little active star formation;  (iii) the southern cloud, which is being eroded and shaped by radiation and winds from the massive young cluster Tr 16, giving rise to numerous dust pillars as well as to a second generation of stars forming inside them.

This paper is the first in a series of studies that aim to explore the properties of the interstellar medium (ISM) in the CNC and surrounding regions at high spatial and spectral resolution.  In this work, we present the line maps that trace the molecular gas components.  We present results for the molecular gas column density and dust temperature of the complex, derived by fitting spectral energy distributions (SED) to the Herschel far-infrared images (\citealt{2010PASP..122..314M}).  The data will be used to characterise the molecular gas, enabling a comparison with other wavelength images.  In particular, through measuring both the $^{12}$CO$(1\rightarrow0)$ and $^{13}$CO$(1\rightarrow0)$ lines (hereafter, $\co$ and $\cother$ respectively) we are able to determine the optical depth and produce an improved determination of the molecular gas mass.  

The paper is organised as follows:  In Section \ref{obs} we detail the main characteristics of the data used in our analysis.  In Section \ref{results} we explain our approach in creating the integrated intensity and velocity field maps of the CO line emission cubes, and the technique for estimating the molecular gas column density using these tracers.  In Section \ref{dust-mod} we present an algorithm for the SED fitting to the infrared data in order to estimate the column density and temperature of the dust.  We discuss the main implications of our work in Section \ref{discuss}, and present a summary of the work in Section \ref{summary}.

\begin{figure*}
\centering
\begin{tabular}{c}
\epsfig{file=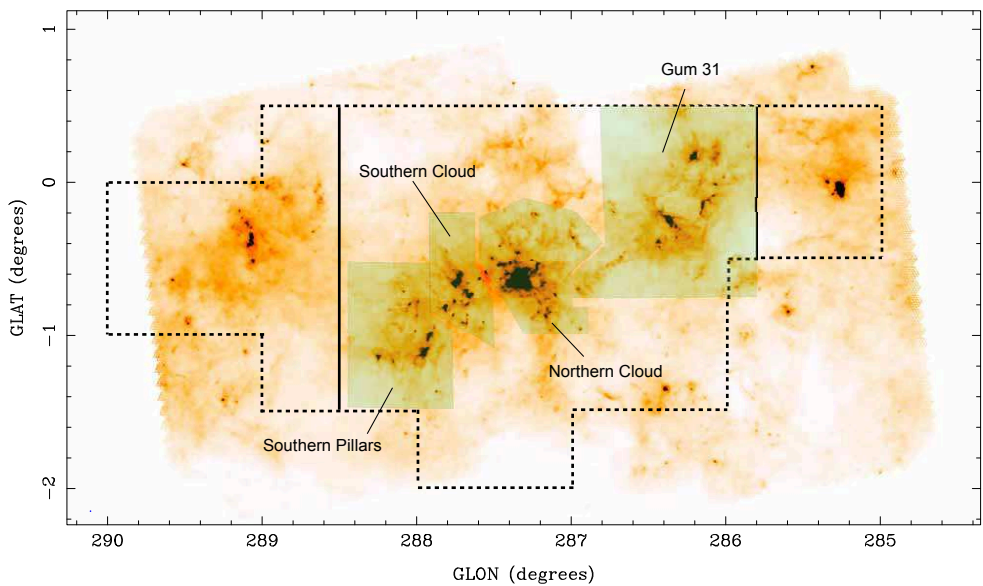,width=0.75\linewidth,angle=0}
\end{tabular}
\caption{Image of {\it Herschel} 500 $\mu$m map of the CNC-Gum 31 region (\citealt{2010PASP..122..314M}).  The black dashed line illustrates the region covered by our Mopra observations.  The solid black lines show the region that encloses the gas associated with the CNC and Gum 31 considered in this paper.  Colour sectors show the region selection in the CNC-Gum 31 region, which includes the Southern Pillars, the Southern Cloud, the Northern Cloud and the Gum 31 region.}
\label{carina.500um.obs}
\end{figure*}

\section{Data}\label{obs}

\subsection{Mopra observations}
The CO observations reported in this paper were conducted using the Mopra telescope as part of the The Mopra Southern Galactic Plane CO Survey (\citealt{2013PASA...30...44B}; \citealt{2015PASA...32...20B}).  Mopra is a 22-m diameter millimeter-wave dish, part of the CSIRO Australia Telescope National Facility and located near Coonabarabran in New South Wales, Australia. We map simultaneously the $\co$ and $\cother$ spectral lines (115.3 and 110.2 GHz, respectively) using fast-on-the-fly mapping.  This technique takes $8\times256$ ms samples in each 2.048s cycle time and uses a series of 60$\arcmin$ long $\times$ 6$\arcmin$ wide mapping grids. Each grid takes $\sim$1 hour, scanning at $\sim$30 \arcsec s$^{-1}$ (in contrast to $\sim$3\arcsec s$^{-1}$ and the square maps used with standard mapping). With a typical $T_\mathrm{sys}\sim 800/300$ K at 115/110 GHz respectively, the 1 $\sigma$ noise in a 0.088 km/s channel is $T_\mathrm{A}^{*} \sim 1.6/0.7$ K, after summing the oversampled pixels and scanning in two orthogonal directions.  For a typical $\sim$ 2 $\kms$ wide line for $\co$ and 1 $\kms$ for $\cother$, the 1 $\sigma$ survey line flux sensitivity will be $\sim$5.8/1.3 K $\kms$ after correction for aperture efficiency.  Twenty grids (ten each in two orthogonal directions) are needed to complete observations of 1 square degree.  In the 2013-2014 season, we have successfully observed an area equivalent to 8 square degrees in both scan directions, centred on $l \sim 287.5\degrees$ and $b \sim -0.5\degrees$.  Figure \ref{carina.500um.obs} illustrates the area covered by our CO observations with Mopra overlaid on a 500 $\mu$m image from {\it Herschel} (\citealt{2010PASP..122..314M}).


\subsection{Infrared dust emission maps}
In order to derive dust properties of the CNC-Gum 31 complex, we use infrared images from The Infrared Galactic Plane Survey (Hi-GAL, \citealt{2010PASP..122..314M}).  Hi-GAL is a {\it Herschel} open-time key project that initially aimed at covering the inner Galaxy ($|l| < 60\degrees$, $|b| < 1$), and was subsequently extended to the entire Galactic disk.  It maps the Galactic plane at $2\degrees \times 2\degrees$ tiles with PACS and SPIRE cameras in parallel mode, in the bands 70, 160, 250, 350, and 500 $\mu$m.  The pixel sizes of the maps are $3\farcs2$, $4\farcs5$, $6\farcs0$, $8\farcs0$, and 11$\farcs5$ at 70, 160, 250, 350, and 500 $\mu$m, respectively.  The noise levels of the maps were estimated in regions with no significant emission located at the edge of the field, and they are 20, 30, 25, 10 and 5 MJy $\mathrm{sr}^{-1}$ for 70, 160, 250, 350, and 500 $\mu$m maps respectively.

The PACS and SPIRE maps of the CNC-Gum31 region have been produced with the software {\sevensize UNIMAP}, which is the 2nd generation map-making software developed within the Hi-GAL consortium (see \citealt{2015MNRAS.447.1471P} for details).  The zero-level offsets in the {\it Herschel} maps were calculated by comparing the flux in the IRAS and Planck data at comparable wavelength bands. This method is fully described in \citet{2010A&A...518L..88B}.


\begin{figure*}
\centering
\begin{tabular}{c}
\epsfig{file=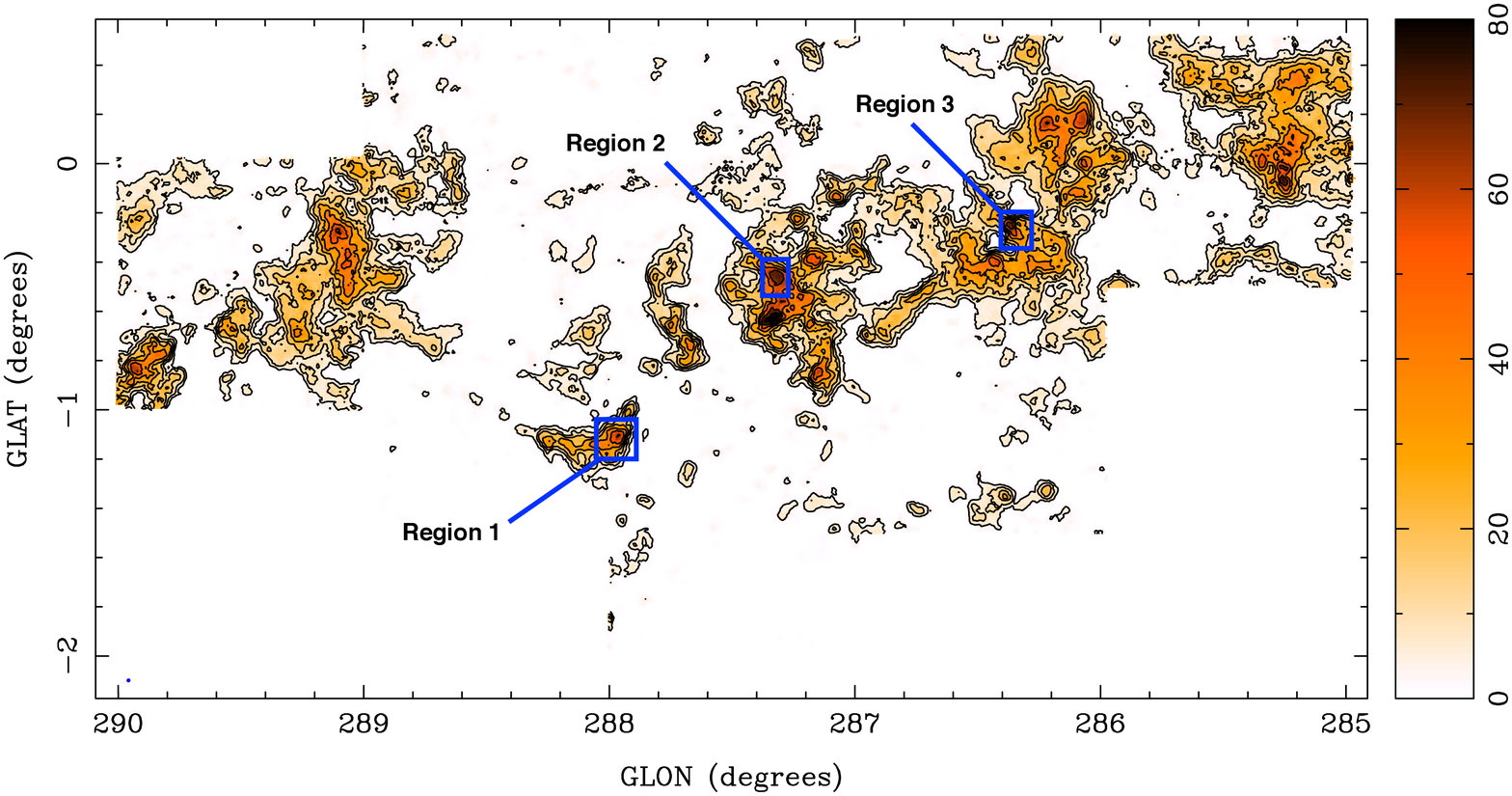,width=0.85\linewidth,angle=0}
\end{tabular}
\caption{$\co$ integrated intensity map of the region covered by our Mopra observations over the velocity range $V_\mathrm{LSR}=-50$ to $50\ \kms$.  Colour bar is in units of K $\kms$.  Contours are spaced by $n^2$ K $\kms$, with $n=2, 3, 4, 5, 6, 7, 8 , 9$.  The line profiles shown in Figure \ref{spect_reg} correspond to the regions inside the blue squares.}  
\label{carina12co}
\end{figure*}

\begin{figure*}
\centering
\begin{tabular}{c}
\epsfig{file=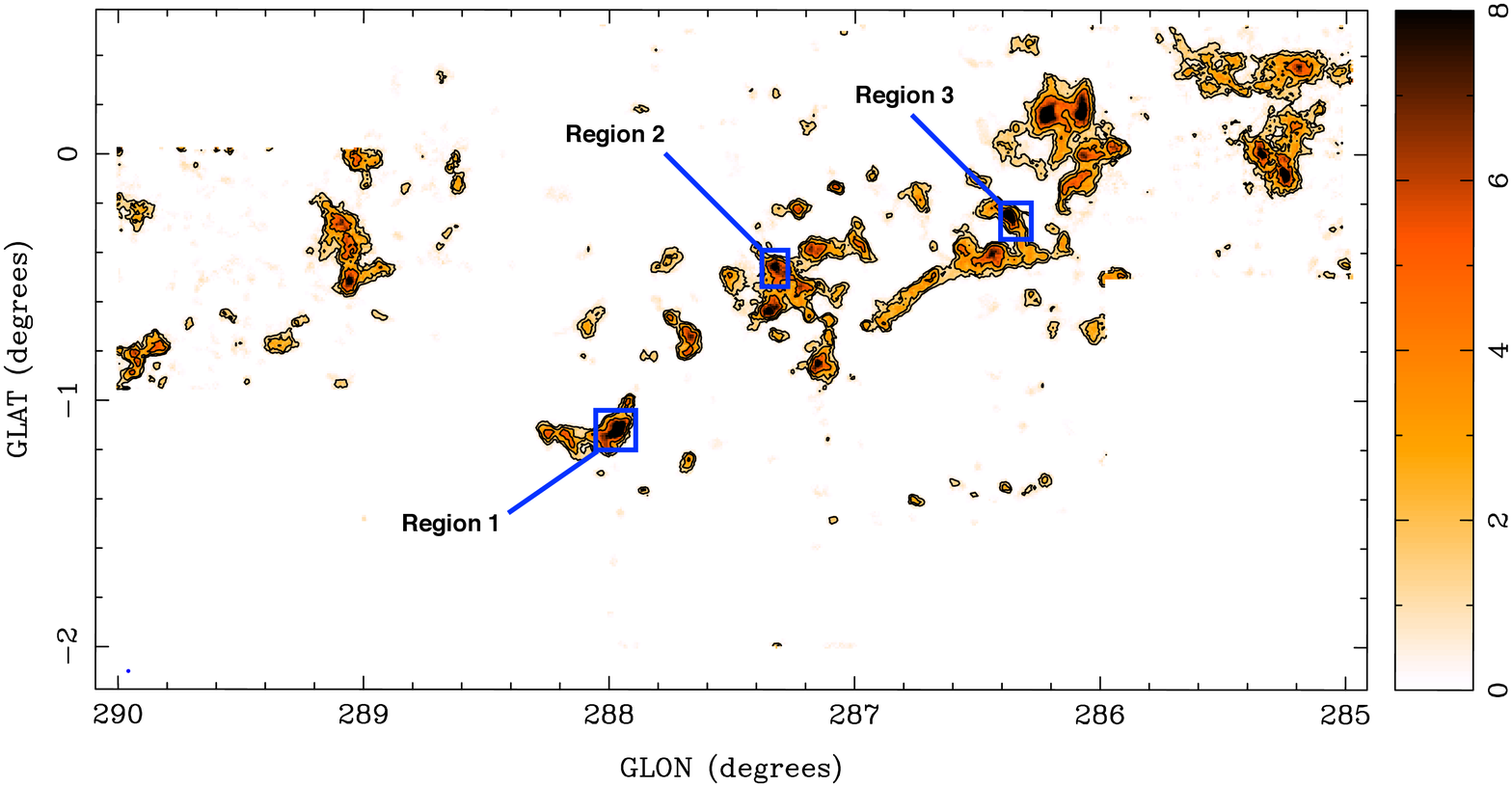,width=0.85\linewidth,angle=0}
\end{tabular}
\caption{$\cother$ integrated intensity map of the region covered by our Mopra observations over the velocity range $V_\mathrm{LSR}=-50$ to $50\ \kms$.  Colour bar is in units of K $\kms$.  Contours are spaced by $2^{n}$ K $\kms$, with $n=0, 1, 2, 3$.    The line profiles shown in Figure \ref{spect_reg} correspond to the regions inside the blue squares.}
\label{carina13co}
\end{figure*}

\begin{figure*}
\centering
\begin{tabular}{c}
\epsfig{file=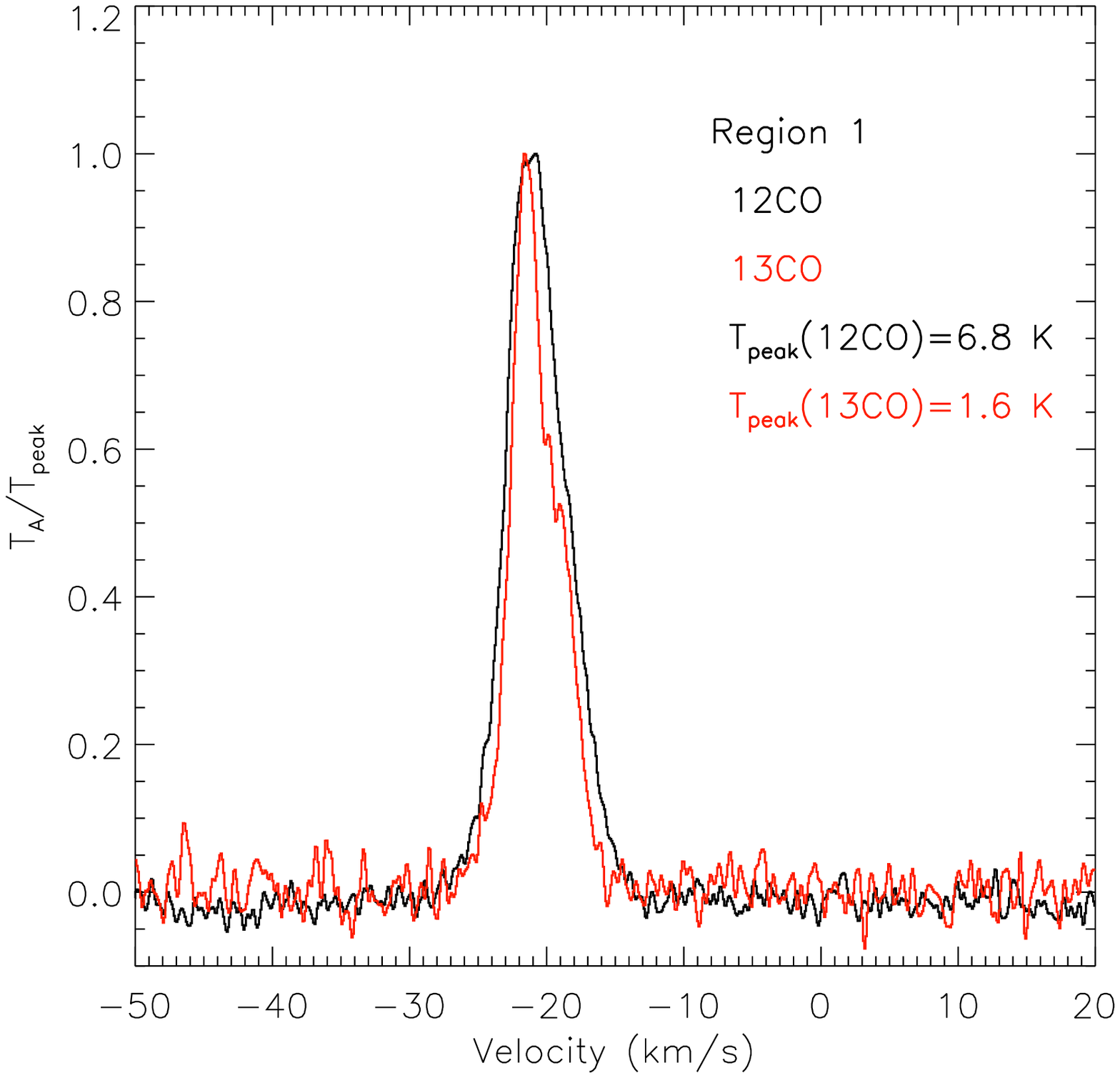,width=0.3\linewidth,angle=90}
\epsfig{file=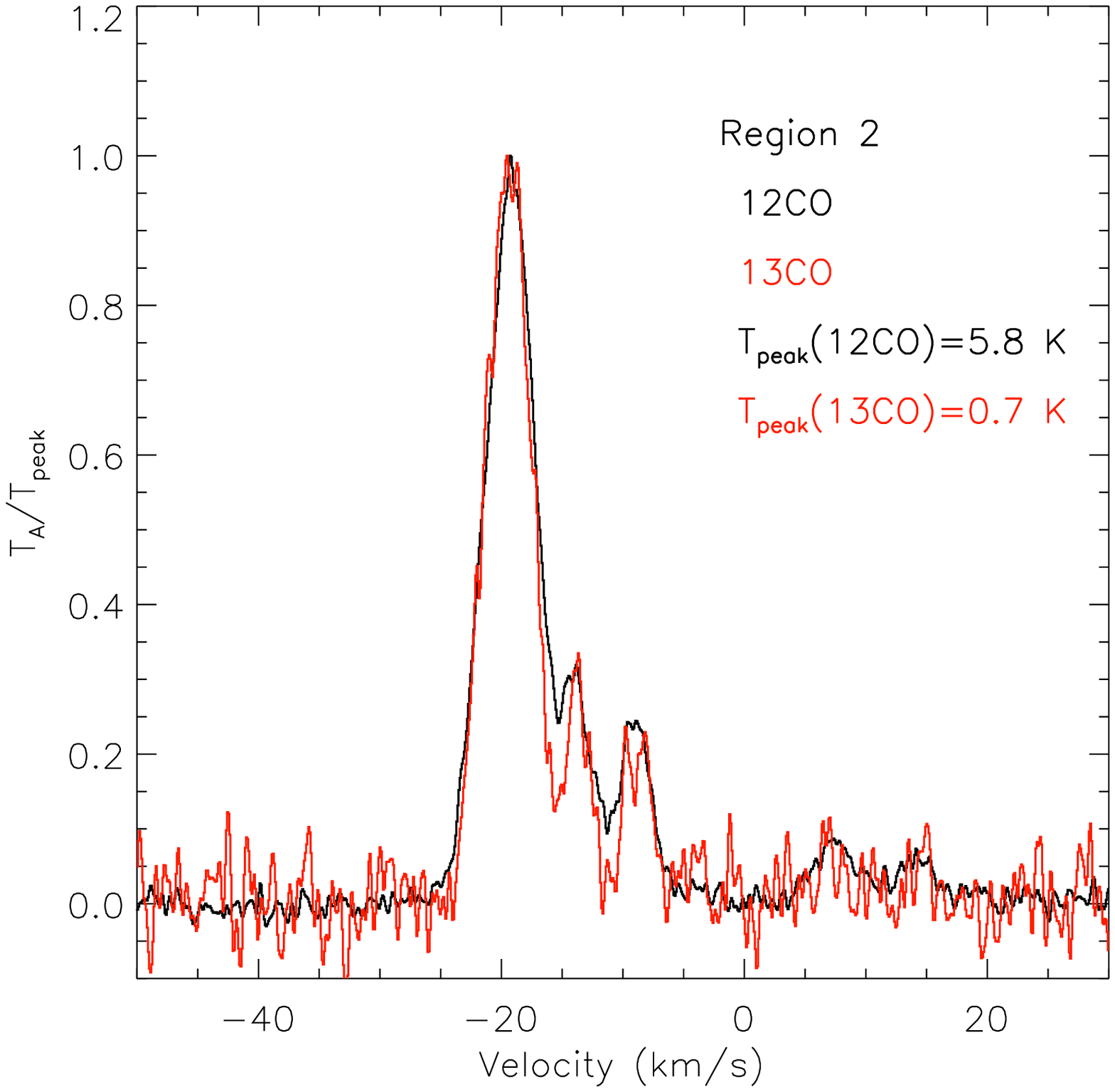,width=0.3\linewidth,angle=90}
\epsfig{file=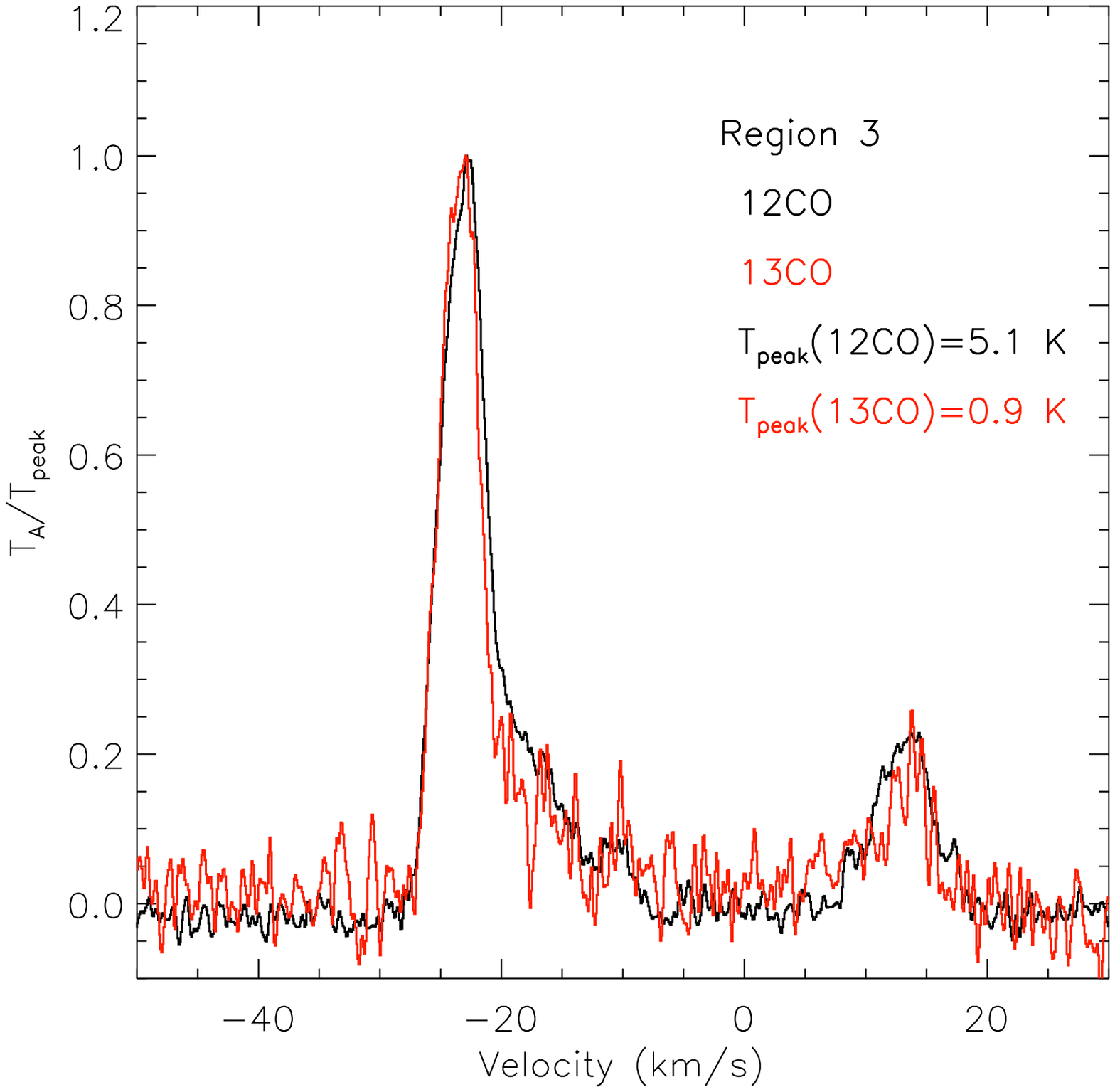,width=0.3\linewidth,angle=90}
\end{tabular}
\caption{Average line profiles of $\co$ and $\cother$ over the regions shown in Figures \ref{carina12co} and \ref{carina13co}.  The profiles have been normalised by the peak of the line.  Black profiles show the $\co$ lines, while red profiles correspond to $\cother$.  Region 1 is located in the Southern Pillars, Region 2 is in the Northern Cloud, and Region 3 is in Gum 31.  In Region 1, the line profile shows a single peak shape.  Multiple peaks are seen in Region 2 and Region 3.}
\label{spect_reg}
\end{figure*}

\section{Results}\label{results}

\subsection{Method to generate integrated intensity maps}\label{moments}
In order to estimate accurate column densities from the CO line emission maps, we need to efficiently identify regions of significant emission in the position-position-velocity cubes.  The method applied here is similar to the method implemented for extragalactic observations of CO (\citealt{2015ApJ...808...99R}).  In order to enhance the signal-to-noise ratio for extended structures, we generate a {\it smoothed} signal mask by degrading the original data cube in angular and velocity resolutions by a factor of 3.  Regions of continuous significant emission are identified for pixels brighter than $n_\mathrm{thresh} \times \sigma_{\rm smo}$, where $ \sigma_{\rm smo}$ is the rms intensity of the smoothed cube, and $n_\mathrm{thresh}$ is a threshold level.  We further expand each region to include any adjacent pixels above $n_\mathrm{edge} \times \sigma_{\rm smo}$, and used the resulting regions to construct a dilated signal mask.  The dilated mask approach was designed to identify extended low brightness emission regions.  For the $\co$ map we used $n_\mathrm{thresh}=7$ and $n_\mathrm{edge}=6$.  For the $\cother$ map we use $n_\mathrm{thresh}=6$ and $n_\mathrm{edge}=5$.  This dilated signal mask was then used to generate integrated intensity images.  The associated uncertainty map for each image was derived from the error propagation through pixels in the signal mask.

\subsection{$\co$ and $\cother$ maps}
\subsubsection{Integrated intensity maps}
Figure \ref{carina12co} shows the integrated intensity map of $\co$ line.  Our high spatial resolution image represents a significant improvement over the CO maps of \citet{2005ApJ...634..476Y}.  The highest peaks of emission are located near the star cluster Tr 14, which has been previously identified as the $\eta$ Car Northern Cloud (\citealt{2003A&A...412..751B}).  Bright peaks of $\co$ are also observed in the vicinity of the \hii\ region Gum 31.

In Figure \ref{spect_reg} we show the line profile of $\co$ over three regions in the CNC-Gum 31 molecular complex.  The position of the regions are shown in Figure \ref{carina12co}.  Region 1 is located in the Southern Pillars, Region 2 is in the Northern Cloud, and Region 3 is in Gum 31.  For Region 1, the $\co$ line shape shows a single peak at $-19\ \kms$ at our velocity resolution.  Region 2 shows a more complex line shape, with a central peak at $-20\ \kms$ and smaller peaks separated by $\sim 5\ \kms$.  The $\co$ line shape in Region 3 shows a clear peak at $-23\ \kms$, and extended emission is clearly seen until $-5\ \kms$.  In contrast to Region 1, Region 2 and 3 show emission at positive velocities which is not related to the CNC-Gum 31 region (see Section \ref{kinematics} for more details).

\begin{figure*}
\centering
\begin{tabular}{c}
\epsfig{file=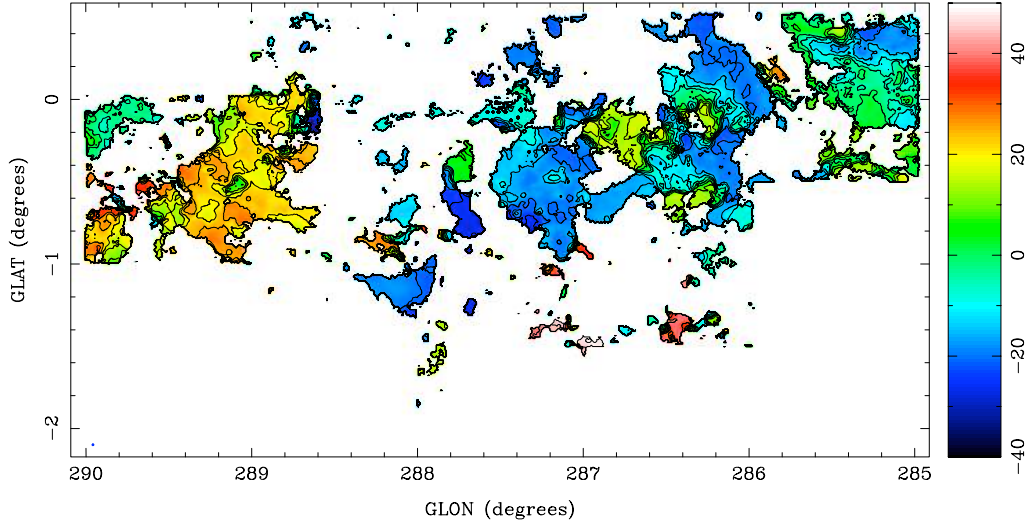,width=0.8\linewidth,angle=0}
\end{tabular}
\caption{$\co$ mean velocity map of the region covered by our Mopra observations.  The colour bar is in units of $\kms$.  The range of the contours is from -25 $\kms$ to 45 $\kms$ spaced by 5 $\kms$.  The red regions located at low latitude correspond to gas not associated with the CNC-Gum 31 molecular complex (see also Figure \ref{carina_pv}).}
\label{carina12co_vel}
\end{figure*}

We identify several apparently independent CO complexes in the area covered by our observations.  At the coordinates $(l,b) \sim (289.1\degrees,-0.4\degrees)$ we observe $\co$ emission extended over $\sim 1\degrees \times 1\degrees$.   Moving away from the direction to the Galactic Center, no significant CO emission is detected at our sensitivity limit for $\sim 0.4\degrees$ (16 pc at a distance of 2.3 kpc) until the Southern Pillars at $(l,b) \sim (288\degrees,-1.1\degrees)$.  The GMC formed by the Southern and Northern clouds surrounding $\eta$ Car covers $\sim 1\degrees \times 1\degrees$, centred at $(l,b) \sim (287.5\degrees,-0.5\degrees)$.  This molecular cloud is spatially and kinematically connected (see Section \ref{kinematics}) to the gas surrounding the \hii\ region Gum 31 at $(l,b) \sim (286.2\degrees,-0.2\degrees)$.  A bridge of material of length $\sim 15$ pc connecting these two regions is observed at $(l,b) \sim (286.7\degrees,-0.5\degrees)$.  The molecular complex of the Carina Nebula and Gum 31 represents the most prominent structure in this part of the Galactic Plane, covering $\sim$ 80 pc in longitude, and $\sim$ 60 pc in latitude.  Finally, our $\co$ maps show emission at $(l,b) \sim (285.3\degrees, 0\degrees)$, which extends over one degree in latitude.

The integrated intensity map of $\cother$ is shown in Figure \ref{carina13co}.  The distribution of $\cother$ is similar to the $\co$ image shown in Figure \ref{carina12co}, with the major features in $\co$ also seen in Figure \ref{carina13co} but less extended due to S/N differences between the two maps.  Significant $\cother$ emission is detected at $(l,b) \sim (289.1\degrees,-0.4\degrees)$, and it is spatially coincident with the strongest $\co$ emission features.  $\cother$ is also detected at the Southern Pillars, the Southern and the Northern clouds and in the bridge of material connecting Carina Nebula and Gum 31.  Extended $\cother$ emission is detected over the region located at $(l,b) \sim (285.3\degrees, 0\degrees)$.

Figure \ref{spect_reg} also shows the line profile of $\cother$ over the three regions shown in Figure \ref{carina13co}.  There is a good correspondence between $\co$ and $\cother$ line shapes in all the targeted regions.  In Region 1, the $\cother$ line shows a single peak at $-19\ \kms$, the same velocity of the peak observed in the $\co$ line.  In the case of Region 2, the $\cother$ line shows multiple peaks at the same velocities observed in the $\co$ line.  In Region 3, the $\cother$ profile traces the emission at positive velocities also observable in the $\co$ emission line.
   
\begin{figure*}
\centering
\begin{tabular}{c}
\epsfig{file=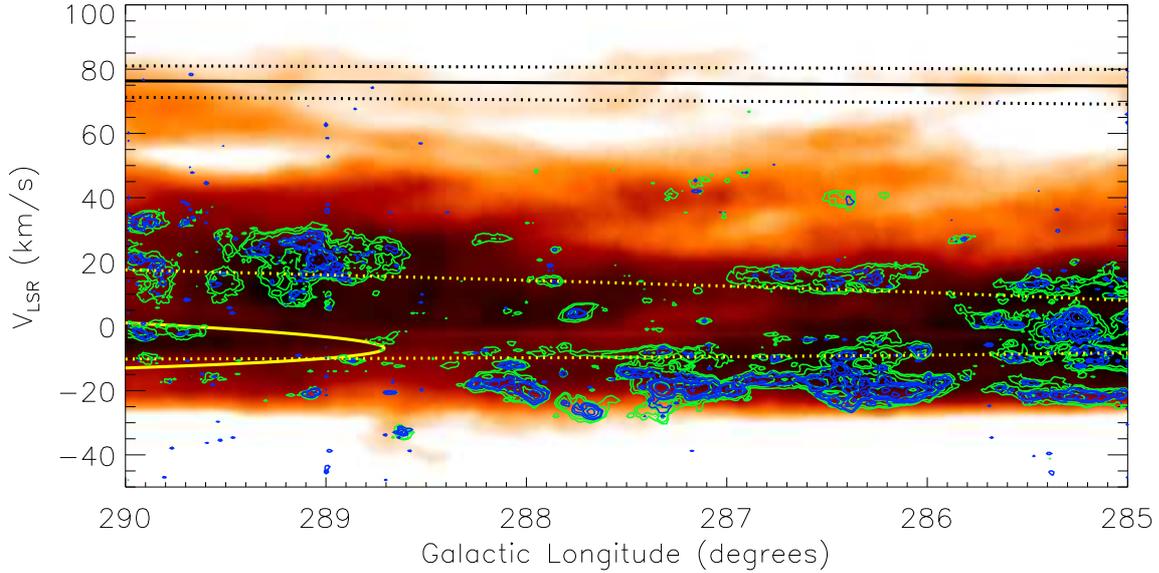,width=0.45\linewidth,angle=90}
\end{tabular}
\caption{Position-velocity diagram of the region covered by our Mopra observations.  The colour map shows the atomic hydrogen distribution.  Green contours illustrate the $\co$, and blue contours show $\cother$.  The data have been averaged over the central degree in latitude.  The yellow solid and dotted line show the mean position and the outer edge, respectively, of the Sagittarius-Carina spiral arm for a four-arm Milky Way model (see Figure \ref{carina_model} for a diagram of the model).  Black solid line shows the mean position of the Perseus arm in the P-V diagram, while the outer and the inner edge are shown by the black dotted lines.}
\label{carina_pv}
\end{figure*}

\begin{figure*}
\centering
\begin{tabular}{cc}
\epsfig{file=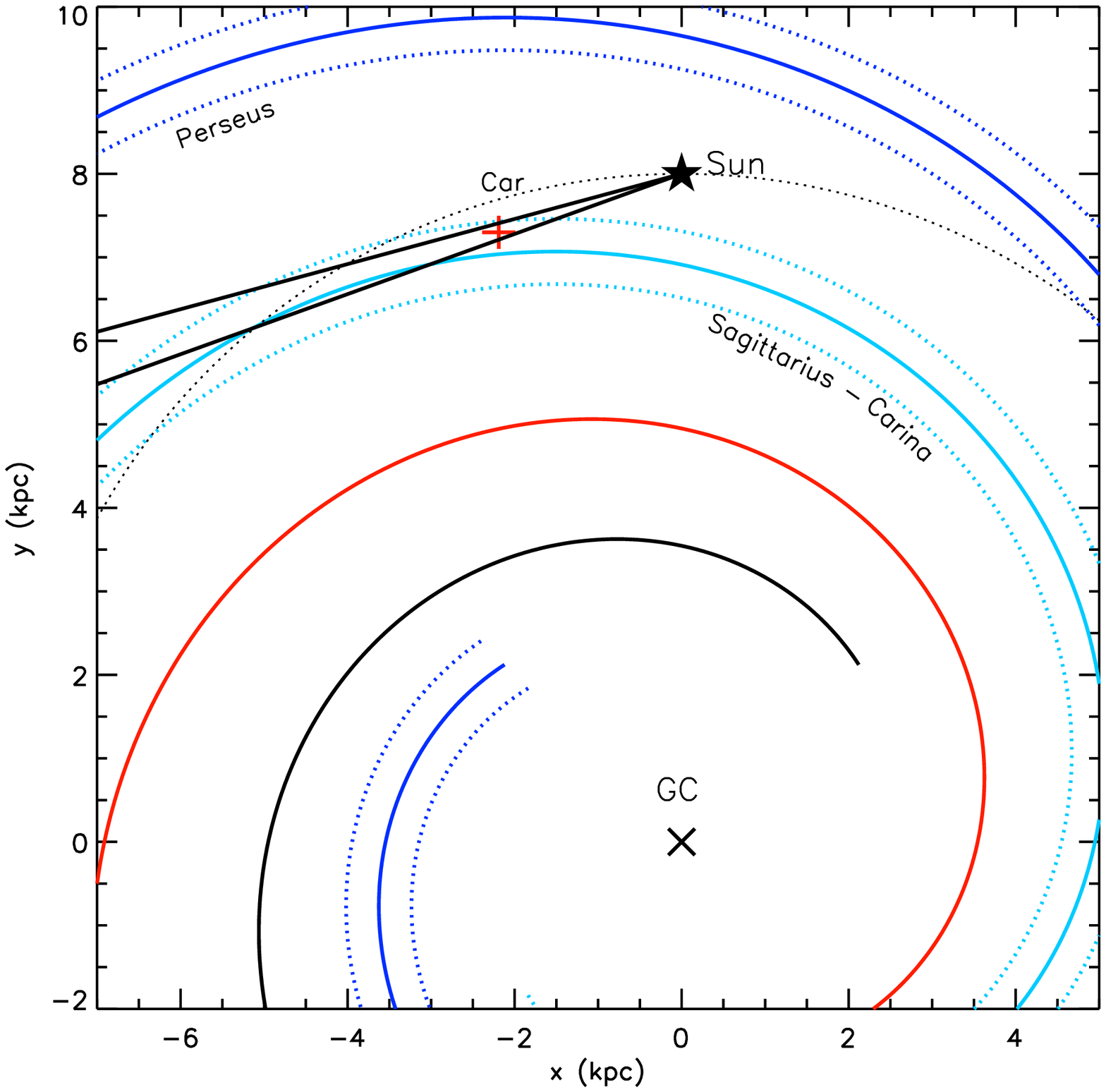,width=0.4\linewidth,angle=90} &
\epsfig{file=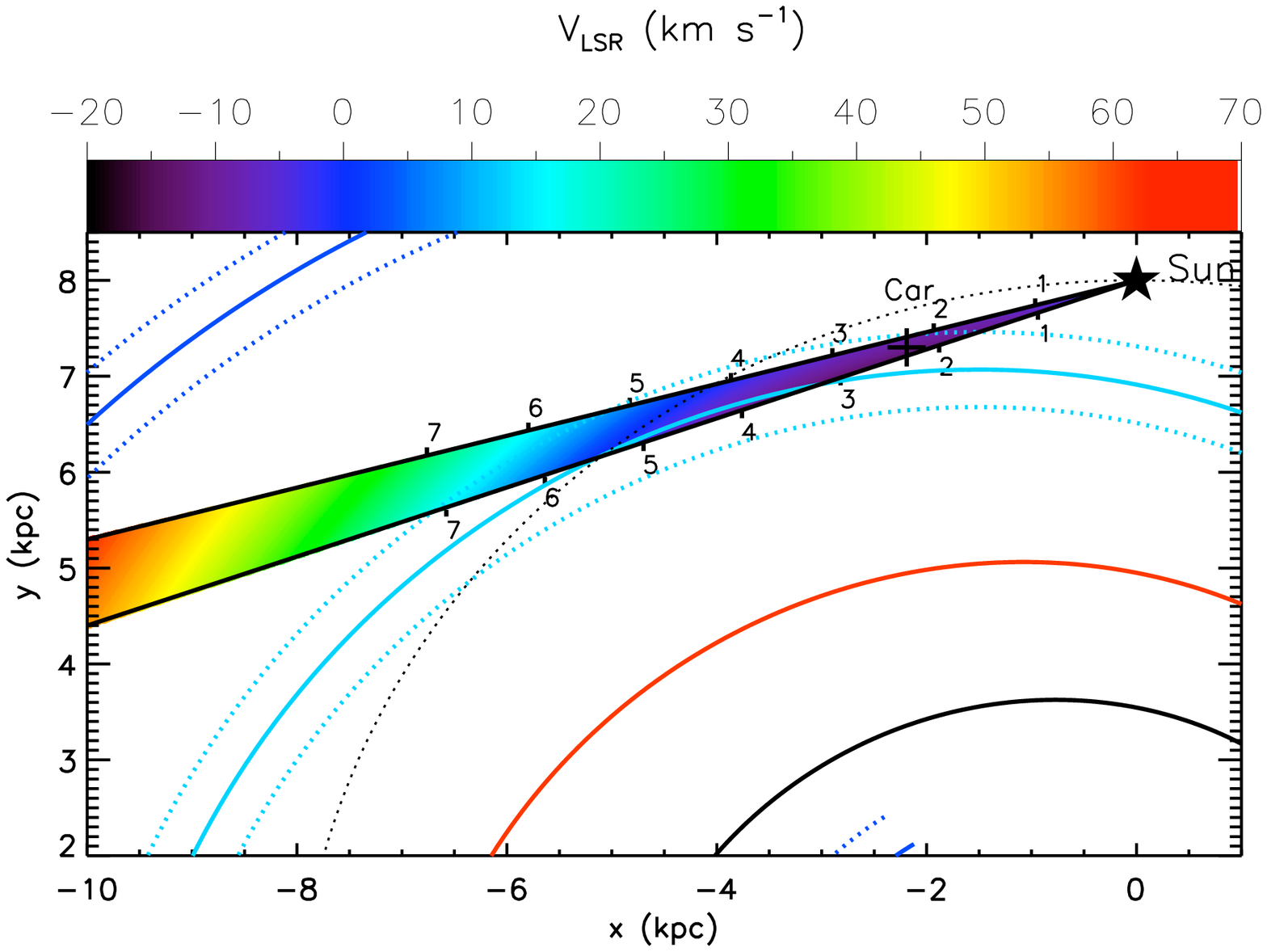,width=0.55\linewidth,angle=0} 
\end{tabular}
\caption{Left: Schematic configuration of a four-arm model of the Milky Way using the parameters from \citet{2014AJ....148....5V}.  The position of the Sagittarius-Carina and the Perseus arms are shown by the light-blue and blue lines respectively.  The black star show the position of the Sun and the black dotted line illustrates the solar circle.  The red cross shows the position of the CNC.  The longitude coverage of our observations is shown by the black lines.  Right:  Closer view of the region that covers the solar circle and the CNC.  The colour map shows the velocity along the line of sight with respect to position of the Sun.  Numbers mark the distance to the Sun in kpc at different positions along the region covered by our observations.}
\label{carina_model}
\end{figure*}

\subsubsection{Velocity field map of $\co$}\label{kinematics}
The intensity weighted mean velocity along the line of sight for the $\co$ line is shown in Figure \ref{carina12co_vel}.  There is a clear progression from positive velocities ($\sim$ 25 $\kms$) at $l\sim 290\degrees$ towards negative velocities ($\sim -20\ \kms$) at the position of the CNC and Gum 31.  However, molecular gas with velocities $\sim$ 20 $\kms$ is also detected in this region.  As we will see below, this molecular gas is probably not related to the CNC-Gum 31 region, and is hence located in the far side of the Sagittarius-Carina spiral arm.  Molecular gas detected at $(l,b) \sim (285.3\degrees,0\degrees)$ has velocities between $-20\ \kms$ to 20 $\kms$.  

In order to provide a better picture of the velocity distribution across the observed region, in Figure \ref{carina_pv} we show the position-velocity (PV) diagram for \HI, $\co$ and $\cother$ maps.  The \HI\ map was obtained from the Southern Galactic Plane Survey (SGPS, \citealt{2005ApJS..158..178M}).  The PV diagrams are created by averaging the line emission maps over 2$\degrees$ in latitude.  While atomic hydrogen is broadly distributed across all the line of sight velocities, the distributions of the molecular gas tracers $\co$ and $\cother$ present a clumpy structure limited to narrower velocity ranges.  

To visualise the position of the different gas tracers, in Figure \ref{carina_pv} we plot the PV diagram of the positions of the spiral arms assuming a four-arm structure model of the Milky Way (\citealt{2014AJ....148....5V}).  Additionally, we include the inner and outer edge of each arm assuming a 400 pc arm width.  As in \citet{2015PASA...32...20B}, we have used the parameters of the Galaxy model from \citet{2014AJ....148....5V}, with a pitch angle of 12.5$\degrees$, a central bar length of 3 kpc, and a distance of the Sun from the Galactic Center of 8 kpc.  The rotation curve model is taken from \citet{2007ApJ...671..427M}.  Figure \ref{carina_model} illustrates the model of the Galaxy as seen from the North Galactic pole (similar to Figure 2 in \citealt{2014AJ....148....5V}).  Moreover, Figure \ref{carina_model} shows the corresponding velocity respect to the local standard rest frame ($V_\mathrm{LSR}$) at each distance to the Sun over the longitude range of our observations.  The two spiral arms that intercept this longitude range are Sagittarius-Carina (the closest to the Sun) and the Perseus spiral arm (in the outer part of the Galactic disk).  The transition between negative to positive velocities over the Sagittarius-Carina spiral arm is also evident, as the spiral arm crosses the solar circle distance from the Galactic Center.

In Figure \ref{carina_pv} it is evident that the location of the $\co$ and $\cother$ emission is consistent with the position of the outer edge of the Sagittarius-Carina spiral arm.  The velocity of molecular gas in the CNC (about $-20\ \kms$) is consistent with the location of the near side, with a mean distance of $\sim 2.3$ kpc.  The molecular gas with velocity $\sim 15\ \kms$ with the same longitude range as the CNC-Gum 31 complex is located on the far side of the outer edge at a distance of $\sim 6$ kpc.  According to the model, the CO complex at $l\sim 289\degrees$ with velocity $\sim$ 25 $\kms$ is located at the far side of the outer edge at a distance of $\sim 6.8$ kpc.  This CO complex has been linked to a star-forming complex dominated by the \hii\ region Gum 35 (\citealt{2000A&A...357..308G}).  Finally, CO emission detected at $l \sim 285.3\degrees$ with velocities between $-20\ \kms$ to 20 $\kms$ is located between 3 and 6 kpc from the Sun.  This molecular structure appears to be a superposition of gas at different distances.  Although we do not detect CO at the position of the Perseus spiral arm, \HI\ line observations show emission at 80 $\kms$ which is consistent with the velocity range of the Perseus arm predicted by the Galaxy model.

Considering the positional picture revealed by the PV diagrams of the CO emission line and the Galaxy model shown in Figure \ref{carina_pv} and Figure \ref{carina_model} respectively, we can now identify the region associated with the CNC and Gum 31 molecular complex.  For the remainder of the paper, we define the region located between $l= 285.85\degrees$ to $288.5\degrees$ as the CNC and Gum 31 molecular complex and we will limit our analysis to this section of the Mopra CO survey.

\begin{figure*}
\centering
\begin{tabular}{cc}
\epsfig{file=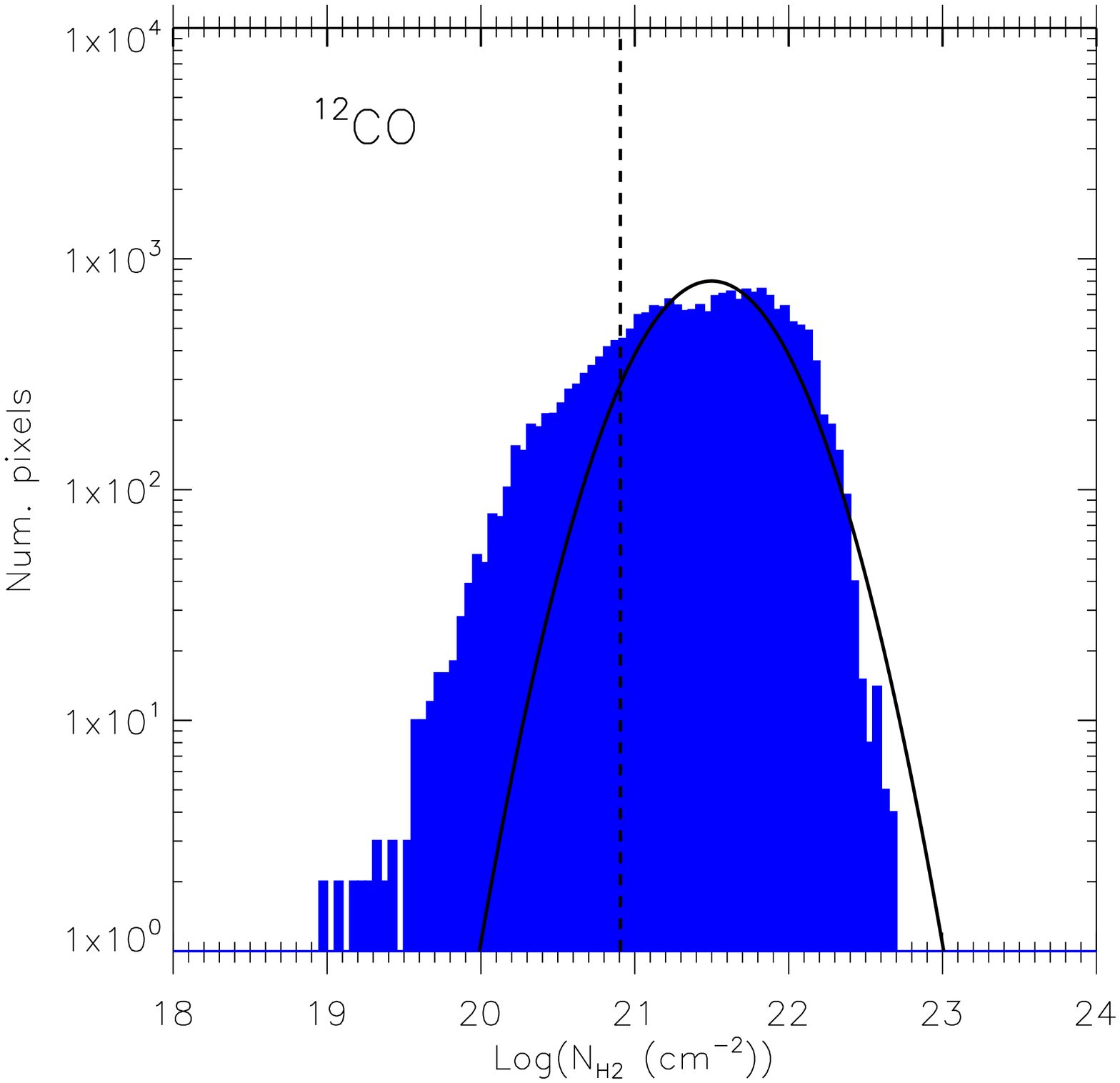,width=0.45\linewidth,angle=90} &
\epsfig{file=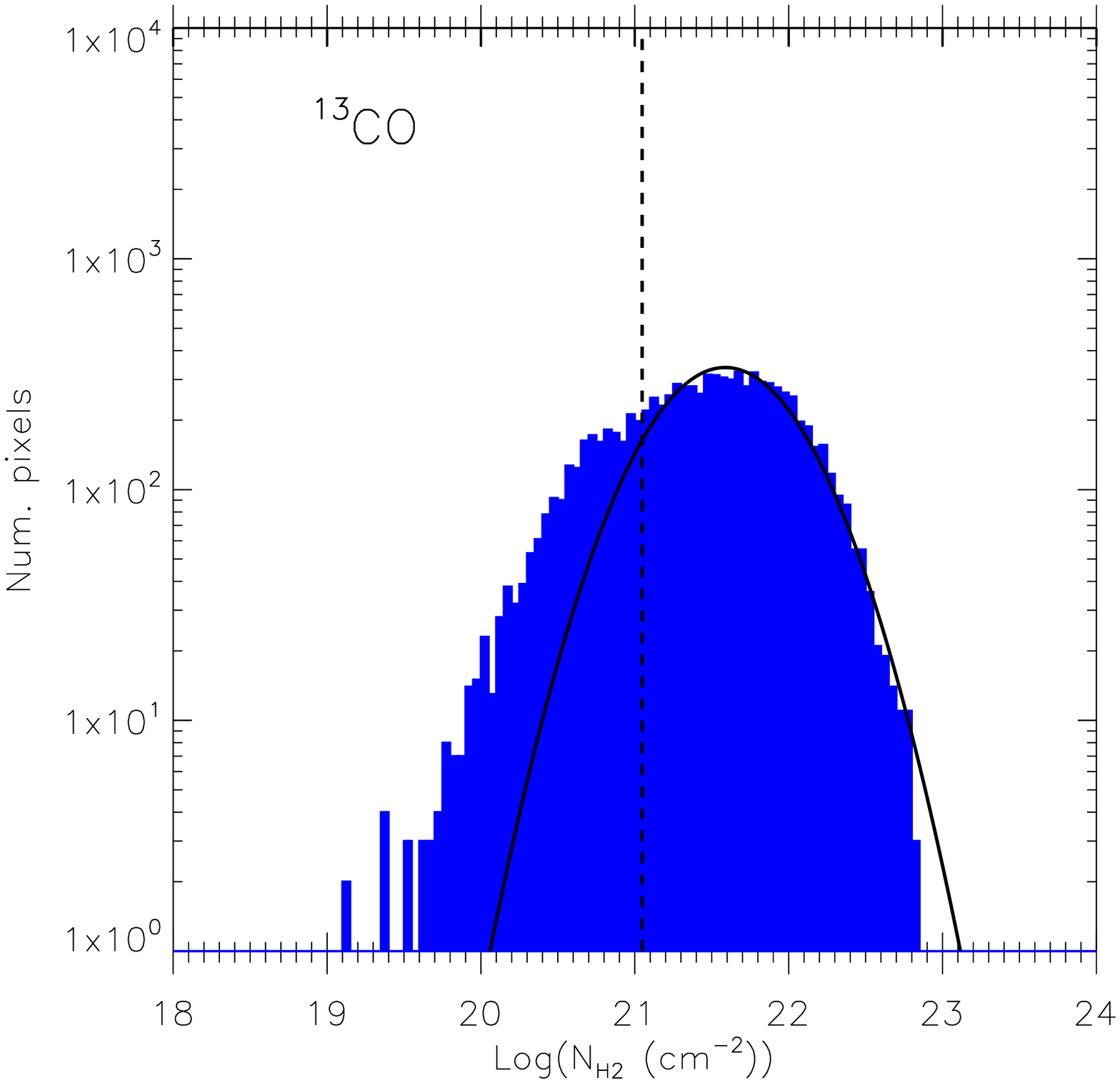,width=0.45\linewidth,angle=90}
\end{tabular}
\caption{Left: Molecular gas column density distribution obtained from the $\co$ integrated intensity map assuming a constant $X_\mathrm{CO}$ factor.  Right: Molecular gas column density distribution derived from the $\cother$ map assuming LTE conditions, and a constant $[\mathrm{H}_2/\mathrm{^{13}CO}]$ abundance ratio.  The vertical black dashed lines illustrate the 3$\sigma$ column density sensitivity limit.  The black solid lines illustrate the log-normal functions fit to the column density distributions.}
\label{hist_12co_13co}
\end{figure*}

Figure \ref{carina_pv} shows CO emission at a velocity $\sim 15-20\ \kms$ that is located at the same projected position of the CNC-Gum 31 region.  This emission is likely to be related to a molecular cloud in the far side of the Sagittarius-Carina arm, and should not be considered as part of the CNC-Gum 31 molecular complex.  In order to remove CO emission from background sources not related to the CNC-Gum 31 region, we have limited the velocity range of the CO emission to be from -45 to 0 $\kms$.

\subsection{Molecular gas column density}\label{ncol-co}

\subsubsection{Column density from $\co$}\label{ncol-12co}

Although H$_{2}$ is the most abundant molecule in the Universe, it is not directly observable in emission at the typical physical conditions present in the ISM of molecular clouds.  On the other hand, the CO molecule is easily observed at the low temperatures of these clouds, and it has become the preferred tracer to estimate the amount of H$_{2}$ in the Universe.  The molecular gas column density can be derived from the $\co$ integrated intensity map, $I_\mathrm{^{12}CO}$, using the CO-to-H$_{2}$ conversion factor, $X_\mathrm{CO}$, according to the equation

\begin{equation}\label{NHI_12co}
\frac{N_\mathrm{H_2}(^{12}\mathrm{CO})}{\mathrm{cm}^2}=\frac{X_\mathrm{CO}}{\mathrm{cm}^{-2} (\mathrm{K}\ \kms)^{-1} }\frac{I_\mathrm{^{12}CO}}{\mathrm{K}\ \kms}.
\end{equation}

\noindent The $X_\mathrm{CO}$ factor has been the subject of several studies of the Milky Way and external galaxies (see \citealt{2013ARA&A..51..207B} and references therein).  In the present paper, we have assumed a constant $X_\mathrm{CO}=2\times 10^{20}$ which is the canonical value for the MW (\citealt{2013ARA&A..51..207B}).  In a complementary study, we will investigate the spatial variation of the  $X_\mathrm{CO}$ factor across the CNC-Gum 31 region (Rebolledo et al. 2015, in preparation).

Figure \ref{hist_12co_13co} shows the distribution of $N_\mathrm{H_2}(^{12}\mathrm{CO})$.  The mean of the distribution is $\sim 4.5\times 10^{21}$ cm$^{-2}$, with a maximum $\sim 5.0\times 10^{22}$ cm$^{-2}$.  A $3 \sigma$ level sensitivity limit is estimated as follows.  We select a region in the $\co$ map with no significant emission.  The method to generate the integrated intensity maps (see Section \ref{moments}) is then applied to this emission-free region, integrating over the typical line width $\sim 2\ \kms$ of  the $\co$ line.  A rms value is estimated from this velocity integrated noise map.  Assuming the same $X_\mathrm{CO}$ factor in Equation \ref{NHI_12co}, the 3$\sigma$ level is found to be $\sim 8.1\times 10^{20}$ cm$^{-2}$.  In order to investigate the underlying shape of the column density distribution, we fit a log-normal function to the $N_\mathrm{H_2}$ distribution derived from the $\co$.  A log-normal distribution has been found for volume densities in several numerical simulations with different prescriptions for the physical conditions present in the ISM (\citealt{1994ApJ...423..681V}; \citealt{2001ApJ...546..980O}).  However, under certain conditions, a log-normal function can also describe the shape of the distribution of the column density (\citealt{2001ApJ...557..727V}).  In order to investigate whether a log-normal shape is a good description of the observed column density distribution, we have fitted a Gaussian function given by

\begin{equation}\label{12co_dist}
\mathrm{Num}(\mathrm{pixels})=\mathrm{Num}_\mathrm{peak}\times \exp(-\frac{(\log(N_\mathrm{H2})-\log(N_\mathrm{H2,peak}))^{2}}{2\times \delta_\mathrm{H2}^2}).
\end{equation}

The parameters of the log-normal function can be related to physical properties of the gas such as the Mach number, and mean magnetic field strength (\citealt{1997MNRAS.288..145P}; \citealt{2001ApJ...546..980O}; \citealt{2009ApJ...692...91G}).  However, the aim of this paper is to investigate how well the observed column density distributions are described by a log-normal function.  Thus, in the present study we do not examine in detail the physical meaning of the log-normal function parameters.  We have used the IDL MPFIT fitting package (\citealt{2009ASPC..411..251M}).  We have excluded $N_\mathrm{H_2}(^{12}\mathrm{CO})$ values below the $3 \sigma$ level from the fitting.  In Table \ref{log-norm} we show the resulting parameters of the fitted log-normal function, and Figure \ref{hist_12co_13co} plots the fit to the $N_\mathrm{H_2}(^{12}\mathrm{CO})$ distribution.  Considering  column density values above the sensitivity limit, we find that a log-normal function is a good approximation for $N_\mathrm{H_2} \lesssim 2.5\times 10^{22}$ cm$^{-2}$, but for larger column densities it over predicts the observed distribution.  Assuming that the $N_\mathrm{H_2}(^{12}\mathrm{CO})$ distribution should be log-normal, this discrepancy is probably because $\co$ is optically thick under the physical conditions found in molecular clouds.  Thus, the high-column density end of the distribution is not properly recovered by our $\co$ observations. 

\begin{table}
\caption{Parameters of the fitted log-normal function to the observed $N_\mathrm{H2}$ distributions.}
\centering
\begin{tabular}{lccc}
\hline\hline
Tracer  &  $\mathrm{Num}_\mathrm{peak}$ & $\log(N_\mathrm{H2,peak})$   &  $\delta_\mathrm{H2}$ \\
\hline
$\co$ CNC-Gum 31       & 801.2  &     21.5   &   0.41  \\ 
$\cother$ CNC-Gum 31 & 338.2 &     21.6   &   0.45 \\
Dust CNC-Gum 31 & 15601.5  &     21.5  &    0.30 \\  
Dust SP & 4410.3  &     21.5  &    0.27 \\  
Dust SC & 2326.5  &     21.6   &   0.18 \\  
Dust NC & 3111.4   &    21.6  &    0.31 \\  
Dust Gum 31 & 6790.5  &     21.5  &     0.30 \\  
\hline
\end{tabular}
\label{log-norm}
\end{table}

\subsubsection{Column density from $\cother$}\label{ncol-13co}
Because the $\co$ line is inherently optically thick, its optically thin isotopologue molecule, $^{13}$CO, provides a valuable alternative method for estimating the molecular gas column density.  If we assume local thermodynamic equilibrium (LTE) at the excitation temperature $T_\mathrm{ex}$, the column density of the $^{13}\mathrm{CO}$ molecule is given by (e. g., \citealt{2004tra..book.....R}; \citealt{2010ApJ...721..686P})

\begin{equation}\label{Nco-equ}
\frac{N(^{13}\mathrm{CO})}{\mathrm{cm}^2}=3.0\times 10^{14} \frac{T_\mathrm{ex}}{1-\exp(-5.3/T_\mathrm{ex})}\int{\tau_{13}}\mathrm{d}v,
\end{equation}

\noindent where $v$ is velocity in $\kms$, and $\tau_{13}$ is the optical depth of the $J=1\rightarrow0$ transition.  In this study we have determined $T_\mathrm{ex}$ from the $\co$ emission line, assumed to be optically thick, by using (e. g., \citealt{2004tra..book.....R})

\begin{equation}\label{tex-equ}
T_\mathrm{ex}=\frac{5.5}{\ln(1+5.5/(T_\mathrm{B}^{12}+0.82))},
\end{equation}

\noindent and assuming a filling factor of unity.  The optical depth of the $\cother$ line is given by

\begin{equation}\label{tau13-equ}
\tau_{13}=-\ln\left[1-\frac{T_\mathrm{B}^{13}}{5.3}\left(\frac{1}{\exp(5.3/T_\mathrm{ex})-1}-0.16\right)^{-1}\right].
\end{equation}

The molecular gas column density can be computed using,

\begin{equation}\label{Mlte-equ}
\frac{N_\mathrm{H_2}(^{13}\mathrm{CO})}{\mathrm{cm}^2}=\left[\frac{\mathrm{H}_2}{\mathrm{^{13}CO}}\right] N(^{13}\mathrm{CO})
\end{equation}

\noindent where $[\mathrm{H}_2/\mathrm{^{13}CO}]$ is the abundance ratio of $^{13}$CO molecule.  In this study we have assumed the abundance ratio to be $7\times10^5$.  This value is close to the ratio determined for the relationship from \citet{2005ApJ...634.1126M}, which yields $[\mathrm{^{12}CO/^{13}CO}]\sim 60$ for $R_\mathrm{gal} \sim 6$ kpc, if we assume an [$\mathrm{H_2/^{12}CO}$] abundance ratio of 1.1 $\times\ 10^4$ (\citealt{1982ApJ...262..590F}; \citealt{2010ApJ...721..686P}).



\begin{figure}
\centering
\begin{tabular}{c}
\epsfig{file=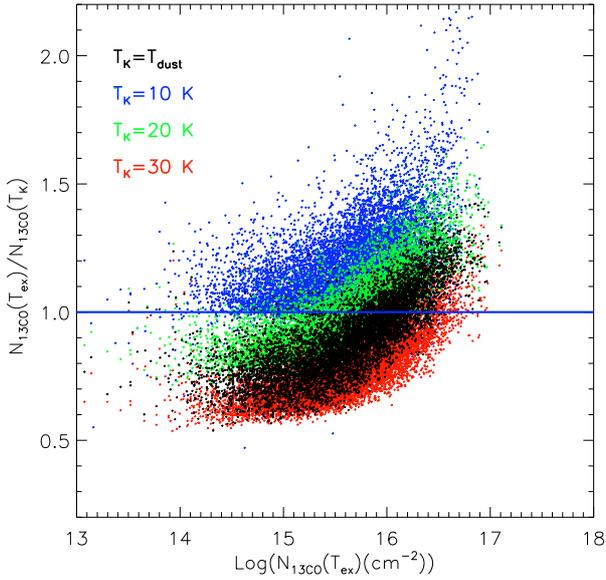,width=0.9\linewidth,angle=90} 
\end{tabular}
\caption{Ratio between the $N(^{13}\mathrm{CO})$ estimated using $T_\mathrm{ex}$ from Equation \ref{tex-equ} and $N(^{13}\mathrm{CO})$ calculated using alternative tracers of $T_\mathrm{ex}$, here labeled as $T_\mathrm{K}$.  Each point correspond to an individual pixel in the map.  The black points show the case $T_\mathrm{K}=T_\mathrm{dust}$.  We also consider the case where $T_\mathrm{K}$ is constant across the map.  The blue points show the case where $T_\mathrm{K}=10$ K, the green points illustrate $T_\mathrm{K}=20$ K and red points show the $T_\mathrm{K}=30$ K case.}
\label{hist_Temp_comp}
\end{figure}

Figure \ref{hist_12co_13co} also shows the $N_\mathrm{H_2}(^{13}\mathrm{CO})$ distribution.  In this case, the $3 \sigma$ level sensitivity limit is $1.1\times 10^{21}$ cm$^{-2}$.  The mean of the distribution is $\sim 5.5\times 10^{21}$ cm$^{-2}$, similar to the mean column density found from the $\co$ integrated map.  Following the same procedure as for the $N_\mathrm{H_2}(^{12}\mathrm{CO})$ distribution, we fit the log-normal function from Equation \ref{12co_dist} to the column density distribution derived from the $\cother$ line map, excluding values below the $3 \sigma$ level sensitivity limit.  The fit is plotted in Figure \ref{hist_12co_13co}, and the resulting parameters of the function are given in Table \ref{log-norm}.  In this case, the column density distribution above the sensitivity limit is well matched by a log-normal function for $N_\mathrm{H_2} \lesssim 6.3\times 10^{22}$ cm$^{-2}$.  This value is a factor of $\sim$ 2.5 larger than the limiting column density for which the $N_\mathrm{H_2}(^{12}\mathrm{CO})$ distribution is well described by a log-normal function.  This can interpreted as evidence for the $\cother$ line becoming optically thick at higher column densities than the $\co$ line.

One key factor in calculating the $N(^{13}\mathrm{CO})$ is the excitation temperature, $T_\mathrm{ex}$ (Equation \ref{Nco-equ}).  By assuming a filling factor of unity in Equation \ref{tex-equ}, we may be underestimating the true value of the excitation temperature due to beam dilution effects, reducing the column densities obtained from the $^{13}$CO map.  In order to assess the robustness of our column density calculations from $^{13}$CO, we use the dust temperature (see Section \ref{dust-mod}) as an alternative tracer of the $T_\mathrm{ex}$ in Equation \ref{Nco-equ}.  We have assumed that the dust and gas have the same temperature, and that the dust temperature estimates are less affected by beam dilution.  We take the ratio between the $N(^{13}\mathrm{CO})$ using $T_\mathrm{ex}$ from Equation \ref{tex-equ} and the $N(^{13}\mathrm{CO})$ estimated using the dust temperature as a tracer of the $T_\mathrm{ex}$.  In Figure \ref{hist_Temp_comp} we show the resulting ratio distribution.  This ratio has a mean of 0.90 and a standard deviation of 0.15.  Thus, in average, there is a $\sim$ 10-20 \% difference between the two approaches, confirming the robustness of the $N(^{13}\mathrm{CO})$ calculations.  For illustrative purposes only, we also consider the case when $T_\mathrm{ex}$ is constant across the map, which is a common assumption in the absence of measurements of $T_\mathrm{ex}$.  We use $T_\mathrm{ex}$ values of 10, 20 and 30 K.  Figure \ref{hist_Temp_comp} also shows the ratio between the $N(^{13}\mathrm{CO})$ estimated using $T_\mathrm{ex}$ from Equation \ref{tex-equ} and when $T_\mathrm{ex}$ is constant across the map.  Among the three $T_\mathrm{ex}$ constant values, the $N(^{13}\mathrm{CO})$ calculated assuming $T_\mathrm{ex}=20$ K provides the best match to the $N(^{13}\mathrm{CO})$ values estimated using $T_\mathrm{ex}$ from Equation \ref{tex-equ}.



\begin{figure*}
\centering
\begin{tabular}{cc}
\epsfig{file=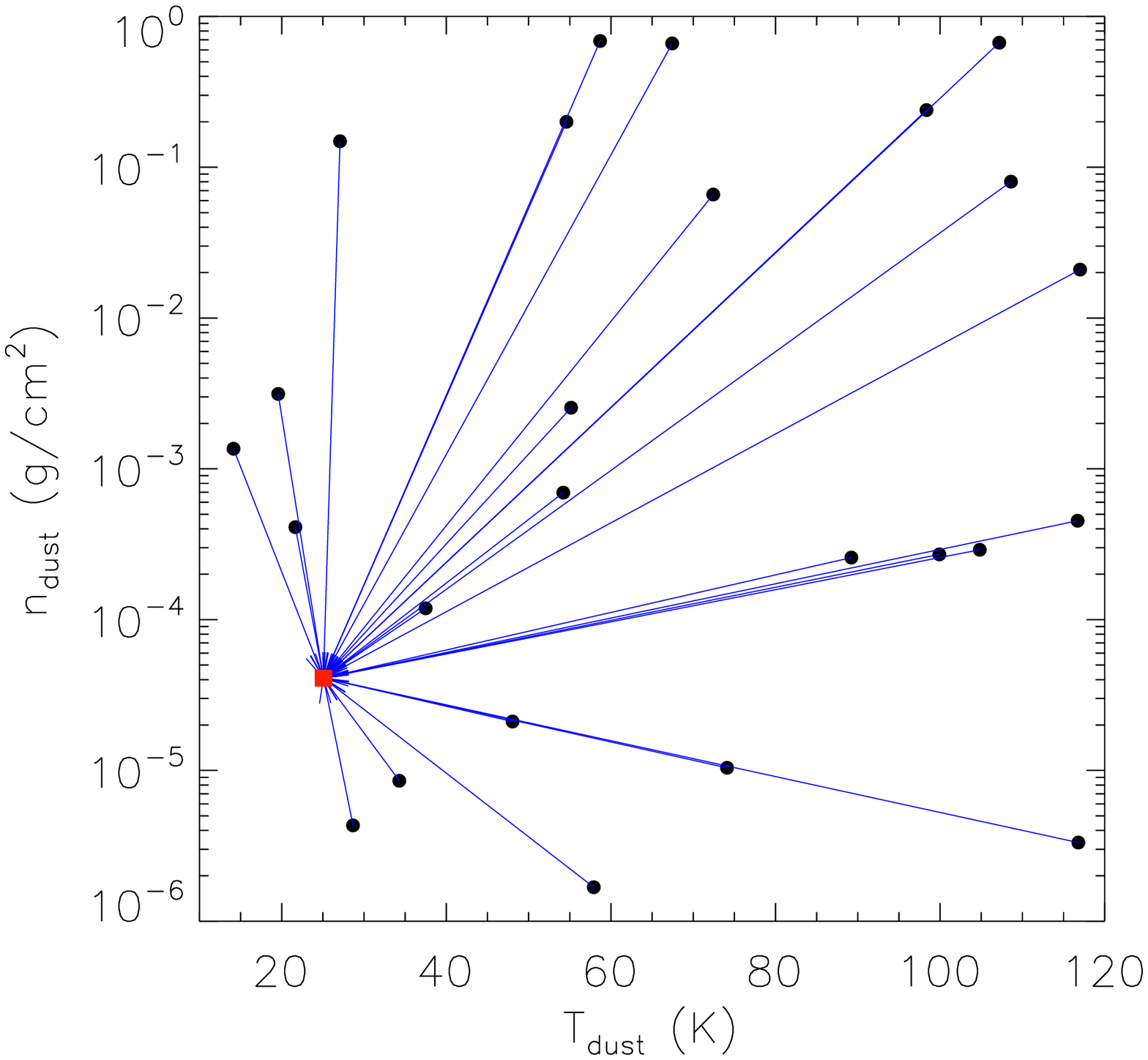,width=0.4\linewidth,angle=90} &
\epsfig{file=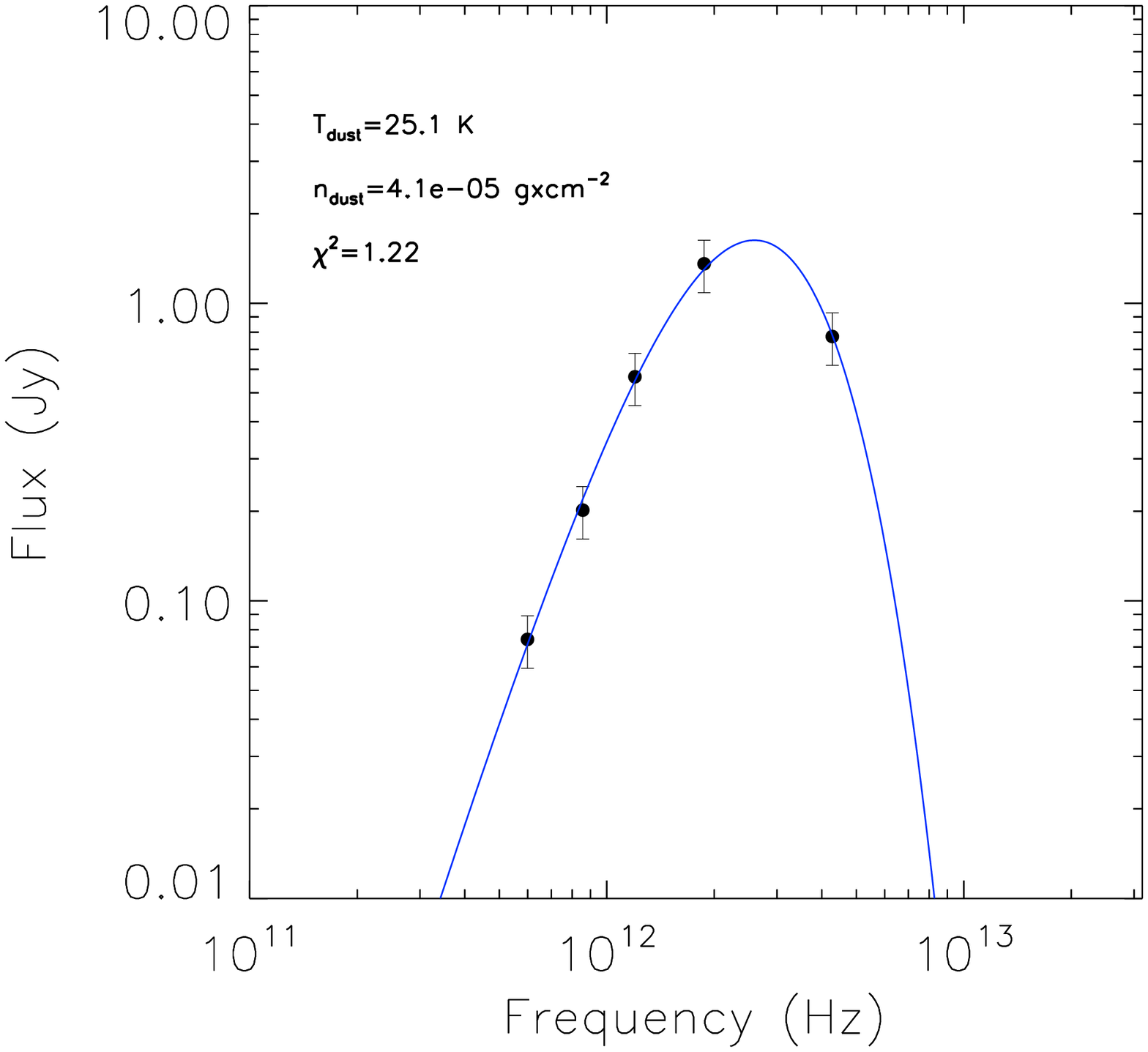,width=0.4\linewidth,angle=90} 
\end{tabular}
\caption{Left:  Diagram that illustrates the convergence of the free parameters ( $T_\mathrm{dust}$, $n_\mathrm{dust}$) in our SED fitting algorithm for a single pixel.  Black dots show the randomly generated initial values of $T_\mathrm{dust}$ and $n_\mathrm{dust}$.  Blue arrows connect each initial guess of each parameter to the final value yielded by the {\it MPFIT} procedure.  Right:  The black dots show the observed flux for each Herschel band.  Error bars are assumed to be 10\% of the total flux.  The blue line shows the resulting grey-body function built from the converged values shown in the upper left of the panel.}
\label{pixel_fit}
\end{figure*}

\section{Spectral Energy Distribution Fitting for the Infrared Images}\label{dust-mod}

\subsection{Dust emission model}
We have modelled the infrared dust emission using a grey-body of the form (\citealt{2004MNRAS.348..638B})

\begin{equation}\label{grey-body}
F_\nu=\Omega\ B_\nu(T_\mathrm{dust})\ \epsilon_\mathrm{dust},
\end{equation} 

\noindent where $T_\mathrm{dust}$ is the temperature of the dust and $\Omega$ is the angular size of the region being observed.  The dust emissivity, $\epsilon_\mathrm{dust}$, and the optical depth $\tau_\nu$ are related by the expression $\epsilon_\mathrm{dust}=(1-e^{-\tau_\nu})$.  The optical depth is proportional to the dust opacity $\kappa_\nu$ and to the mass column density $n_\mathrm{dust}$, i.\ e., $\tau_{\nu}=\kappa_\nu n_\mathrm{dust}$, where $\kappa_\nu$ is in units of cm$^2$ g$^{-1}$ and $n_\mathrm{dust}$ in units of g cm$^{-2}$.  Our approach assumes a power law relation between $\kappa_\nu$ and frequency given by

\begin{equation}\label{kappa-freq}
\kappa_\nu=\kappa_0 \left(\frac{\nu}{\nu_0}\right)^\beta,
\end{equation} 

\noindent where $\kappa_0$ is the dust opacity at the reference frequency $\nu_0$ and $\beta$ is the dust emissivity index.  Several authors have investigated the dependency of $\beta$ on several factors such as grain size and shape, dust temperatures and grain mixtures (\citealt{1983QJRAS..24..267H}; \citealt{1981ApJ...244..483M};  \citealt{1997ApJ...491..615G}).  In particular, $\beta$ is observed to be influenced by the temperature of the dust, with the dust emissivity decreasing with increasing dust temperature (\citealt{2001ApJ...553..604D}).  

 
If we further assume that the dust emission is optically thin at infrared wavelengths, the dust emissivity can be expressed as $\epsilon_\mathrm{dust}=\tau_\nu$.  Combining Equations \ref{grey-body} and \ref{kappa-freq}, the expression for the grey-body is given by 
 
\begin{equation}\label{grey-body2}
F_\nu=\Omega\ B_\nu(T_\mathrm{dust})\ \kappa_0 \left(\frac{\nu}{\nu_0}\right)^\beta\ n_\mathrm{dust}.
\end{equation} 

The gas column density is estimated by

\begin{equation}\label{ngas}
N_\mathrm{H_2}(\mathrm{dust})= \frac{n_\mathrm{dust}}{\mu_\mathrm{H_2}\ m_\mathrm{H}\ R_\mathrm{dg}} .
\end{equation} 

\noindent where $R_\mathrm{dg}$ is the dust-to-gas ratio assumed to be 0.01, and $\mu_\mathrm{H_2}$ is the mean molecular weight in units of the hydrogen atom mass equal to 2.72.  Equation \ref{ngas} assumes that all the gas traced by the dust emission is in molecular form.  This approximation has been adopted in order to compare the gas column density calculated from the dust emission with the column density derived from the molecular gas tracers $\co$ and $\cother$.  This assumption is certainly problematic in regions of low column density, especially at the the edge of the clouds where the gas is likely to be atomic dominated.

\begin{figure*}
\centering
\begin{tabular}{c}
\epsfig{file=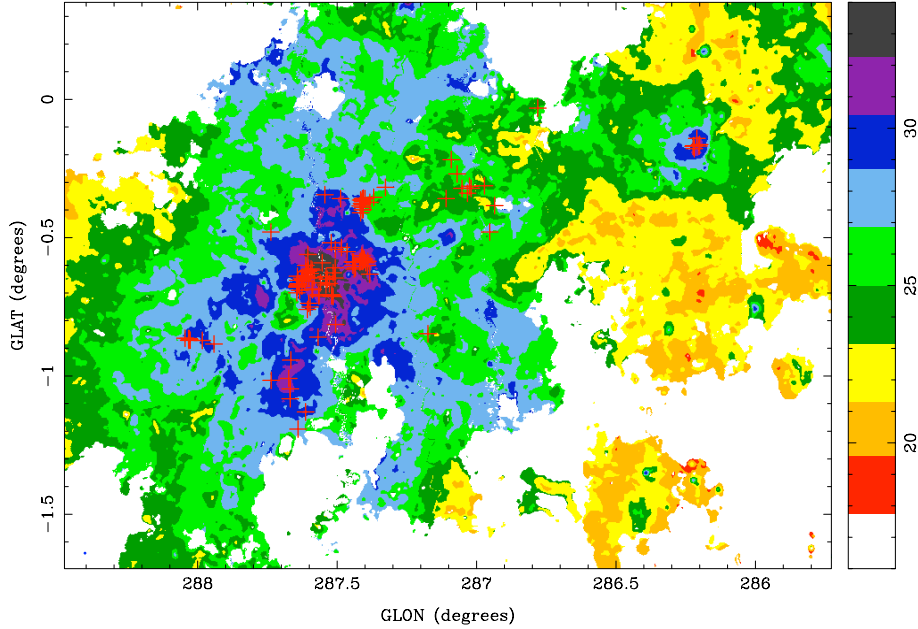,width=0.7\linewidth,angle=0}
\end{tabular}
\caption{Dust temperature map of the CNC and Gum 31 regions from the SED fitting of the Herschel maps.  Colour bar is in units of K, from 16 K to 36 K, with steps of 2 K for each bar colour.  The red crosses illustrate the positions of the high-mass stellar members of clusters present at the CNC-Gum 31 region as listed in \citet{2006MNRAS.367..763S} and \citet{2001A&A...371..107C}.  The influence of the massive stars is clearly seen in the variation of the dust temperature across the observed region.}
\label{carina_temp}
\end{figure*}

\subsection{Fitting approach}\label{sed-fit}
Equation \ref{grey-body2} is fitted to each pixel of the infrared images using the IDL MPFIT fitting package.  We exclude from our analysis pixels with emission less than $3 \sigma$, with $\sigma$ estimated for regions with no significant emission.  Masking at the $3 \sigma$ level gives estimates of the column density sensitivity limit to be $\sim 1.3 \times 10^{21}$ cm$^{-2}$.  All the Herschel images have been degraded to the grid and resolution of  the $500\ \mu$m image, which corresponds to $11\farcs5$ and $36\farcs5$ respectively.  The convolution of the Herschel maps was achieved using a gaussian kernel.  We use the function {\it mpfitfun} to fit Equation \ref{grey-body2} to each pixel, leaving $T_\mathrm{dust}$ and $n_\mathrm{dust}$ as free parameters.  In this work, we have used the value of the dust opacity from \citet{1994A&A...291..943O} for the standard MRN distribution without ice mantels (\citealt{1977ApJ...217..425M}).  At 500 $\mu$m, they estimate $\kappa$=1.77 cm$^2$ g$^{-1}$, which is used in our analysis.  For simplicity, we also assume a fixed value of $\beta=2$.  

To avoid dependence on the initial guesses of the free parameters, we defined an initial range of values for $T_\mathrm{dust}$ and $n_\mathrm{dust}$.  For the dust temperature, the range is 10 K $< T_\mathrm{dust} < $ 120 K.  The dust mass column density range is restricted to $10^{-6}\ \mathrm{g\ cm}^{-2} <  n_\mathrm{dust} < 1\ \mathrm{g\ cm}^{-2}$.  Randomly selected initial values for dust temperature and dust mass column density within the corresponding ranges were used and the convergent solution recorded for each parameter.  After iterating 25 times, a distribution of the resulting $T_\mathrm{dust}$ and $n_\mathrm{dust}$ parameters is generated.  The peak of the distribution for each parameter is then assigned to the corresponding pixel.  For example, Figure \ref{pixel_fit} shows the distribution in the parameter space of the randomly generated set of initial values of $T_\mathrm{dust}$ and $n_\mathrm{dust}$ along with the converged values for one pixel.  Figure \ref{pixel_fit} also shows the resulting fitted SED function.

\subsection{Masking procedure for the dust maps}\label{mask-dust}
Besides the $3 \sigma$ level masking that is applied before the SED fitting, we also mask out pixels based on the goodness of the fit of our SED model.  We only include pixels in our analysis where the SED fitting algorithm yields $\chi^{2}$ values below 20.  In this way, noisy pixels are removed from the fitting procedure, and simultaneously pixels are removed if the SED dust model is not a good description of the infrared flux spectral distribution.

\begin{figure*}
\centering
\begin{tabular}{c}
\epsfig{file=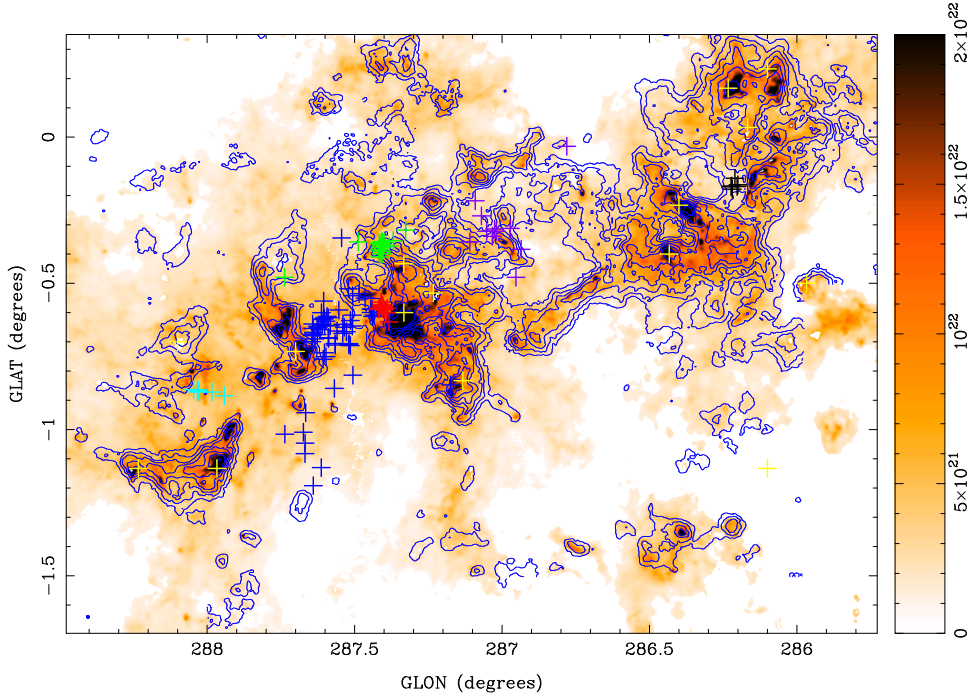,width=0.6\linewidth,angle=-90}
\end{tabular}
\caption{Gas column density map of the CNC and Gum 31 region from the SED fitting of the Herschel maps.  Colour bar is in units of $\mathrm{cm}^{-2}$.  For comparison, contours show the integrated intensity of the $\co$ line, and they are spaced by $n^2$ K $\kms$, with $n=2, 4, 6, 8$.  The blue crosses show the Trumpler 16 and Collinder 228 stellar clusters, red crosses show Trumpler 14, green show Trumpler 15, violet shows Bochum 10 and cyan crosses show Bochum 11 cluster, all of them listed in \citet{2006MNRAS.367..763S}.  The black crosses show the star member of the NGC 3324 cluster listed in \citet{2001A&A...371..107C}.  The yellow crosses show the positions of the C$^{18}$O cores found by \citet{2005ApJ...634..476Y}.  A good correlation between the regions with the highest gas column density from dust SED fitting and the $\co$ emission map is found.}
\label{carina_ngas}
\end{figure*}

\subsection{Dust temperature}\label{ncol-tdust}
Figure \ref{carina_temp} shows the resulting dust temperature map from the SED fitting.  We have included the positions of the high-mass stellar members of the star clusters present in the CNC region listed in \citet{2006MNRAS.367..763S} and in the Gum 31 region listed in \citet{2001A&A...371..107C}.  A clear temperature gradient is seen across the observed region, with the highest temperatures (> 30 K) corresponding to the dust located near the star clusters Trumpler 14, Trumpler 15 and Trumpler 16, and in the vicinity of the star cluster NGC 3324 in the Gum 31 region.  The regions of colder dust (< 26 K) are coincident with areas of high column density, except for the eastern section of the Northern Cloud which is heated by the star clusters Trumpler 16 and Trumpler 14.

\begin{figure*}
\centering
\begin{tabular}{cc}
\epsfig{file=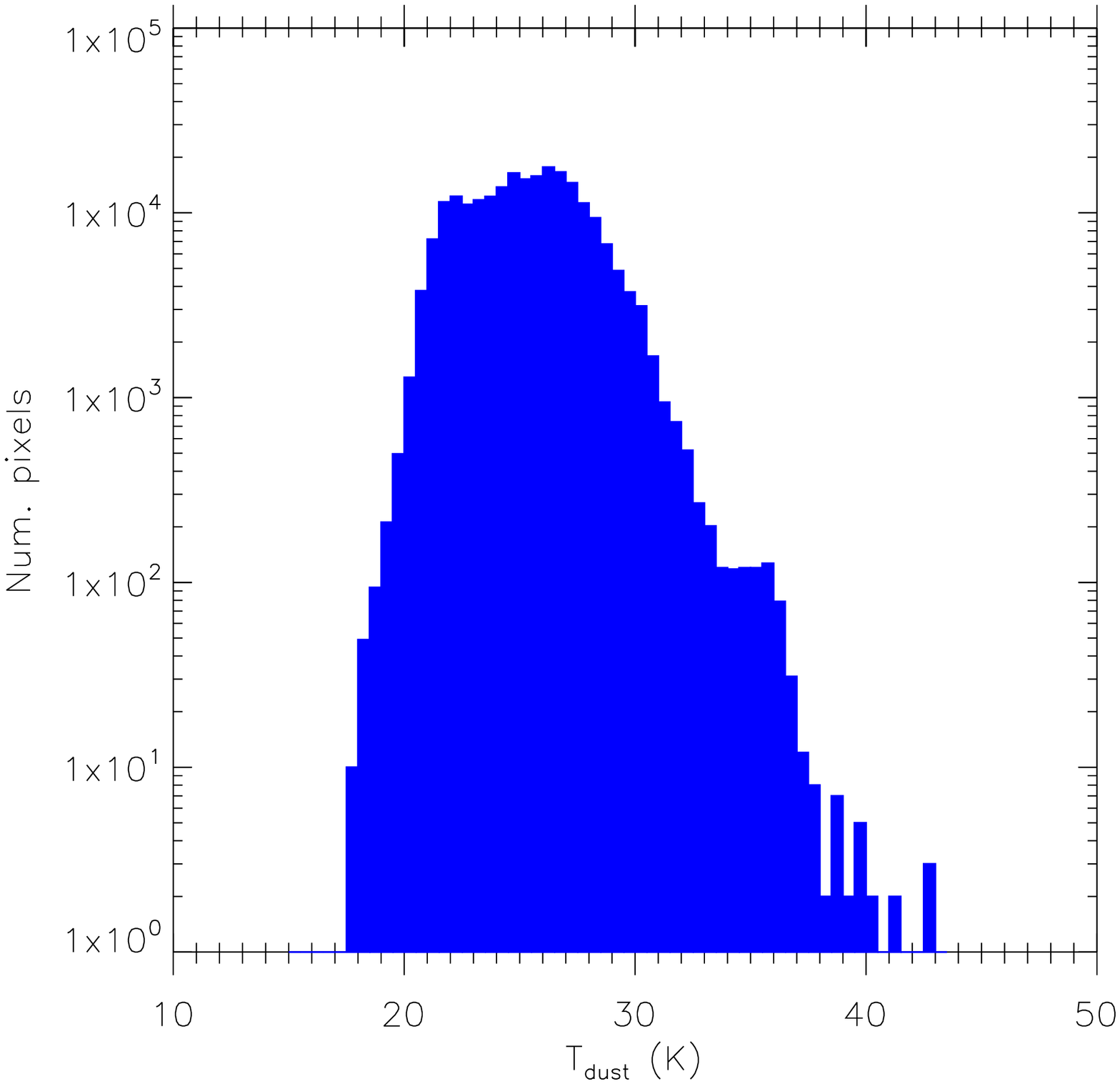,width=0.4\linewidth,angle=90} &
\epsfig{file=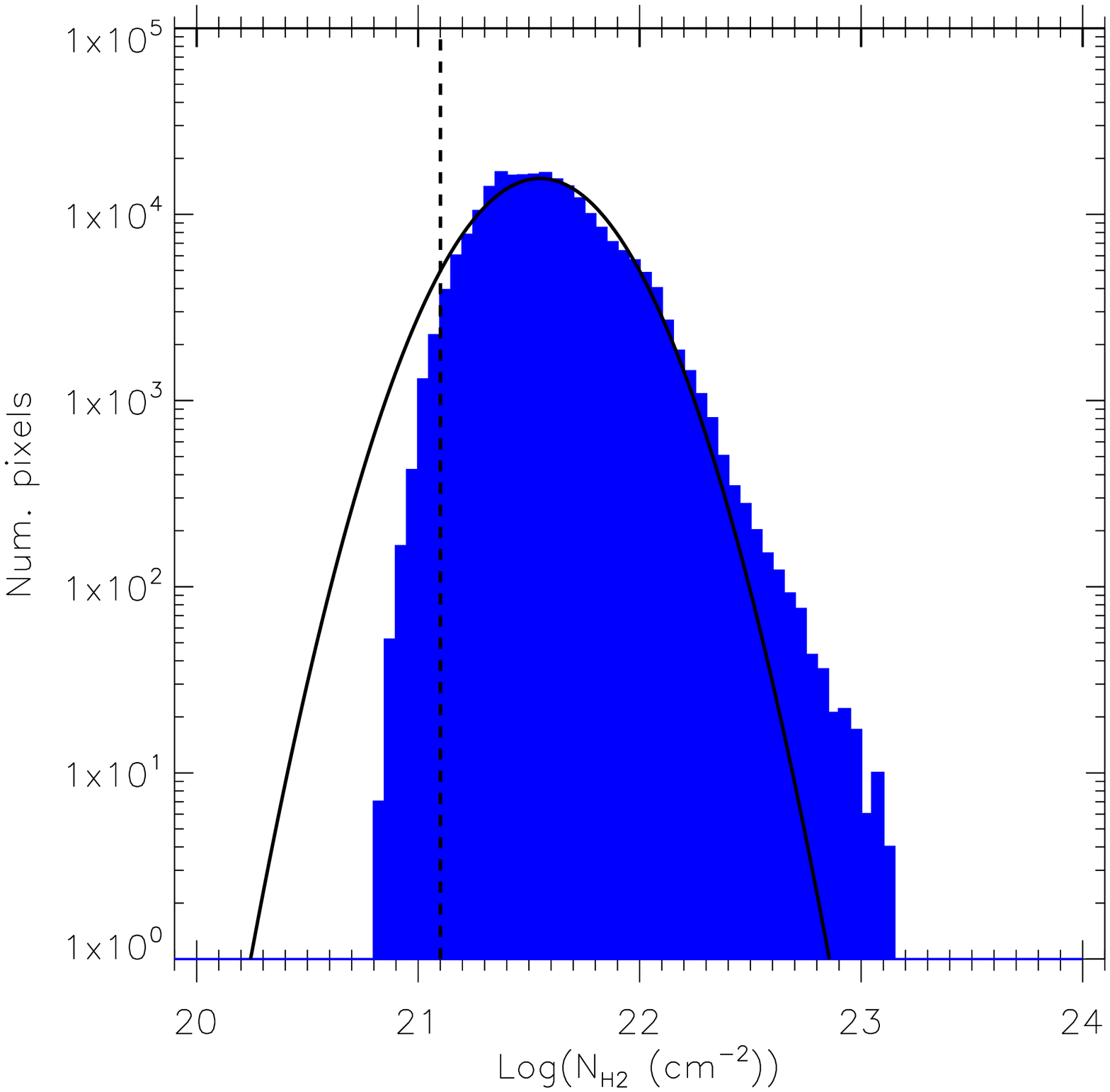,width=0.4\linewidth,angle=90}
\end{tabular}
\caption{Distributions of the dust temperature (left) and the gas column density (right) obtained from our SED fitting algorithm.  The black solid line shows the log-normal function fit to the column density distribution.  The vertical black dashed line illustrates the column density sensitivity limit of our SED fitting algorithm.}
\label{hist_temp_ndust}
\end{figure*}

Figure \ref{hist_temp_ndust} shows the dust temperature distribution for the CNC-Gum 31 complex.  The distribution has values in the range from $17$ K to $43$ K, and shows a multiple-peak shape, with peaks at $T_\mathrm{dust} \sim 22$, 27, and 36 K.  It is similar to the dust temperature distribution found by \citet{2012A&A...541A.132P} using colour temperatures derived from the ratio between the 70 $\mu$m and the 160 $\mu$m Herschel maps.  However, we do not find the high-temperature tail for $T_\mathrm{dust} > 40$ K they detect in the CNC region.  \citet{2013A&A...554A...6R} performed a SED fitting to Herschel maps to obtain detailed maps of the dust temperature and gas column density of the CNC, in an approach similar to the method used in this paper.  Their dust temperature map is basically consistent with the results presented in Figure \ref{carina_temp}, and the temperatures they find are from $20$ K to $40$ K, similar to the range recovered by our algorithm.

\subsection{Gas column density from dust emission}\label{ncol-colden}
Figure \ref{carina_ngas} shows the resulting gas column density map from our SED fitting approach.  For comparison, we have overlaid the contours of the $\co$ integrated intensity.  A clear spatial correlation is observed between regions of high density column density and CO emission.  This correlation is stronger for regions with $N_\mathrm{H_2} > 3\times\ 10^{21}\ \mathrm{cm}^{-2}$.  The densest regions are found in the eastern part of the Northern Cloud and the Southern Pillars, with column density $N_\mathrm{H_2} > 10^{22}\ \mathrm{cm}^{-2}$.  

Figure \ref{hist_temp_ndust} shows the distribution of the gas column density derived from the dust emission SED.  The gas column density has a range from $6.3\times\ 10^{20}\ \mathrm{cm}^{-2}$ to $1.4\times 10^{23}\ \mathrm{cm}^{-2}$.  The $N_\mathrm{H_2}(\mathrm{dust})$ distribution has a mean of $\sim 4.8 \times 10^{21}$ cm$^{-2}$, similar to the values derived from $\cother$.  As in Sections \ref{ncol-12co} and \ref{ncol-13co} for the molecular gas column density distributions derived from $\co$ and $\cother$ respectively, we have fitted a log-normal function to the $N_\mathrm{H_2}(\mathrm{dust})$ distribution.  We have excluded gas column density values below the sensitivity limit in the fitting precedure.  The fitted function is shown in Figure \ref{hist_temp_ndust}, and the resulting parameters are shown in Table \ref{log-norm}.  Overall, the distribution is not well described by a log-normal shape at the low and high column density ends.  For low column densities, the log-normal function overpredicts the number of pixels in the observed distribution.  This difference is likely to be an effect of the sensitivity limit of our gas column density map which is $\sim 1.3 \times 10^{21}$ cm$^{-2}$, close to the value where the log-normal function and the observed distribution start to differ.  For the high column density regime, the $N_\mathrm{H_2}$ distribution deviates from the log-normal shape at $\sim 1.0 \times 10^{22}$ cm$^{-2}$.  For $N_\mathrm{H_2}$ larger than this value, the observed column density distribution clearly shows a tail with respect to the log-normal distribution.  Similar tails at high column density have been observed for other molecular clouds (\citealt{2010ApJ...721..686P}), and it is thought that such high-column density tails are present in regions of active star formation (\citealt{2009A&A...508L..35K}).

\section{Discussion}\label{discuss}

In order to investigate the variation of the gas properties across the CNC-Gum 31 molecular complex, several distinct regions are analysed.  Similar to the spatial CO distribution described by \citet{2007MNRAS.379.1279S}, we have identified the regions as follows: the Southern Pillars (SP), the Southern Cloud (SC), and the Northern Cloud (NC).  Additionally, we have included the Gum 31 region which was identified by \citet{2005ApJ...634..476Y} as {\it region 4} and {\it region 5}.  In Figure \ref{carina.500um.obs} we show the extent of each region considered in this study.

\begin{figure}
\centering
\begin{tabular}{c}
\epsfig{file=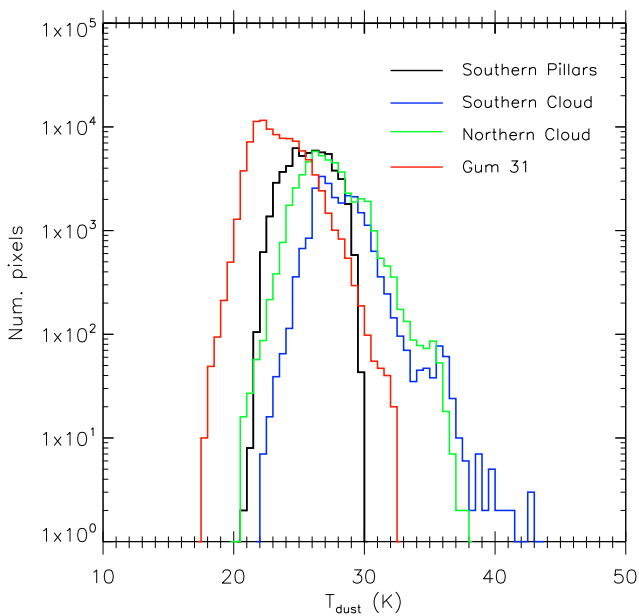,width=0.95\linewidth,angle=90} 
\end{tabular}
\caption{Distributions of the dust temperature for each of the regions identified in this study.  The black line shows the temperature distribution for the Southern Pillars, the blue line shows the Southern Cloud, the green line shows the Northern Cloud and the red histogram is for the Gum 31 region.}
\label{hist_tdust_reg}
\end{figure}

\subsection{Regional variation of the dust temperature}\label{tdust-var}
Figure \ref{hist_tdust_reg} shows the distribution of the dust temperature for each region identified in this study.  In the case of the SP, the dust temperature shows values from 21 to 30 K, with a mean temperature of 26 K.  In Figure \ref{carina_temp} the region of warmer dust is dominated by the star clusters Trumpler 14 and 16 which are located at the centre of the CNC.  Because the SC is located between the SP and the position of these massive star clusters, the material of the SP is less affected by the strong radiation field from the massive OB stars inside clusters Trumpler 14 and 16, when compared to the SC and NC which are located near these clusters.  In the inner part of the dust pillars, the dust temperature can reach values as cold as $\sim 22$ K.    

The dust temperature in the SC ranges from 22 K to 43 K with a mean of  28 K, a couple of degrees higher than the mean of the SP.  In the case of the NC, the dust temperature has a mean of 27 K, ranging from 20 K to 38 K.  The tail in the dust temperature distribution for $T_\mathrm{dust} >  30$ K for both the SC and NC is the result of being located on the vicinity of the massive star clusters Trumpler 16 and 14. 

The Gum 31 region has dust temperatures ranging from 18 K to 32 K, with a mean of 23 K.  The mean dust temperature in this region is the lowest of all the regions identified in this study.  Because the gas in Gum 31 is located further from the massive star cluster present at the centre of the CNC, the material constituting this part of the molecular cloud complex has $T_\mathrm{dust}$ below 23 K in the densest regions.  For the material immediately surrounding the Gum 31 \hii\ region, the dust is being radiated by the OB stars inside the star cluster NGC 3324, and the $T_\mathrm{dust}$ is found to be > 30 K.

\begin{figure*}
\centering
\begin{tabular}{cc}
\epsfig{file=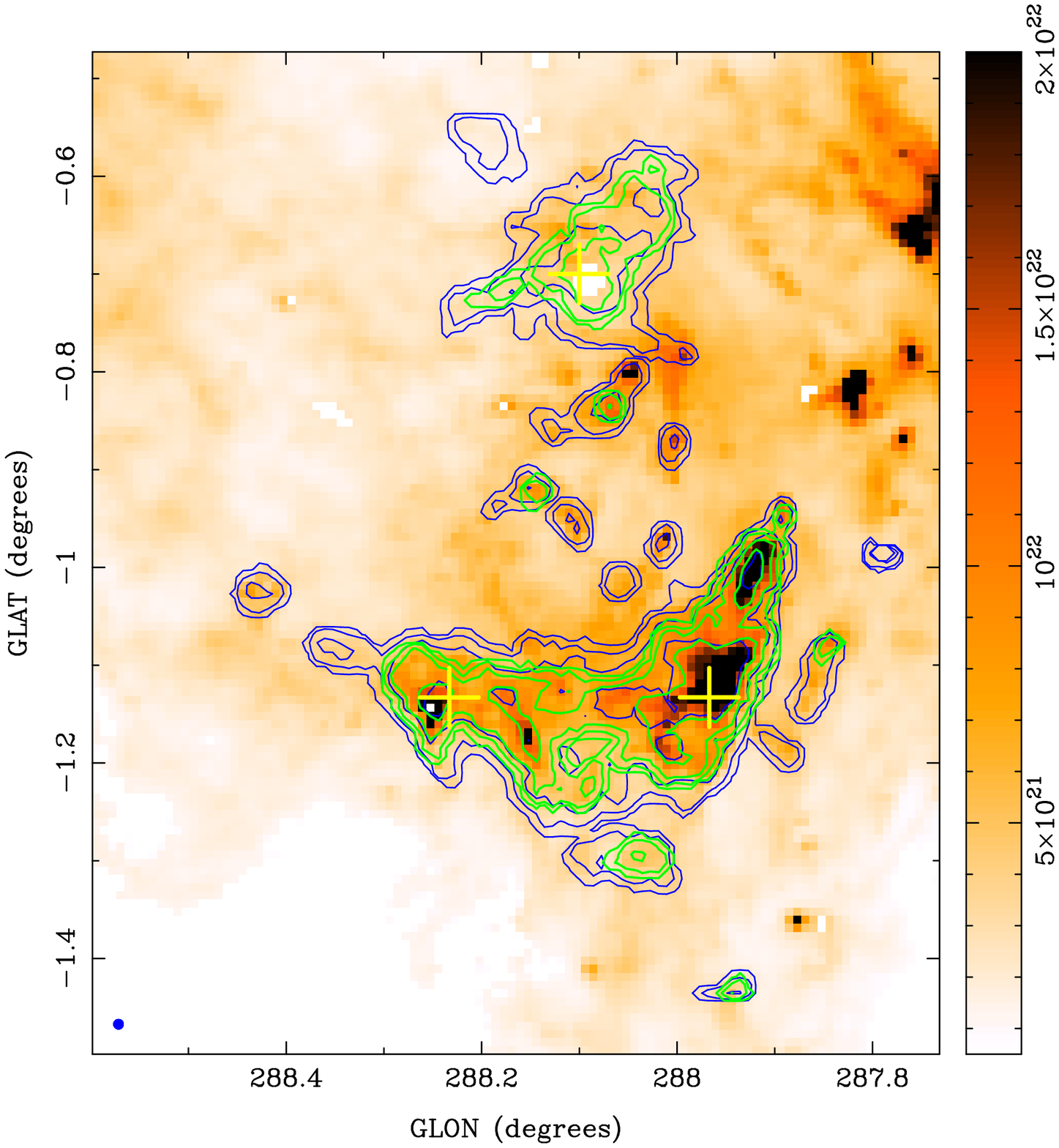,width=0.4\linewidth,angle=0} &
\epsfig{file=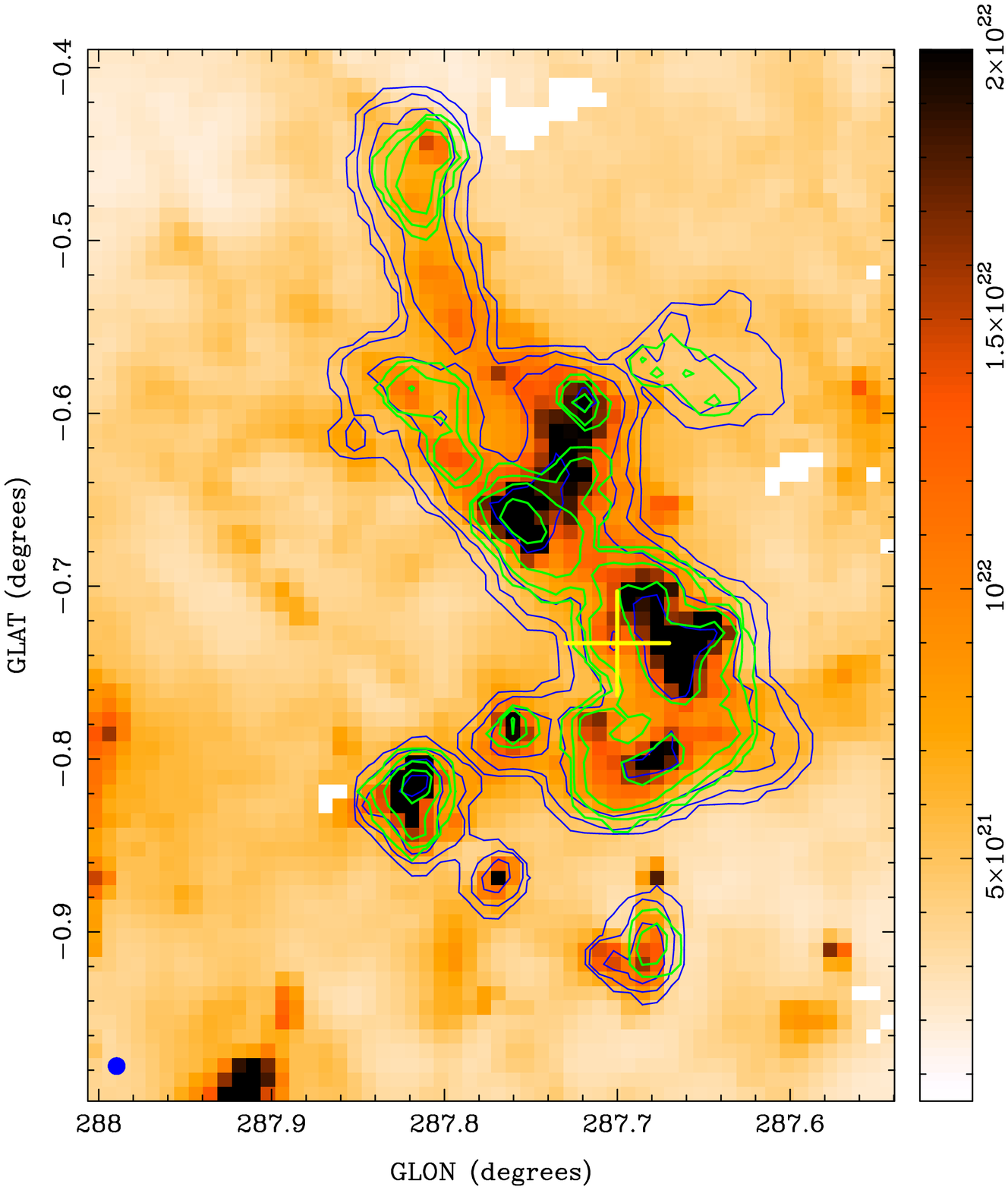,width=0.37\linewidth,angle=0} 
\end{tabular}
\caption{Molecular gas column density map of the Southern Pillars (left) and the Southern Cloud (right).  The colour map is the $N_\mathrm{H_2}(\mathrm{dust})$, the blue contours show $N_\mathrm{H_2}(^{12}$CO), and the green contours show $N_\mathrm{H_2}(^{13}$CO).  Contours levels are $N_\mathrm{H_2} = 1\times 10^{20}, 8\times 10^{20}, 3.2\times 10^{21}, 1.28\times 10^{22}$ cm$^{-2}$.  The yellow crosses show the positions of the C$^{18}$O cores found by \citet{2005ApJ...634..476Y}.}
\label{southp_reg}
\end{figure*}

\begin{figure*}
\centering
\begin{tabular}{cc}
\epsfig{file=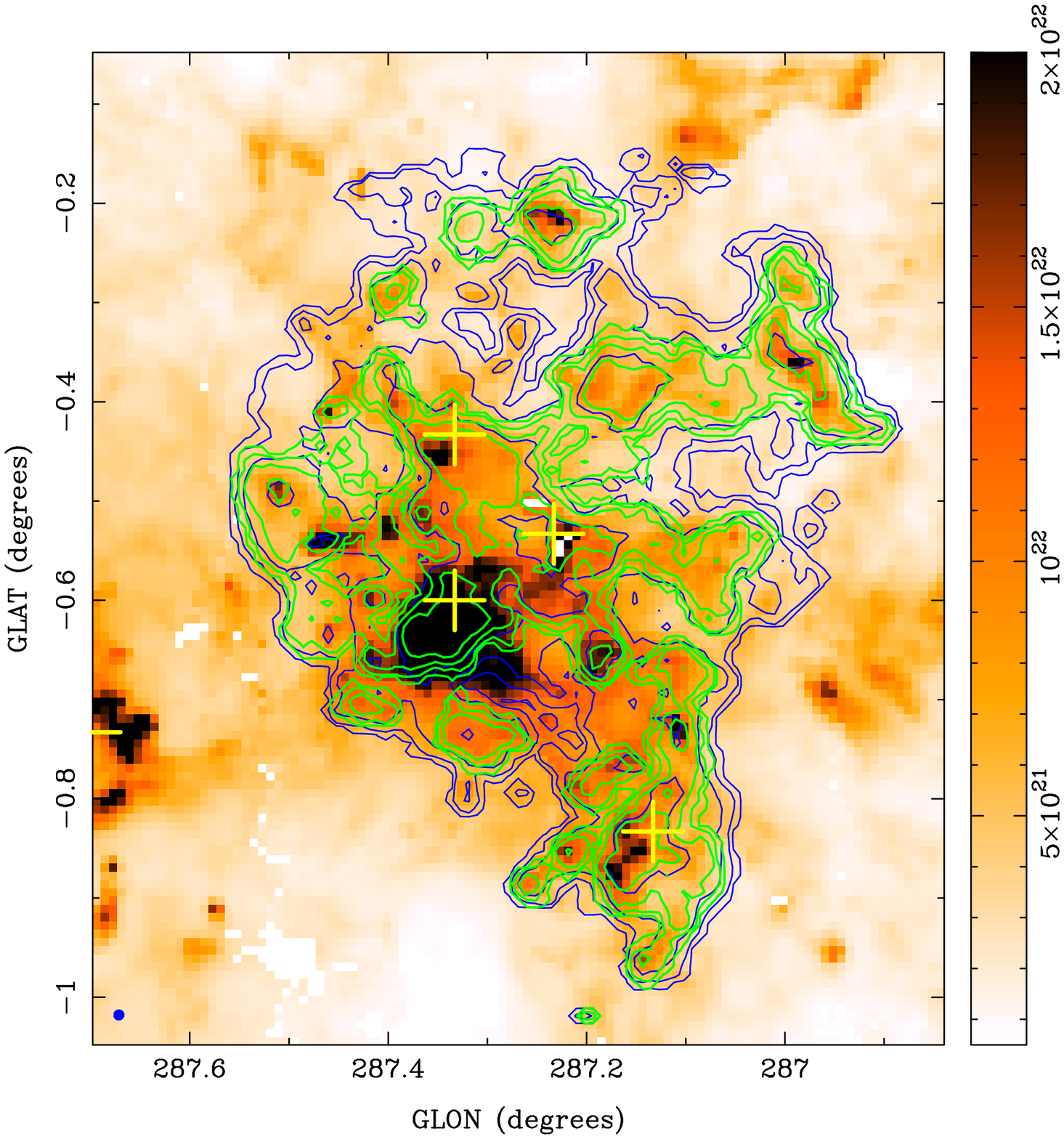,width=0.37\linewidth,angle=0} &
\epsfig{file=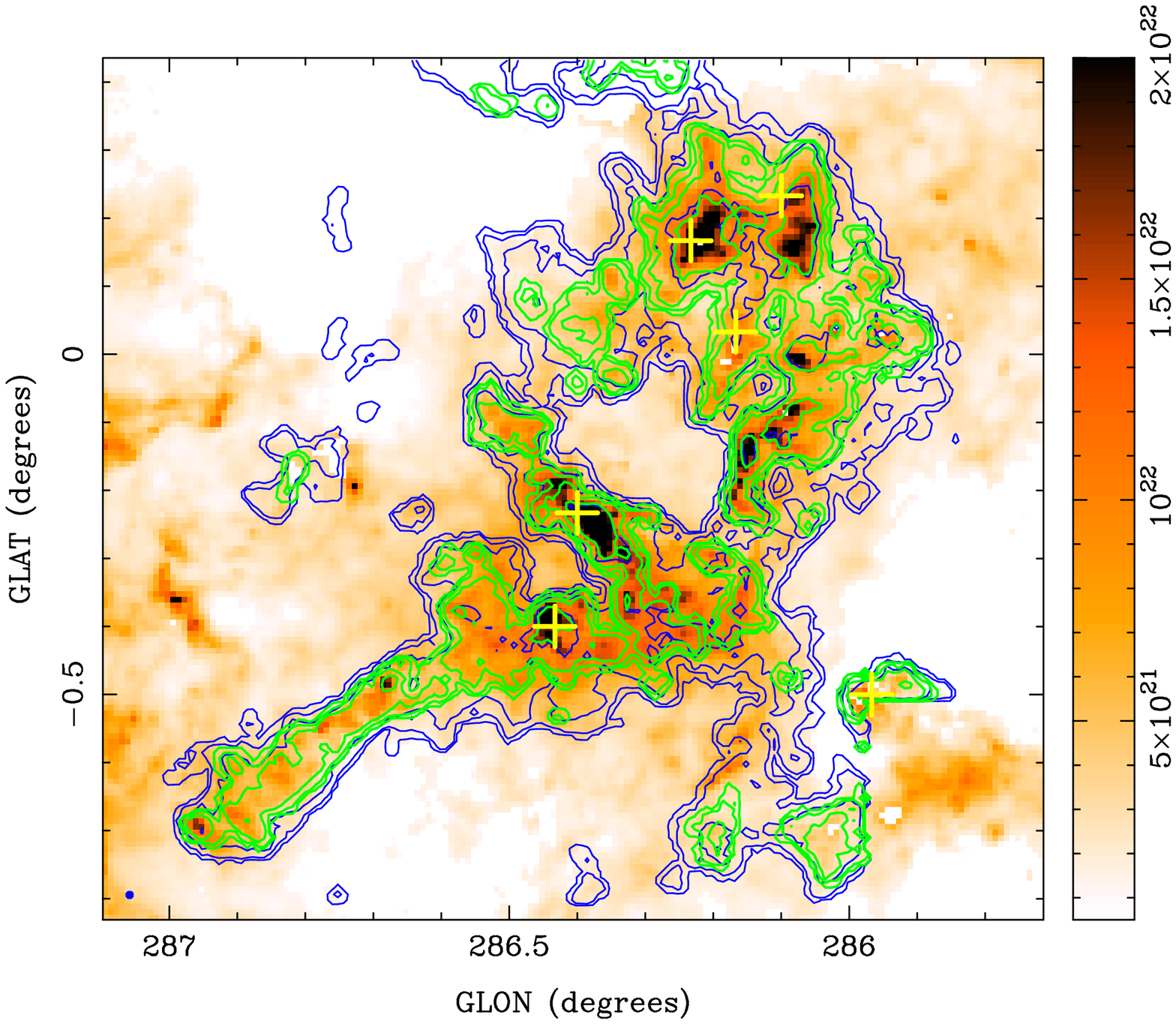,width=0.45\linewidth,angle=0} 
\end{tabular}
\caption{Molecular gas column density map of the Northern Cloud (left) and Gum 31 (right).  The colour map, the contour colours, the contour levels and the symbols are the same as Figure \ref{southp_reg}.}
\label{northc_reg}
\end{figure*}

\subsection{Regional variation of the gas column density}\label{nh2-var}

\subsubsection{Southern Pillars}\label{southp}
Figure \ref{southp_reg} shows the gas column density maps of the SP region.  Together with the gas column density map derived from the dust emission, $N_\mathrm{H_2}$(dust), we have also included the $N_\mathrm{H_2}(^{12}\mathrm{CO})$ and $N_\mathrm{H_2}(^{13}\mathrm{CO})$ maps.  The densest regions, with $N_\mathrm{H_2}> 2\times 10^{22}$ cm$^{-2}$, are located on the inner part of the pillars, and they are traced by both $\co$ and $\cother$.  There is a region at $(l,b) \sim (288.1\degrees,-0.7\degrees)$, which shows extended emission in the molecular line maps but a similar structure is not seen in the dust emission map.  That we do not observe a corresponding extended dust emission feature could be related to a local variation of the dust-to-gas ratio.  However, in the centre of this region, \citet{2013A&A...549A..67G} identified the far-infrared compact source J104813.4-595845 using Herschel maps, and also \citet{2005ApJ...634..476Y} identified one of the C$^{18}$O cores.  Thus, it is evident that a compact source of dense gas is present in the centre of this region.  Unfortunately, this compact source has been removed from our dust column density map because the SED fitting algorithm yields $\chi^{2}$ values above 20 in those pixels.  The other two C$^{18}$O cores found by \citet{2005ApJ...634..476Y} in the SP are coincident with regions of high column density detected by the three tracers in this study.

\begin{figure*}
\centering
\begin{tabular}{cc}
\epsfig{file=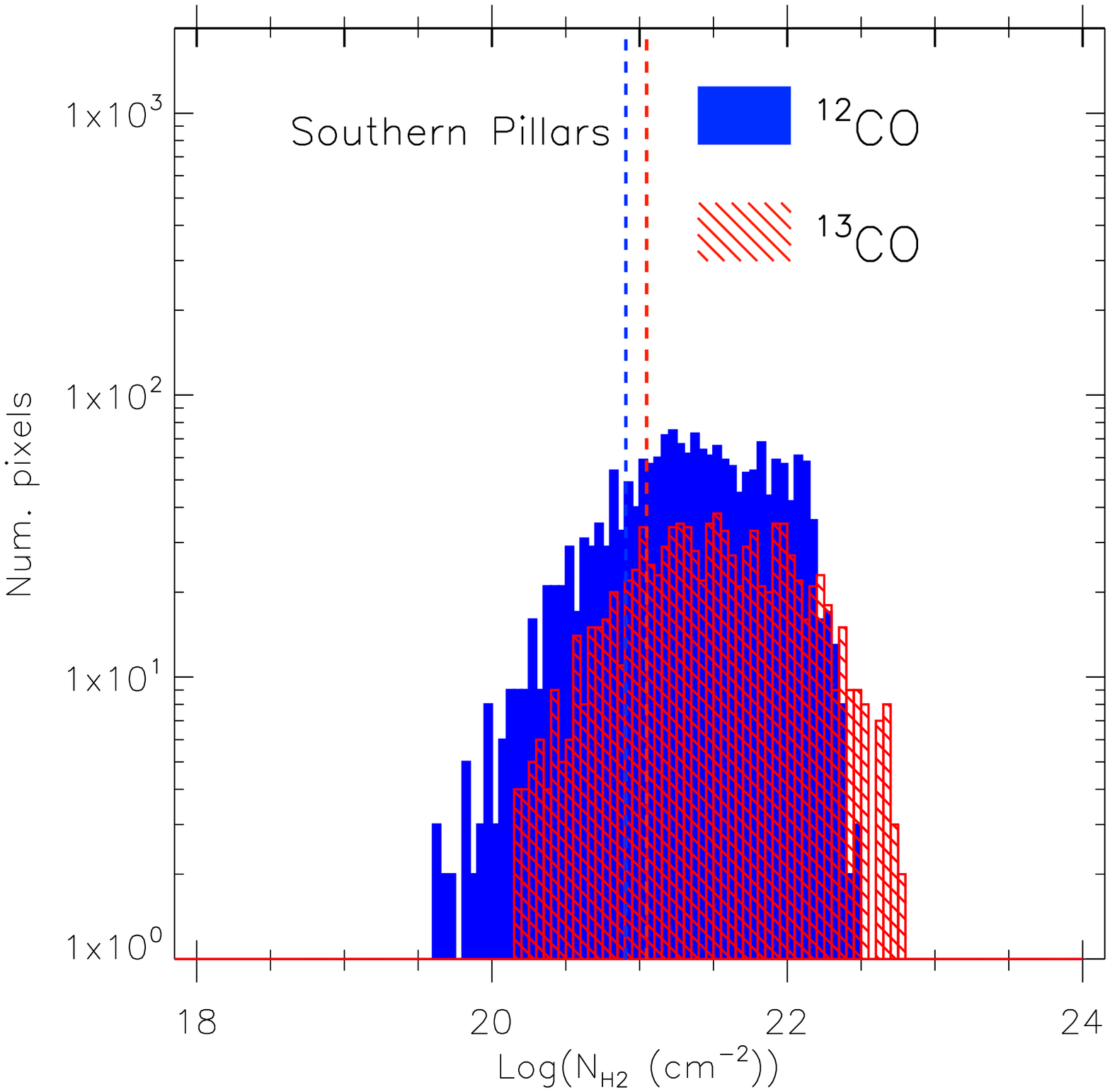,width=0.35\linewidth,angle=90} &
\epsfig{file=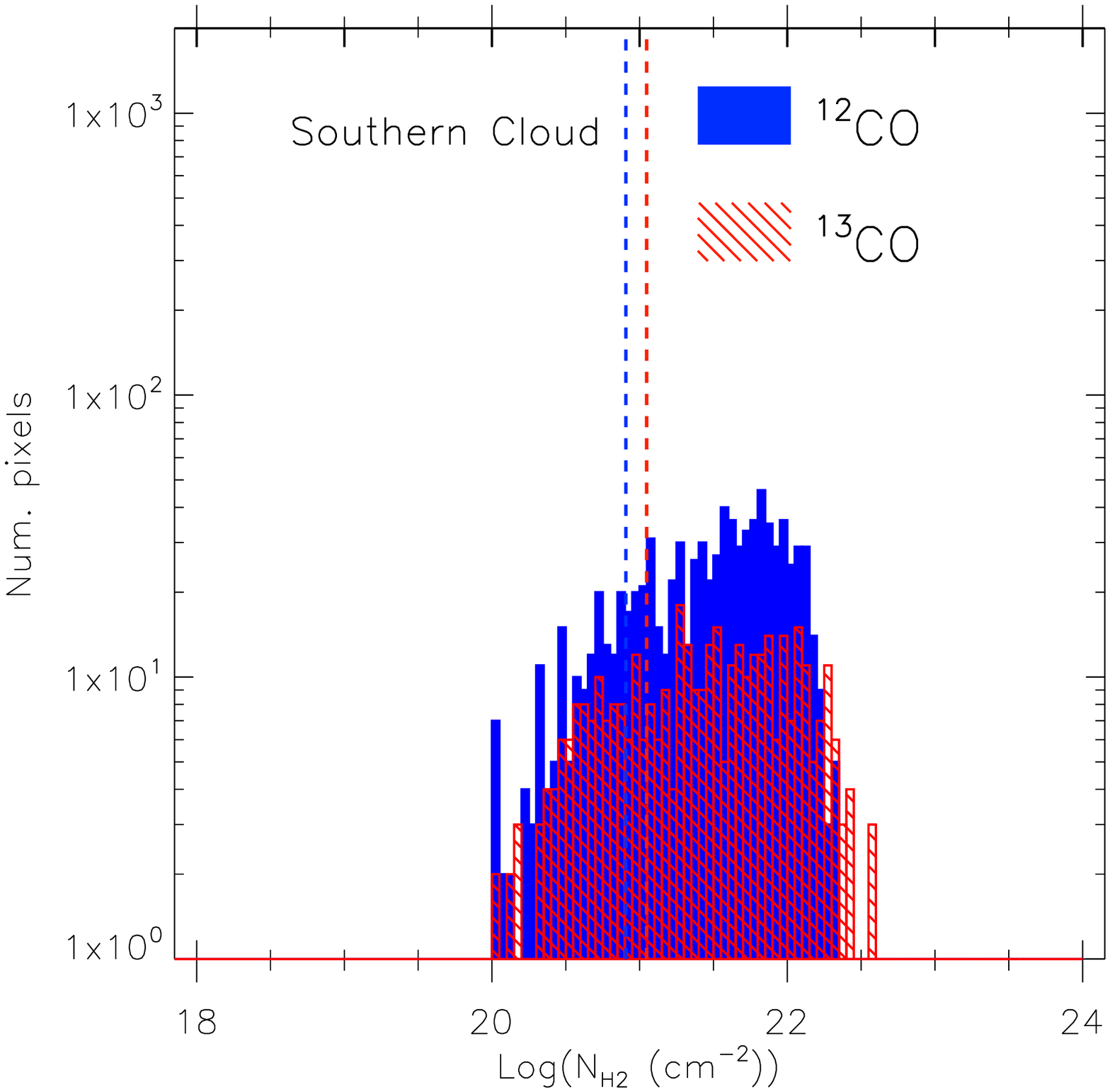,width=0.35\linewidth,angle=90}\\
\epsfig{file=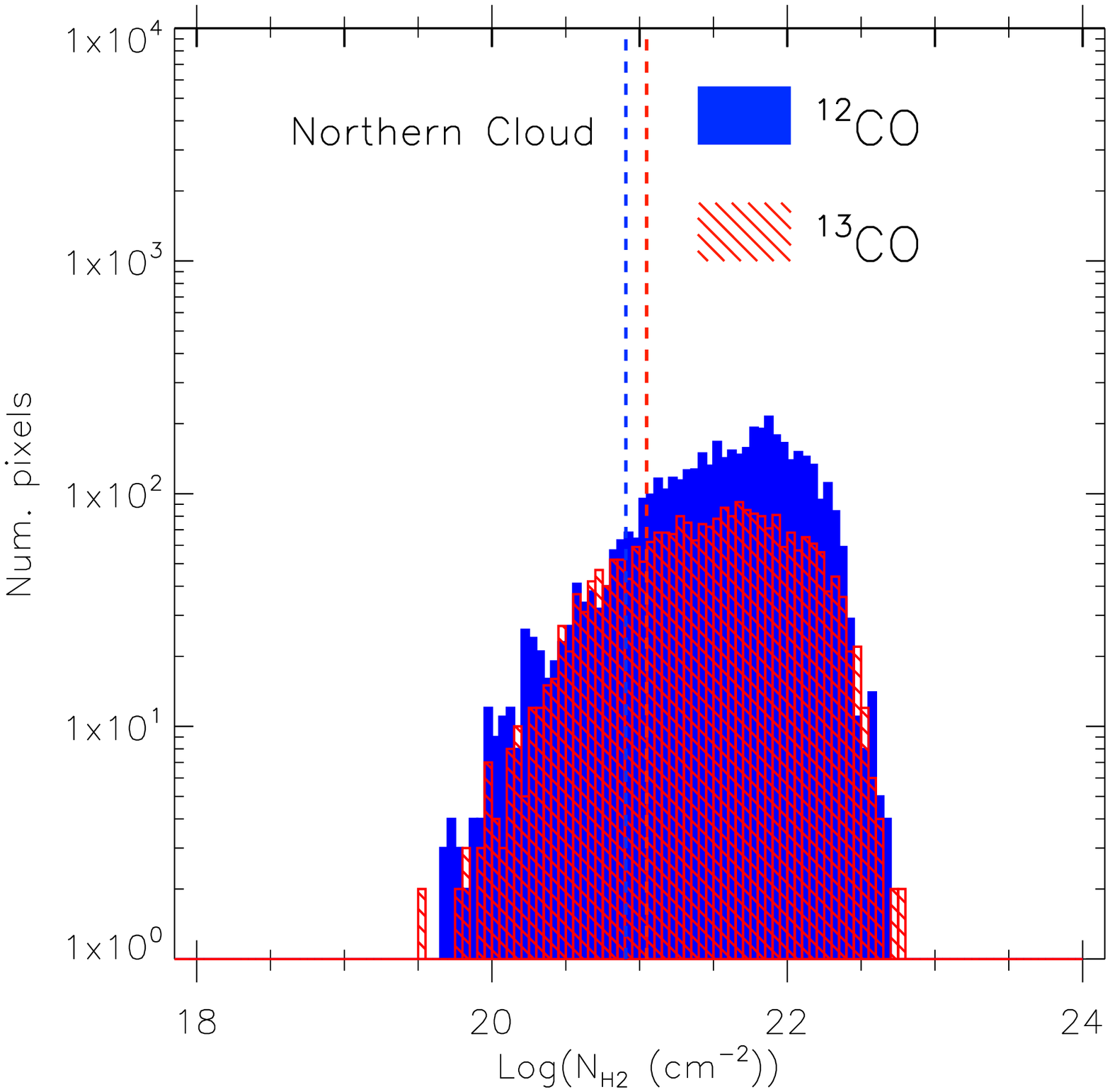,width=0.35\linewidth,angle=90} &
\epsfig{file=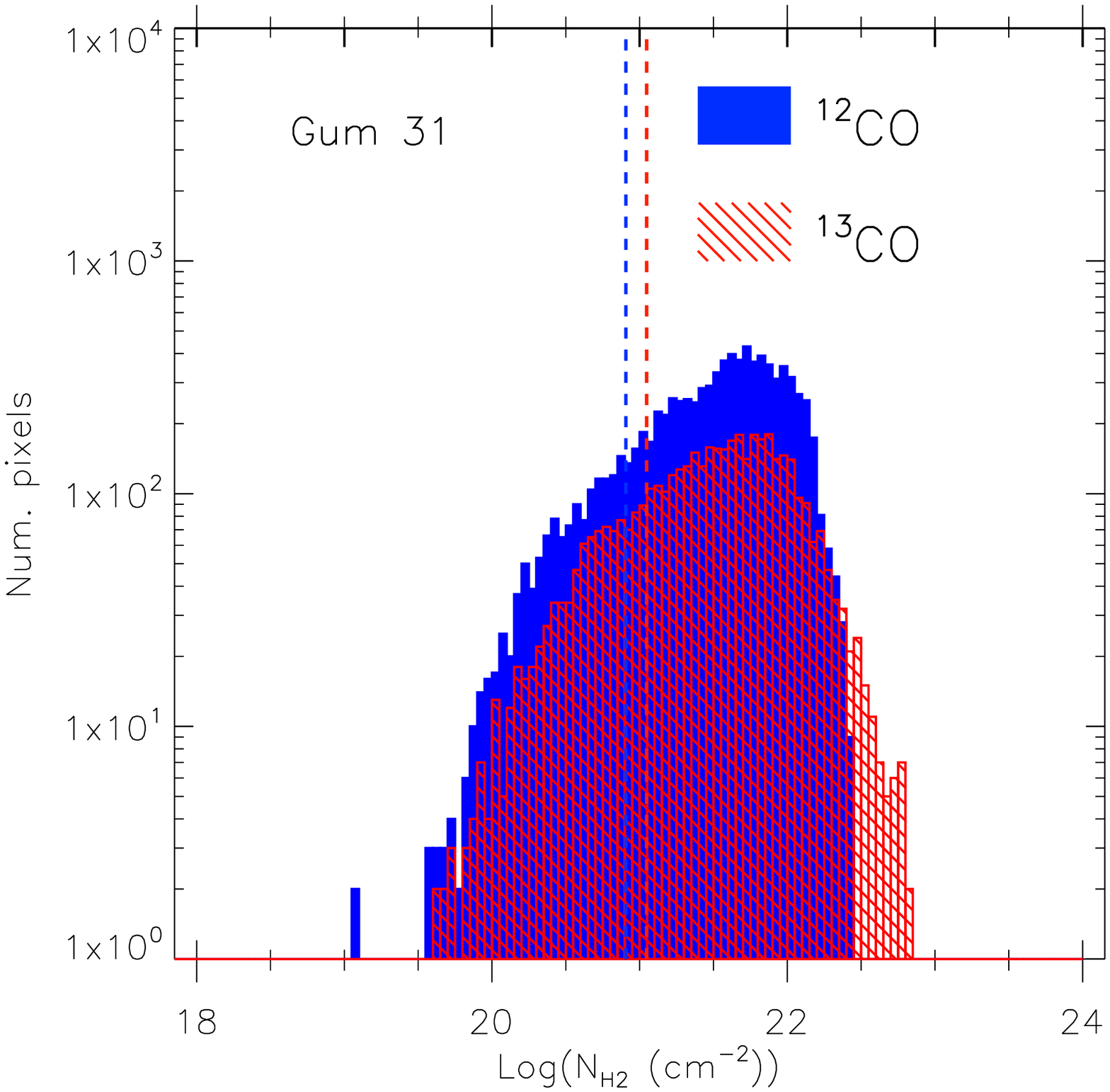,width=0.35\linewidth,angle=90}
\end{tabular}
\caption{Molecular gas column density distribution derived from the $\co$ and $\cother$ line emission for each of the regions identified in the CNC and Gum 31 regions.  The dashed vertical lines illustrate the sensitivity limit of the $\co$ (blue) and $\cother$ (red) maps.}
\label{hist_co_regions}
\end{figure*}

Figure \ref{hist_co_regions} shows the molecular gas distribution derived from $\co$ and $\cother$ for each of the identified regions.  In the case of the SP, there are some differences between the $N_\mathrm{H_2}(^{12}\mathrm{CO})$ and $N_\mathrm{H_2}(^{13}\mathrm{CO})$ distributions.  While the $N_\mathrm{H_2}(^{13}\mathrm{CO})$ distribution traces column densities up to $\sim 6.0\times 10^{22}$ cm$^{-2}$, the $N_\mathrm{H_2}(^{12}\mathrm{CO})$ distribution traces column density only to $\sim 3.0\times 10^{22}$ cm$^{-2}$.

In Figure \ref{hist_gas_regions}, we show the $N_\mathrm{H_2}(\mathrm{dust})$ distribution of the SP.  The dust traces gas from $7.7 \times 10^{20}$ cm$^{-2}$ to $5.1\times 10^{22}$ cm$^{-2}$, with a mean of $3.8 \times 10^{21}$ cm$^{-2}$.  Similar to the global distribution of the entire CNC-Gum 31 region shown in Figure \ref{hist_temp_ndust}, the $N_\mathrm{H_2}(\mathrm{dust})$ in the SP is not well described by a log-normal function.  The fitted parameters can be found in Table \ref{log-norm}.  The log-normal function is a reasonable approximation to the observed distribution only for column densities between $1.0 \times 10^{21}$ to $1.0 \times 10^{22}$ cm$^{-2}$.  For column densities below this range, the observed distribution decreases faster than the log-normal distribution, due to the sensitivity limit in the infrared emission maps.  As before, the observed distribution shows a tail with respect to the log-normal distribution for column densities larger than this range.  

\subsubsection{Southern Cloud}\label{southc}
Figure \ref{southp_reg} also shows the molecular gas column density maps of the SC.  Overall, the distributions of the different tracers are similar, with $N_\mathrm{H_2}(^{13}\mathrm{CO})$ coincident with the regions of high column density seen in  $N_\mathrm{H_2}(^{12}\mathrm{CO})$  and dust emission maps.  One C$^{18}$O core is located inside this cloud, but it is slightly shifted from the closest region of high column density.  

Given the reduced number of pixels with significant emission in the SC region, especially in the $\cother$ map, it is not possible to draw any conclusion about the overall shape of the $N_\mathrm{H_2}(^{12}\mathrm{CO})$ and $N_\mathrm{H_2}(^{13}\mathrm{CO})$ distributions as seen in Figure \ref{hist_co_regions}.  However, we see that both tracers cover a similar range in column density from $\sim 1.0 \times 10^{20}$ cm$^{-2}$ to $\sim 3.0 \times 10^{22}$ cm$^{-2}$.

Figure \ref{hist_gas_regions} also shows the $N_\mathrm{H_2}(\mathrm{dust})$ distribution for the SC region.  The mean gas column density is $5.3 \times 10^{21}$ cm$^{-2}$, ranging from $1.5 \times 10^{21}$ to $7.0 \times 10^{22}$ cm$^{-2}$.  Again, we have fitted a log-normal function to the observed distribution, finding a poor correspondence, especially for $N_\mathrm{H_2} > 1.0 \times 10^{22}$ cm$^{-2}$.

\subsubsection{Northern Cloud}\label{northc}
Figure \ref{northc_reg} shows the $N_\mathrm{H_2}$ maps derived for the NC.  In general, the distribution of the different tracers are similar, with the densest region being located at $(l,b) \sim (287.35\degrees,-0.64\degrees)$.  Located inside this region of dense gas, there is another C$^{18}$O core identified by \citet{2005ApJ...634..476Y}.  Three other cores are located inside the NC, in regions of high column density as traced by the molecular line and the dust map.  

\begin{figure*}
\centering
\begin{tabular}{cc}
\epsfig{file=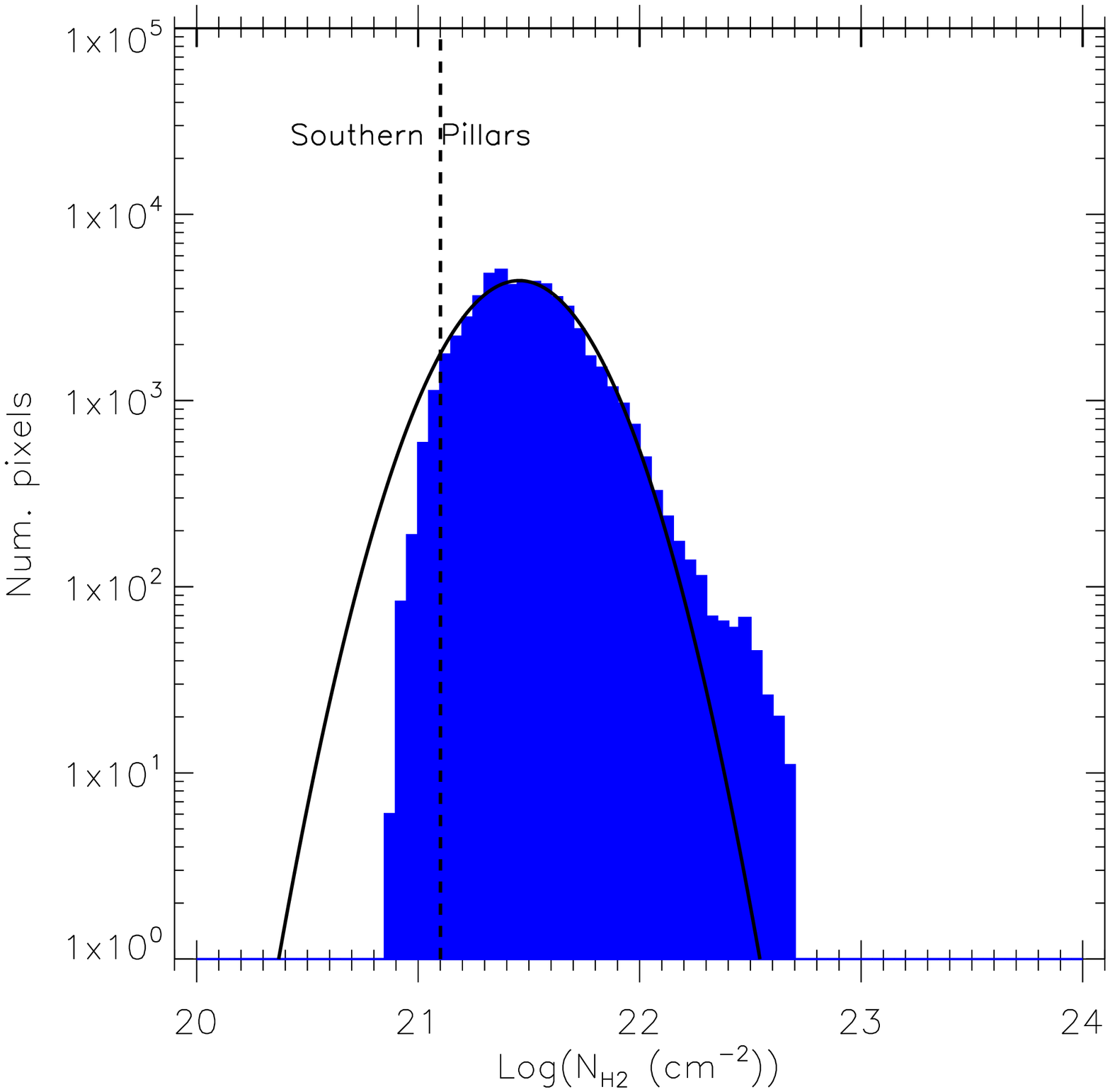,width=0.35\linewidth,angle=90} &
\epsfig{file=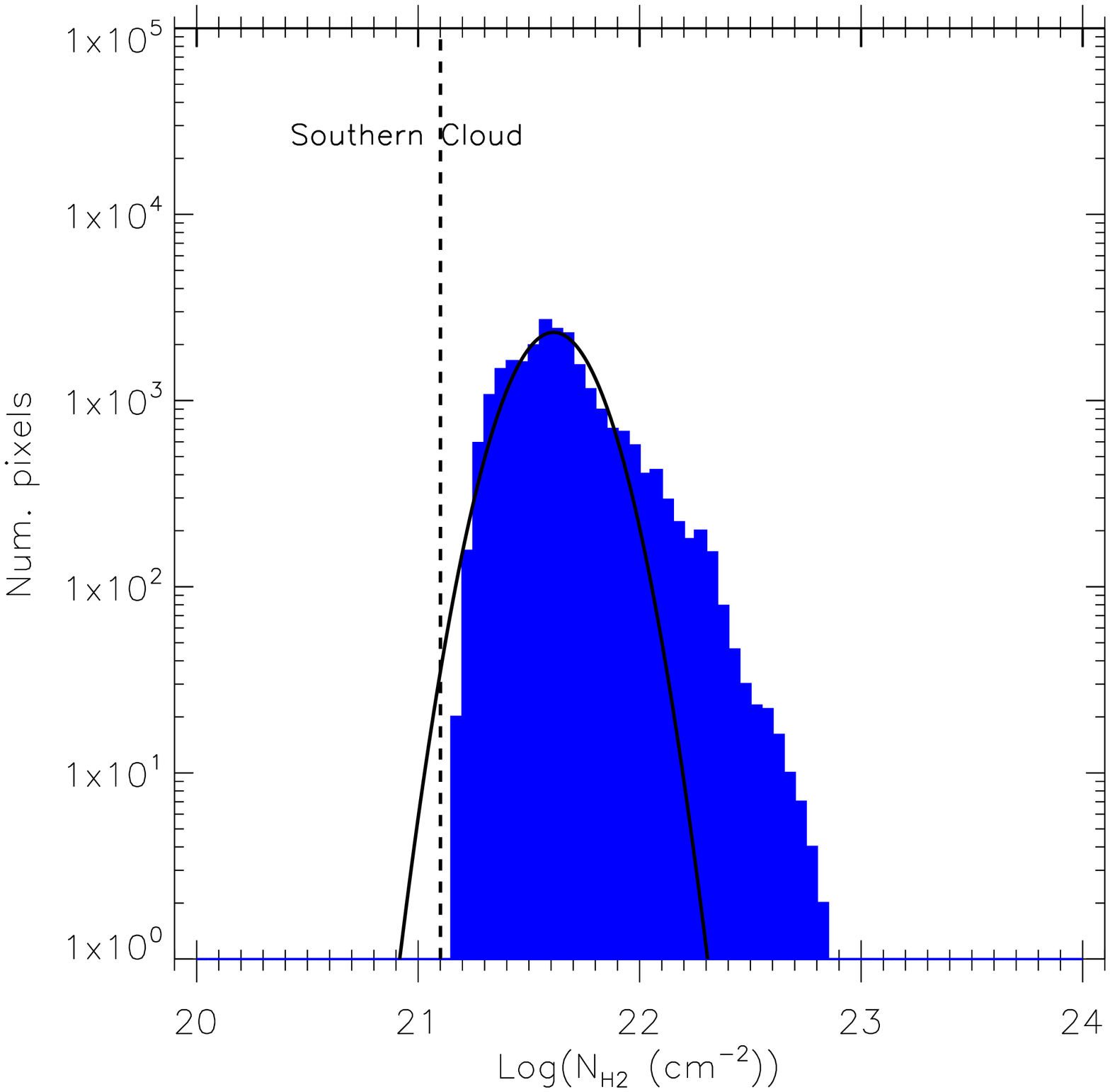,width=0.35\linewidth,angle=90}\\
\epsfig{file=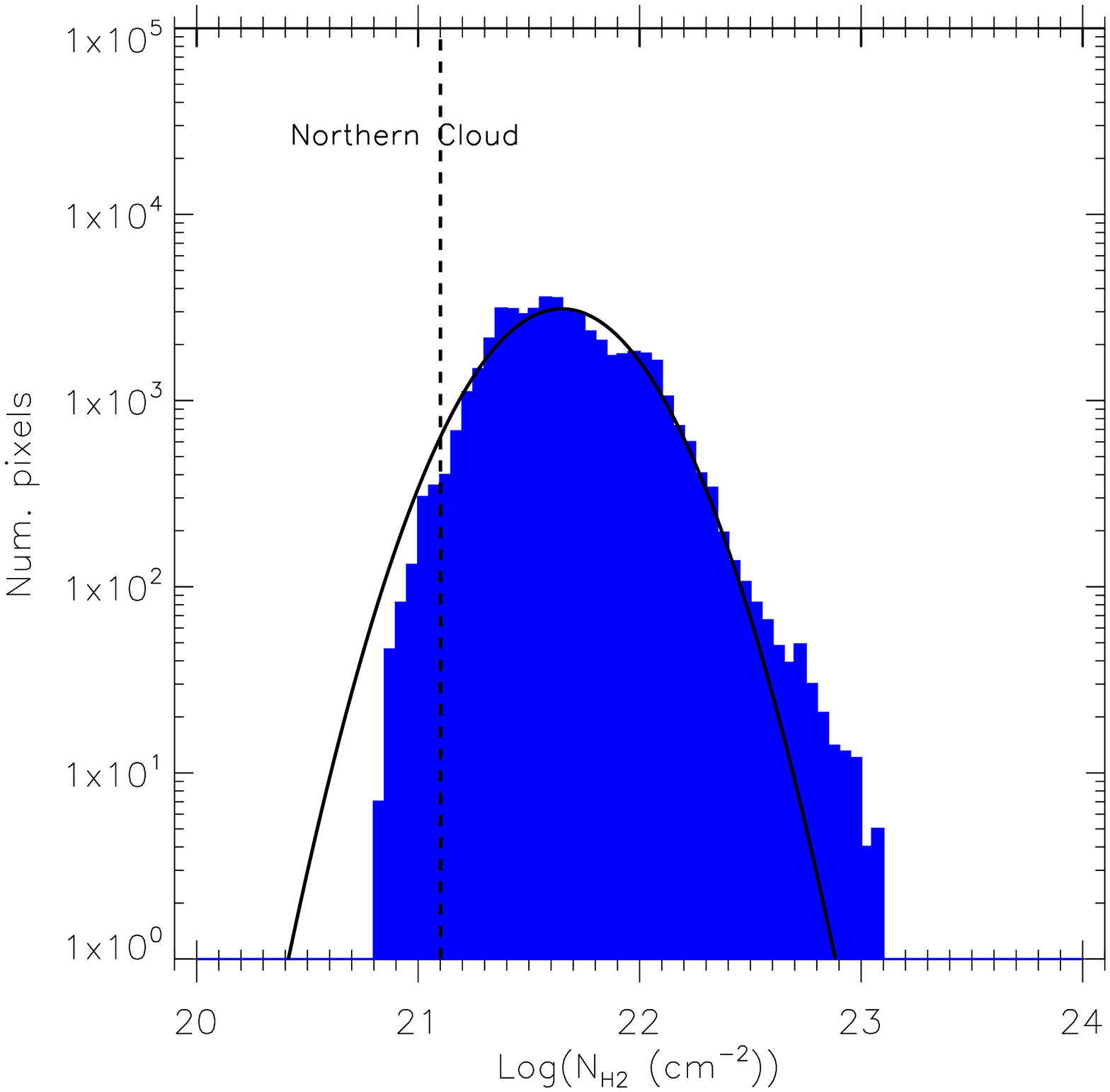,width=0.35\linewidth,angle=90} &
\epsfig{file=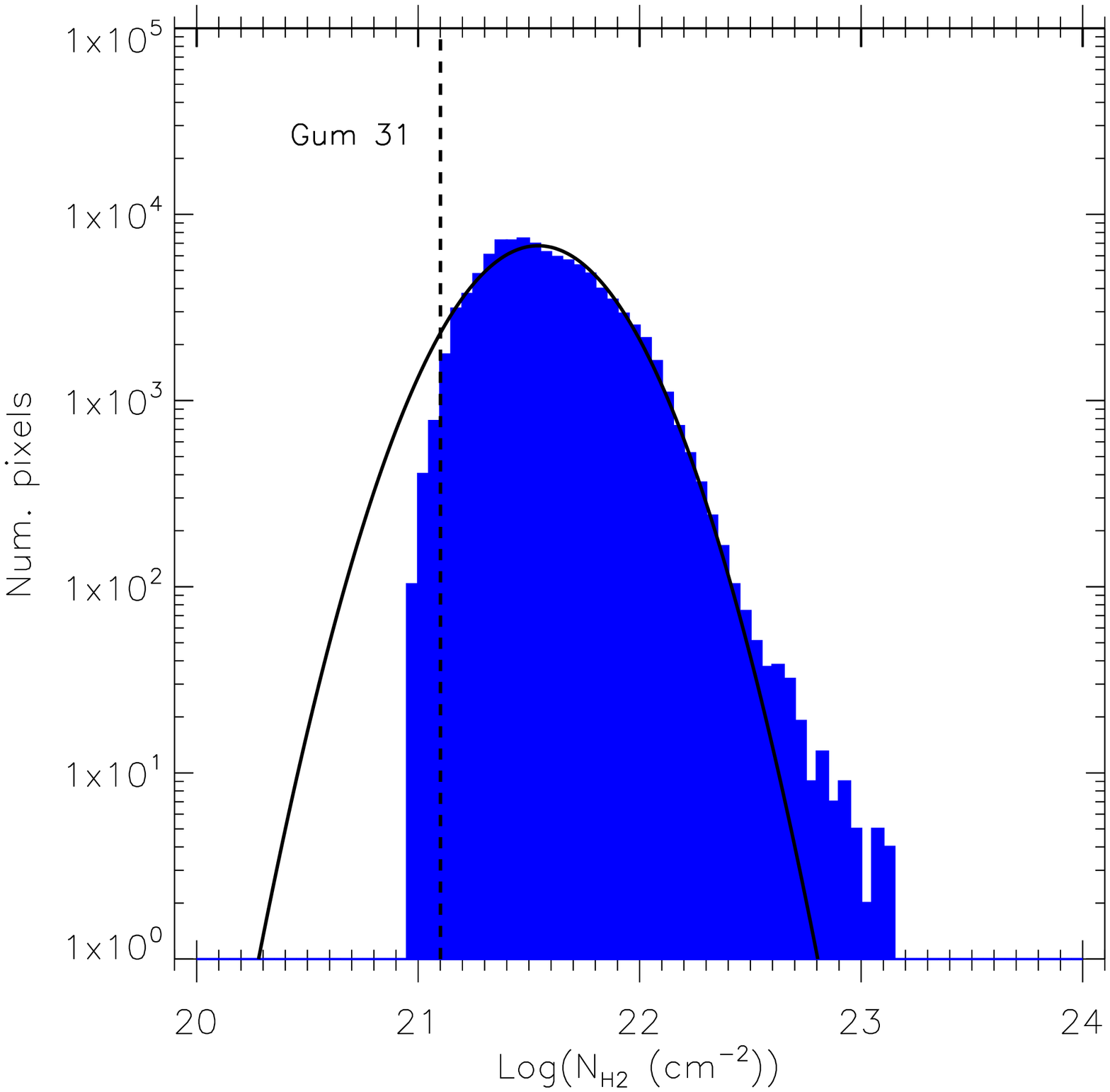,width=0.35\linewidth,angle=90}
\end{tabular}
\caption{Gas column density distribution derived from the dust emission for each of the regions identified in the CNC and Gum 31 regions.  The black solid lines show the log-normal function fit to the column density distributions. The vertical black dashed lines illustrate the column density sensitivity limit of our SED fitting algorithm.}
\label{hist_gas_regions}
\end{figure*}

In Figure \ref{hist_co_regions} we observe that both $N_\mathrm{H_2}(^{12}\mathrm{CO})$ and $N_\mathrm{H_2}(^{13}\mathrm{CO})$ distributions cover the same range from $\sim 4.0 \times 10^{19}$ cm$^{-2}$ to $\sim 5.0 \times 10^{22}$ cm$^{-2}$.  The $N_\mathrm{H_2}(^{12}\mathrm{CO})$ distribution shows a mean of $\sim 6.5 \times 10^{21}$ cm$^{-2}$, and the $N_\mathrm{H_2}(^{13}\mathrm{CO})$ shows a mean of $\sim 6.0 \times 10^{21}$ cm$^{-2}$.  The mean column density of the $N_\mathrm{H_2}(^{12}\mathrm{CO})$ distribution is driven by an apparent ``excess'' of pixels with $1.0 \times 10^{21}$ cm$^{-2}$ $ < N_\mathrm{H_2} < 3.2 \times 10^{21}$ cm$^{-2}$ with respect to the $N_\mathrm{H_2}(^{13}\mathrm{CO})$ distribution.  This can be interpreted as local variations of the factors used to convert the $\co$ integrated intensity and the $N(^{13}\mathrm{CO})$ to the total molecular gas column density $N_\mathrm{H_2}$.  For example, If the $[\mathrm{H}_2/\mathrm{^{13}CO}]$ abundance ratio of $^{13}$CO molecule increases by a factor of two at the edge of the cloud, then the overall shape of the $N(^{13}\mathrm{CO})$ distribution would be more similar to the $N_\mathrm{H_2}(^{12}\mathrm{CO})$ distribution.  Alternatively, if the $X_\mathrm{CO}$ factor increases at regions of high column density due to the optically thick regime of the line emission, the $N_\mathrm{H_2}(^{12}\mathrm{CO})$ distribution will be more similar to the  $N_\mathrm{H_2}(^{13}\mathrm{CO})$ distribution over the column density range traced by both lines.  In Section \ref{factor-var} we will discuss the variation of the conversion factors for each individual region.  

The $N_\mathrm{H_2}(\mathrm{dust})$ distribution for the NC is shown in Figure \ref{hist_gas_regions}.  This region has the largest mean column density among all the regions identified in this study, with a value of $\sim 6.0 \times 10^{21}$ cm$^{-2}$.  The dust emission map traces gas over the range $6.9 \times 10^{20}$ to $1.2 \times 10^{23}$ cm$^{-2}$.  The shape of the observed distribution is well described by a log-normal function for column densities between $1.0 \times 10^{21}$ and $3.2 \times 10^{22}$ cm$^{-2}$.  As before, a high-column density tail is evident in the observed distribution with respect to a log-normal shape.  See Table \ref{log-norm} for the parameters of the fitted function.

\subsubsection{Gum 31}\label{gum31}
Figure \ref{northc_reg} also shows the gas column density maps derived from the three tracers for the Gum 31 region.  A shell-like structure surrounding the Gum 31 \hii\ region is seen in the CO and dust emission maps, and the densest regions are located on this ring of material.  Five C$^{18}$O cores are located on this region of high column density.  There is another core located at the edge of the CO maps, but a similar feature in the dust emission map is not seen. 

\begin{table*}
\caption{Gas mass budget for each region in the CNC-Gum 31 molecular complex as defined in Figure \ref{carina.500um.obs}.}
\centering
\begin{tabular}{lrrrccc}
\hline\hline
Region  &  $M$(dust)$^\mathrm{a}$  &  $M_{12}$$^\mathrm{b}$  &   $M_{13}$$^\mathrm{c}$ & $M_{12}/M$(dust) & $M_{13}/M$(dust) & $M_{13}/M_{12}$ \\
& $\Msun$ & $\Msun$  &  $\Msun$  &  & &  \\ 
\hline
SP & 76041 &  21725 & 16072 &   0.29  & 0.21 & 0.74   \\ 
SC & 45098 &  10031 &  5368  &  0.22  & 0.12 &  0.54 \\
NC & 106217 &  71356 &  36790  &  0.67  & 0.35 & 0.52 \\  
Gum 31 & 165354 & 107244   & 59615 & 0.65 & 0.36  & 0.56   \\
\hline
Total & 392709 & 210355  & 117844 &  0.54 & 0.30  & 0.56  \\
\hline
\multicolumn{7}{l}{{\bf Notes.}} \\
\multicolumn{7}{l}{$^\mathrm{a}$ $M$(dust) is the gas mass derived from dust.}\\
\multicolumn{7}{l}{$^\mathrm{b}$ $M_{12}$ is the gas mass derived from $\co$.}\\
\multicolumn{7}{l}{$^\mathrm{c}$ $M_{13}$ is the gas mass derived from $\cother$.}\\
\end{tabular}
\label{mass-budget}
\end{table*}

The molecular column density distributions derived from $^{12}$CO and $^{13}$CO maps for the Gum 31 region are shown in Figure \ref{hist_co_regions}.  Both distributions have similar mean column density $\sim 5.0 \times 10^{21}$ cm$^{-2}$.  However, while the  $^{12}$CO map traces gas up to $\sim 2.8 \times 10^{22}$ cm$^{-2}$, the $N_\mathrm{H_2}(^{13}\mathrm{CO})$ distribution traces molecular gas up to $\sim 7.0 \times 10^{22}$ cm$^{-2}$.  These pixels with high column density correspond to dense gas located on the surrounding region of the star cluster NGC 3324.

In Figure \ref{hist_gas_regions} we show the $N_\mathrm{H_2}(\mathrm{dust})$ distribution for the Gum 31 region.  The minimum, maximum and mean gas column density is $8.9 \times 10^{20}, 1.5 \times 10^{23}$ and $4.7 \times 10^{21}$ cm$^{-2}$ respectively.  Similar to the observed behaviour for the $N_\mathrm{H_2}(\mathrm{dust})$ distributions of the SP, SC and NC, the gas column density distribution derived from the dust emission for the Gum 31 region is well described by a log-normal function only for a limited column density range.  In this case, this range covers $N_\mathrm{H_2}$ from $1.0 \times 10^{21}$ to $2.5 \times 10^{22}$ cm$^{-2}$.  The high-column density tail is also clearly observable in the $N_\mathrm{H_2}(\mathrm{dust})$ distribution for the Gum 31, which is also traced by the $^{13}\mathrm{CO}$ maps.  Table \ref{log-norm} shows the parameters of the fitted function.

\subsection{The mass budget in the CNC-Gum31 region}\label{masses}
In Table \ref{mass-budget} we summarize the masses for each of the regions as defined in Figure \ref{carina.500um.obs}, and for the entire CNC-Gum31 complex.  The gas mass traced by the dust emission ($M$(dust)) is $7.6 \times 10^{4}\ \Msun$ for the SP.  The molecular gas mass is $2.2 \times 10^{4}\ \Msun$ and $1.6 \times 10^{4}\ \Msun$ as traced by $\co$ ($M_\mathrm{12}$) and $\cother$ ($M_\mathrm{13}$) respectively.  For the SC, which is the smallest region considered in this study, the total $M$(dust) is $4.5 \times 10^{4}\ \Msun$, while $M_\mathrm{12}$ is $1.0 \times 10^{4}\ \Msun$.  The $M_\mathrm{13}$ is a factor of 2 smaller.  In the case of the NC, the $M$(dust) is $10.6 \times 10^{4}\ \Msun$.  The $M_\mathrm{12}$ is $7.1 \times 10^{4}\ \Msun$, and the $M_\mathrm{13}$ is $3.7 \times 10^{4}\ \Msun$.  The $M$(dust) for the Gum 31 region defined in this paper is $16.5 \times 10^{4}\ \Msun$, while the molecular component is $10.7 \times 10^{4}\ \Msun$ and $6.0 \times 10^{4}\ \Msun$ as traced by $\co$ and $\cother$ respectively.

We compare the molecular gas masses found here to the masses reported by \citet{2005ApJ...634..476Y} using poorer spatial resolution CO maps.  They define seven different regions over the CNC-Gum 31 molecular complex similar to our region selection:  region 1 and 7 correspond to our SP region, region 2 corresponds to the SC, region 3 is the NC, and regions 4 and 5 include the region defined as Gum 31 in this study.  They assume that the CNC is located at 2.5 kpc, and they use $X_\mathrm{CO}=1.6\times 10^{20} \mathrm{cm}^{-2} (\mathrm{K}\ \kms)^{-1}$ to calculate the molecular gas masses, which are different from the values assumed in this work.  After correcting for these different values, they find $2.8\times 10^{4}\ \Msun$ for the SP, $1.8\times 10^{4}\ \Msun$ for the SC, $10.4\times 10^{4}\ \Msun$ for the NC and $15.4\times 10^{4}\ \Msun$ for Gum 31.  These values are larger than the values reported here.  These differences are likely to be related to the differences in the area considered to estimate the masses and differences in the pixel masking procedure applied to the CO maps.

There are differences between the molecular gas mass estimates traced by $\cother$ reported here and the corresponding values reported by \citet{2005ApJ...634..476Y}.  They use $[\mathrm{H}_2/\mathrm{^{13}CO}]=5\times 10^5$, which is $\sim 30$ \% smaller than the value used here.  After correction, the molecular mas traced by $\cother$ is $2.0\times 10^{4}\ \Msun$ for the SP, $0.9\times 10^{4}\ \Msun$ for the SC, $4.5\times 10^{4}\ \Msun$ for the NC and $7.3\times 10^{4}\ \Msun$ for the Gum 31 region.  Except for the SC where the molecular mass reported by \citet{2005ApJ...634..476Y} is 60\% larger than our value, the other regions present $\sim$ 20\% larger values than to the masses we find in this study.  Again, this discrepancy is likely to be related to the different pixel masking procedure used to create the integrated intensity maps, and also to the different extent of the area assigned to each region. 

The fraction of the mass recovered by the CO emission line maps with respect to the dust map is shown in Table \ref{mass-budget}.  We estimate this fraction by taking the ratio between the molecular gas mass traced by $\co$ (and $\cother$) and the total gas traced by the dust emission.  Regarding the molecular gas mass traced by $\co$, there is a clear difference between the SP and SC with respect to the NC and Gum 31 regions.  While the fraction of the mass recovered by $\co$ is $\sim$ 20-30 \% for the southern regions, this fraction increases to $\sim 65$\% for the NC and Gum 31, a factor $\sim$ 2.5 larger.  Similar molecular gas fraction differences are seen when the molecular gas is traced by $\cother$.  For the SP, this fraction is 21 \%, for the SC $\sim$ 12 \%, and $\sim$ 35 \% for the NC and Gum 31.  This difference can be interpreted as due to the radiation field from the massive star clusters located at the centre of the CNC.  The southern regions (SP and SC) are heavily affected by the Trumpler 14, 16 and Bochum 11 star clusters radiation field.  This result was already observed in the dust temperature map (Figure \ref{hist_tdust_reg}).  In this region, the molecular gas has been disassociated by the strong radiation field, reducing the amount of molecular gas with respect to the total gas estimated from the dust maps.  The northern regions (NC and Gum 31) appear to be more effective in shielding the CO molecules.  Between these two regions, the most remarkable case is the NC.  The cloud is located just besides the star cluster Trumpler 14, but the molecular cloud mass fraction is still comparable to the fraction observed in Gum 31.  This may be interpreted as the result of the NC being in a younger evolutionary stage than the SP and SC, with very dense material that helps the shielding of the CO molecules, even though the radiation field is comparable to the radiation field received by the southern regions.

\begin{table*}
\caption{Gas mass budget of the CNC-Gum 31 region using the masks from the $^{12}$CO and $^{13}$CO maps.}
\centering
\begin{tabular}{lrrrrcc}
\hline\hline
Region  &  $M$(dust)$^\mathrm{a}$  &  $M_{12}$$^\mathrm{a}$  & $M$(dust)$^\mathrm{b}$ &  $M_{13}$$^\mathrm{b}$ & $M_{12}$$^\mathrm{a}$$/M$(dust)$^\mathrm{a}$ & $M_{13}$$^\mathrm{b}/M$(dust)$^\mathrm{b}$ \\
& $\Msun$ & $\Msun$  &  $\Msun$  & $\Msun$  & &  \\ 
\hline
SP &           32056 &   21482 & 20665  & 15793 &   0.67 & 0.76   \\ 
SC &           21198 &   10031 & 12227  &   5368 &  0.47 &  0.44 \\
NC &           86568 &   71070 & 59467  & 36431 &   0.82 & 0.61 \\  
Gum 31 & 122451 & 102095 & 82212 & 57911 &  0.83  & 0.70   \\
\hline
Total & 262273 & 204678  & 174571 &  115503 & 0.78  & 0.66  \\
\hline
\multicolumn{7}{l}{{\bf Notes.}} \\
\multicolumn{7}{l}{$^\mathrm{a}$ Mass is calculated summing over the pixels defined by the $^{12}$CO mask.}\\
\multicolumn{7}{l}{$^\mathrm{b}$ Mass is calculated summing over the pixels defined by the $^{13}$CO mask.}\\
\end{tabular}
\label{mass-budget-mask}
\end{table*}

\subsubsection{Mass over the masks defined by $^{12}$CO and $^{13}$CO maps}\label{mask-dust-mass}
The masses reported in Table \ref{mass-budget} were estimated by adding the column densities inside the regions defined in Figure \ref{carina.500um.obs}.  Although these values are useful when estimating the molecular gas fraction over extended regions, the uncertain definition of the extent of the relevant area can also produce large variations in the masses for the more extended gas tracer.  Thus, if one wants to estimate the fraction of gas in molecular form over the {\it same} region, we need to sum the column density over the same pixels.  We have also estimated the gas masses for each region only considering pixels of significant emission in the $^{12}$CO and the $^{13}$CO maps.  In Table \ref{mass-budget-mask} the masses and the molecular gas fraction are estimated for each region.  The first mask is built by considering the pixels with significant emission in the $^{12}$CO map and with significant $N_\mathrm{H_2}(\mathrm{dust})$ values.  For the SP and SC, the ratio between $M_\mathrm{12}$ and $M$(dust) is a factor $\sim$ 2 larger than the values reported in Table \ref{mass-budget}.  For the NC and Gum 31 regions, the $M_\mathrm{12}$ to $M$(dust) ratio increases from $\sim$ 0.65 when the regions are defined as in Figure \ref{carina.500um.obs}, to $\sim$ 0.82 when the regions are defined by the $^{12}$CO mask.  This variation is expected as the $M$(dust) is reduced by the smaller number of pixels in the $^{12}$CO mask, while the $M_\mathrm{12}$ is practically unchanged.  Despite this variation in the molecular gas fraction when the $^{12}$CO mask is considered, the southern regions still present smaller molecular gas fractions ($\sim$ 67\% and 47\% for SP and SC respectively), with respect to the northern regions ($\sim$ 82\% for NC and Gum 31).  

The second mask is constructed by selecting pixels with significant $N(^{13}\mathrm{CO})$ and $N_\mathrm{H_2}(\mathrm{dust})$ values.  The values of the gas masses and the gas mass ratios using this mask are also shown in Table \ref{mass-budget-mask}.  The ratios between $M_\mathrm{13}$ and $M$(dust) increase in each region with respect to the ratios reported in Table \ref{mass-budget} due to the smaller $M$(dust) yielded by the $^{13}\mathrm{CO}$ mask.  The most remarkable change is observed in the SP region: it goes from 21\% when the target area is taken from the definition in Figure \ref{carina.500um.obs} to 76\% when $^{13}\mathrm{CO}$ mask is applied.  When we use the $^{13}\mathrm{CO}$ mask to estimate masses, the SC shows the smallest molecular gas fraction (44\%) as traced by $\cother$ among the CNC-Gum 31 region.

\begin{figure*}
\centering
\begin{tabular}{cc}
\epsfig{file=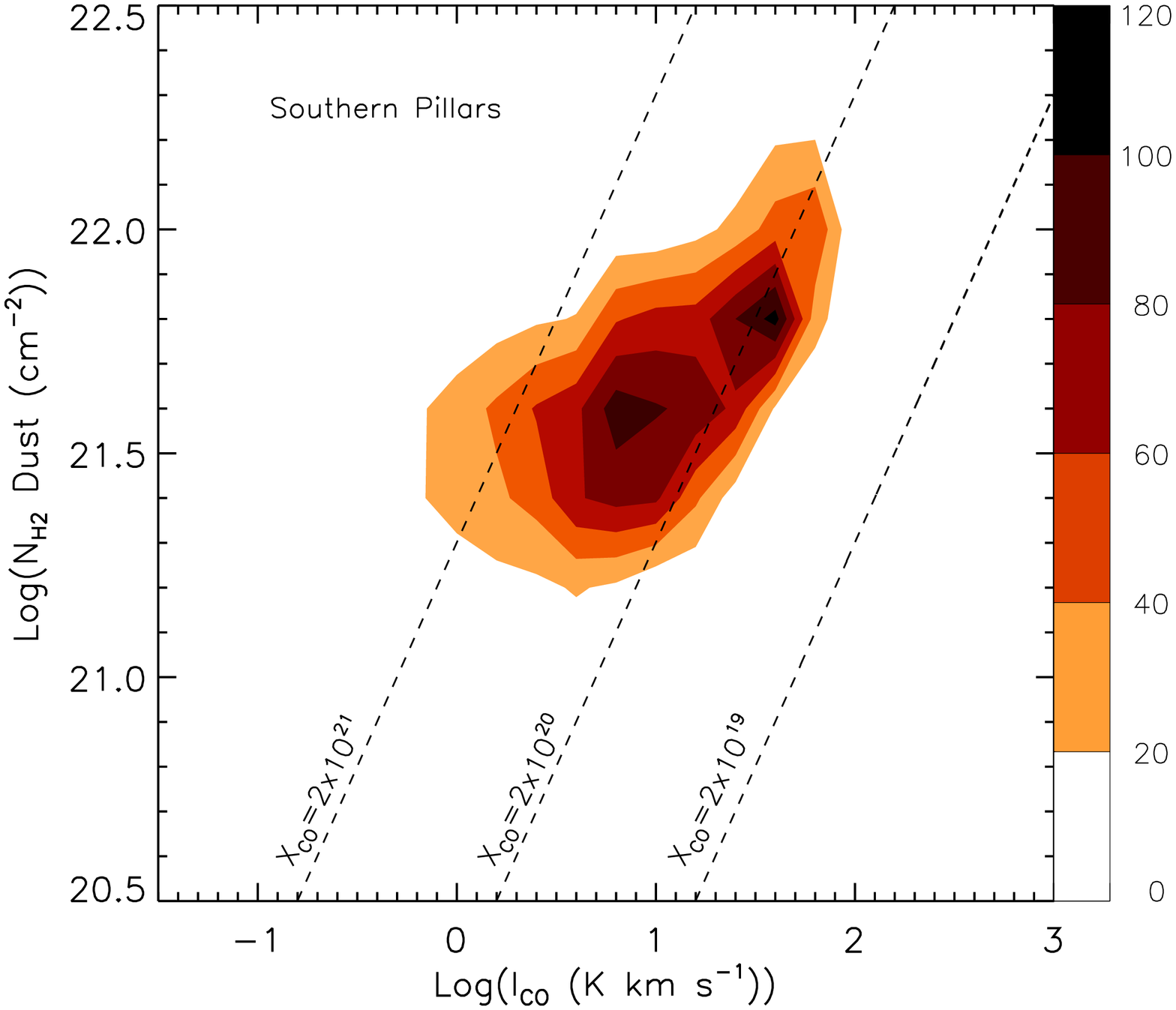,width=0.3\linewidth,angle=90} &
\epsfig{file=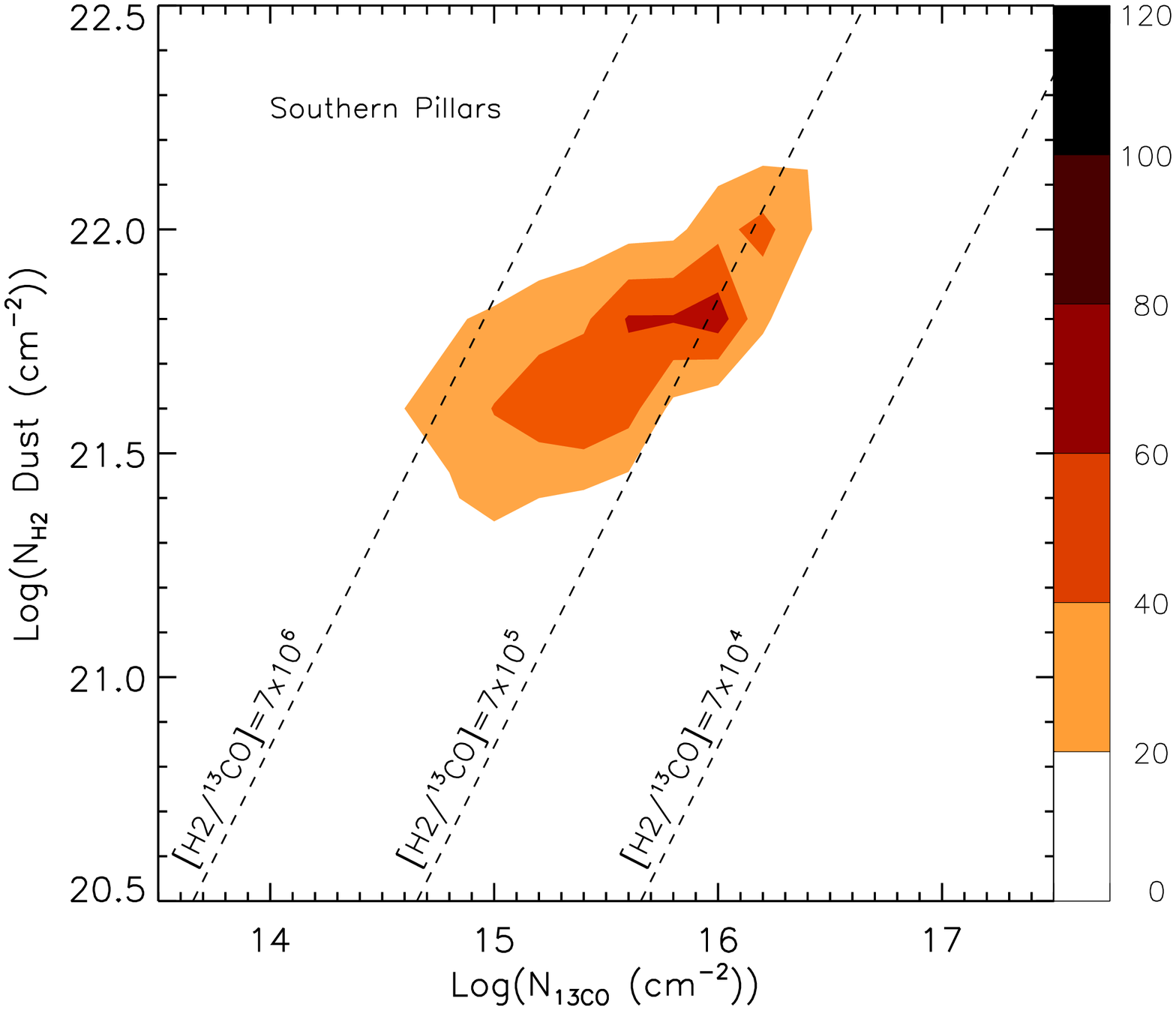,width=0.3\linewidth,angle=90} \\
\epsfig{file=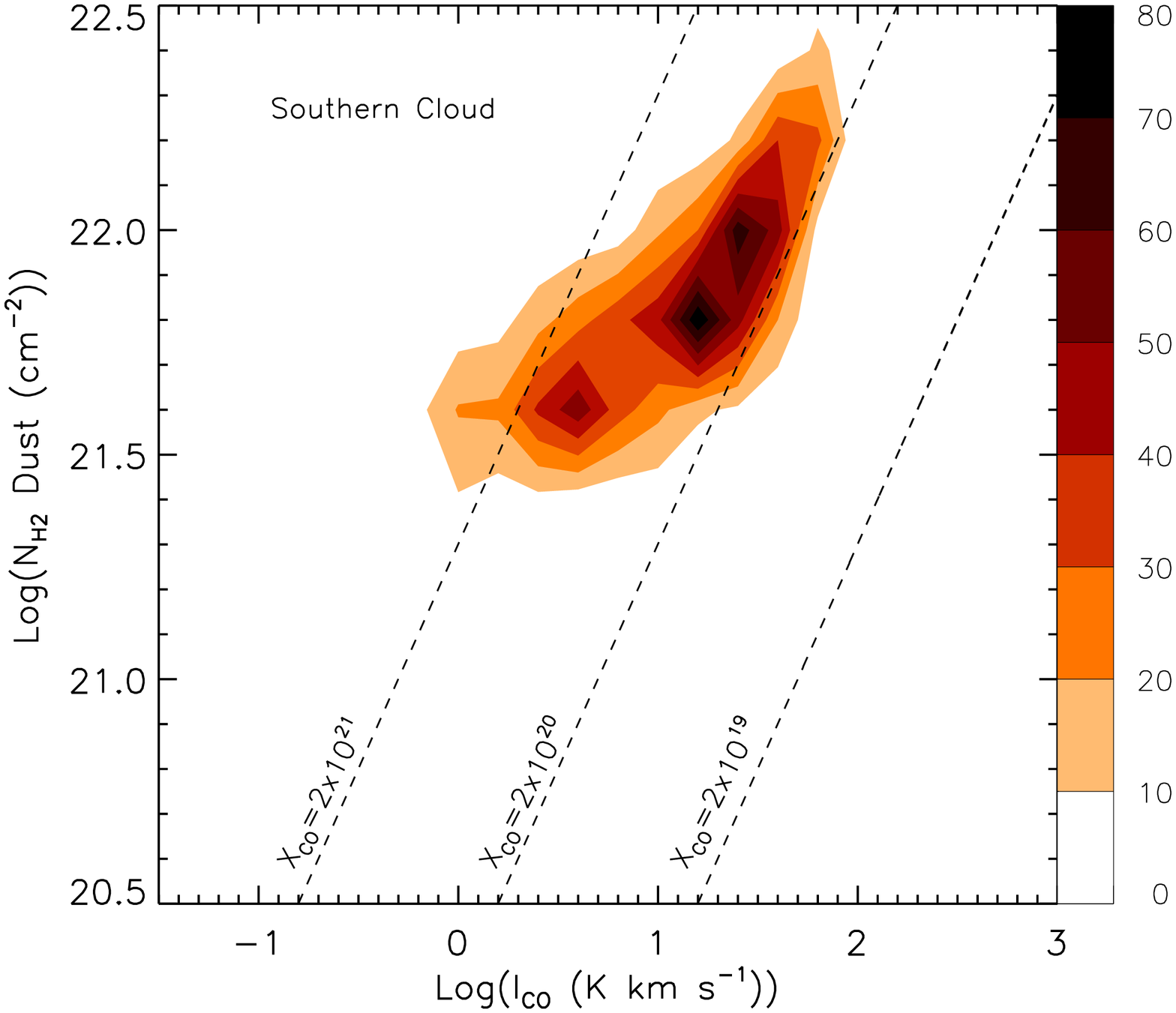,width=0.3\linewidth,angle=90} &
\epsfig{file=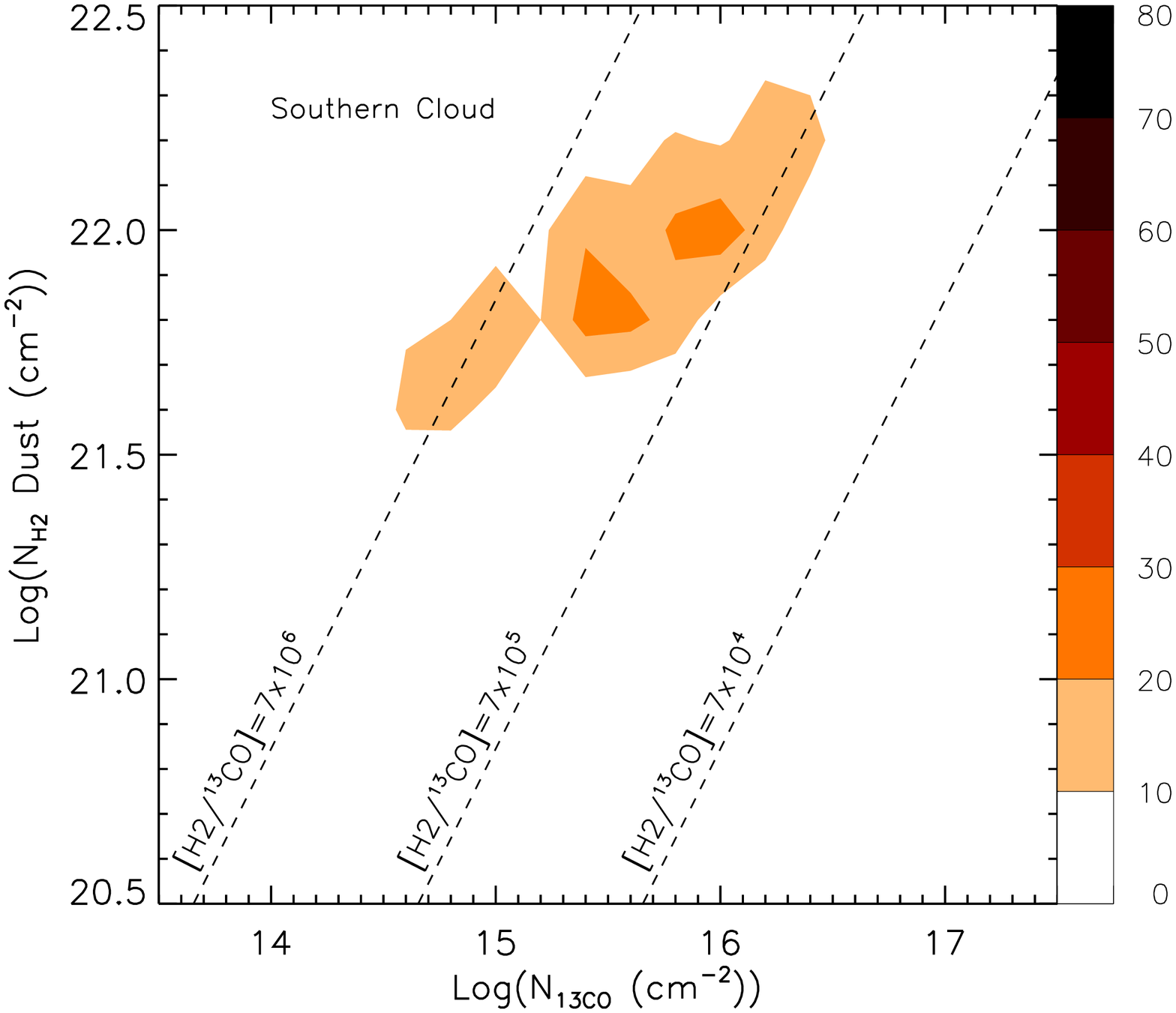,width=0.3\linewidth,angle=90} \\
\epsfig{file=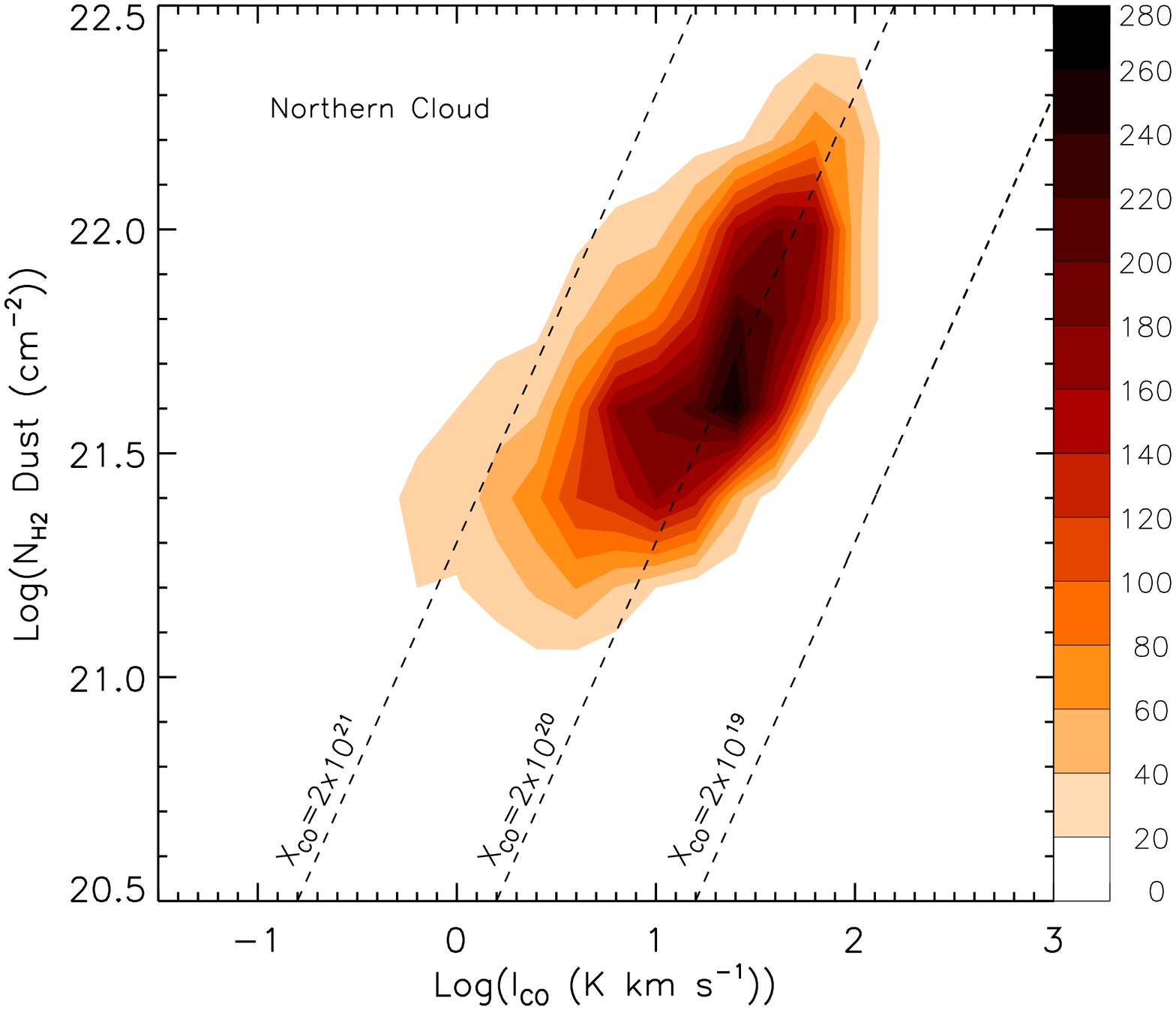,width=0.3\linewidth,angle=90} &
\epsfig{file=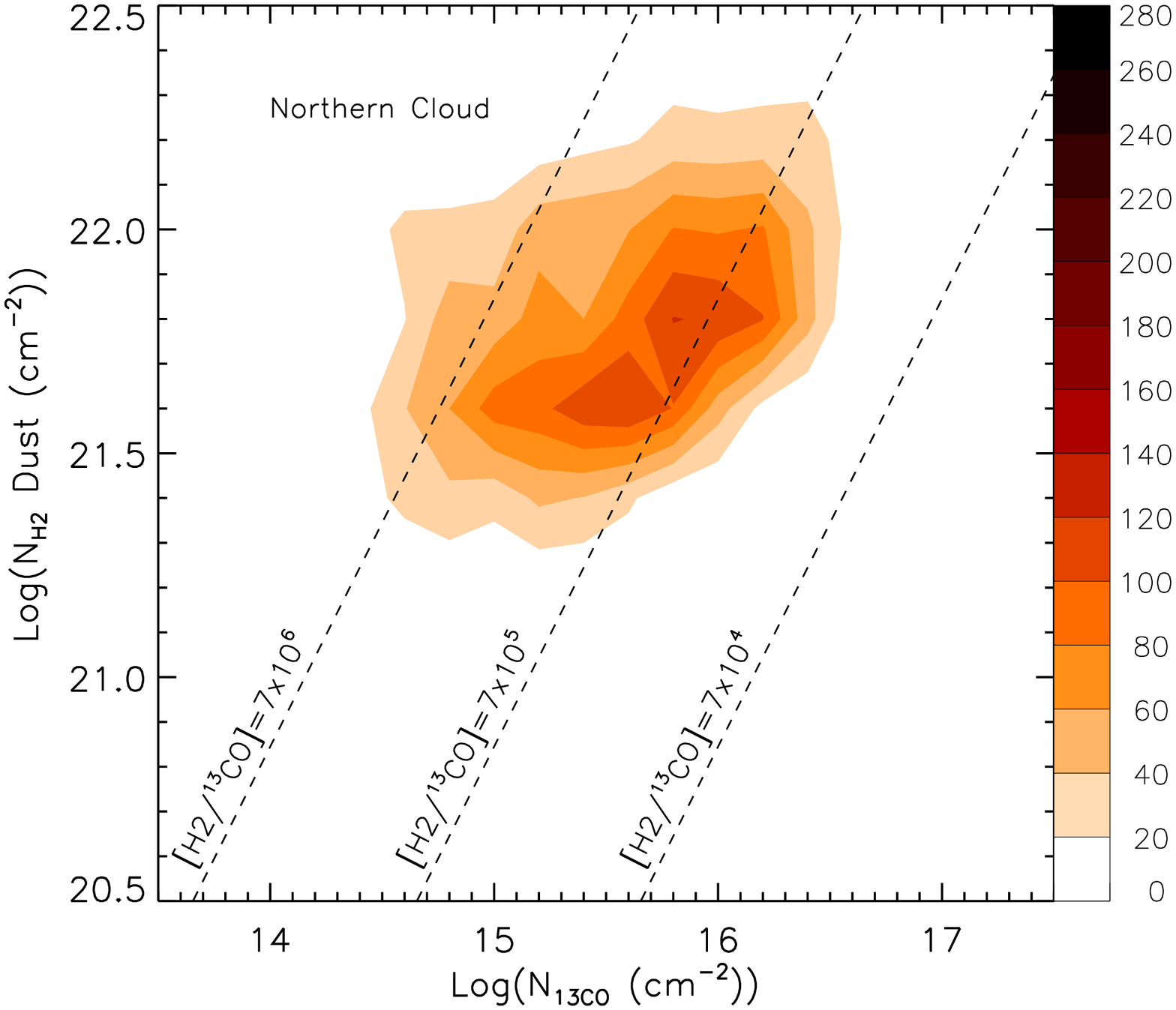,width=0.3\linewidth,angle=90} \\
\epsfig{file=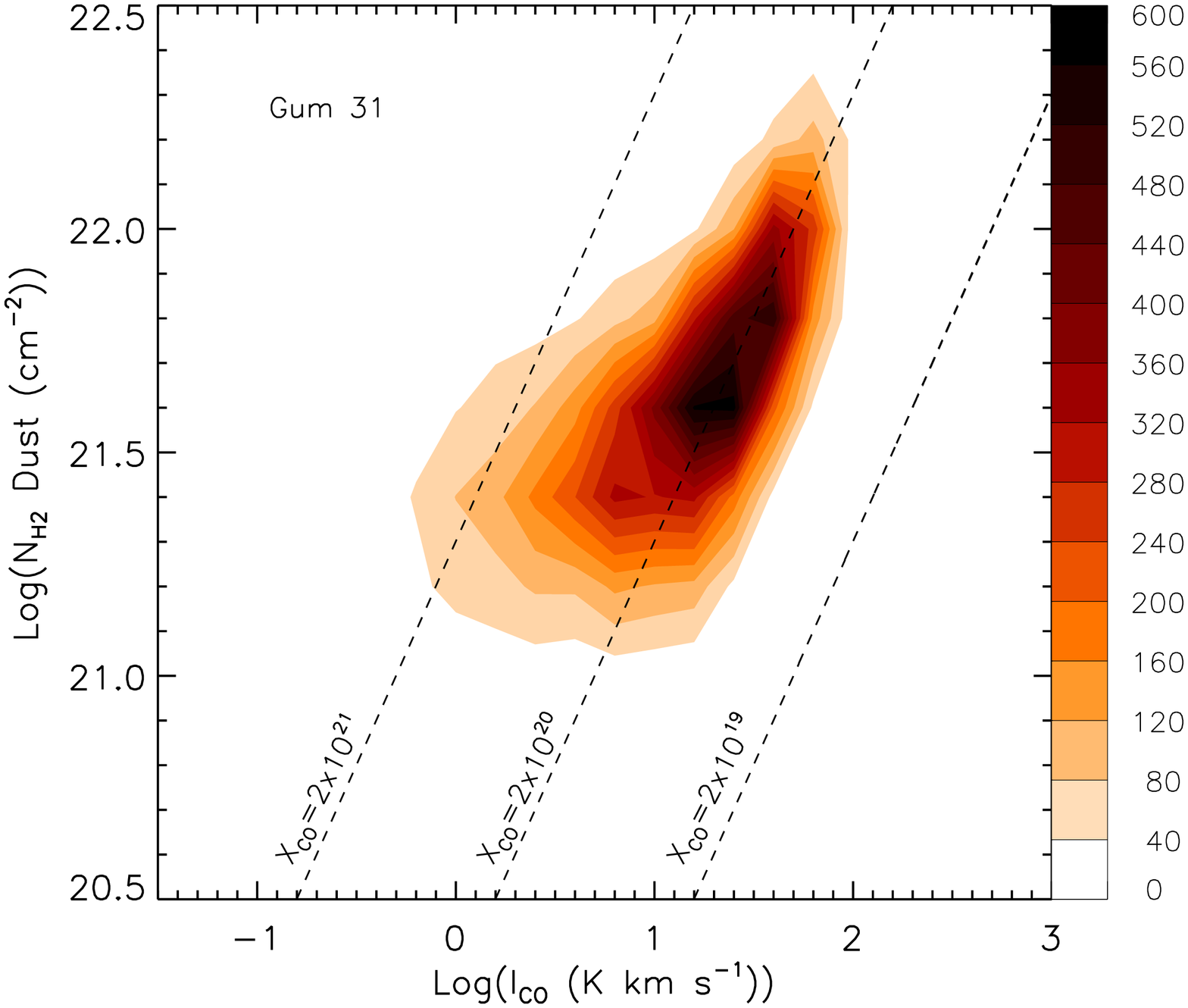,width=0.3\linewidth,angle=90} &
\epsfig{file=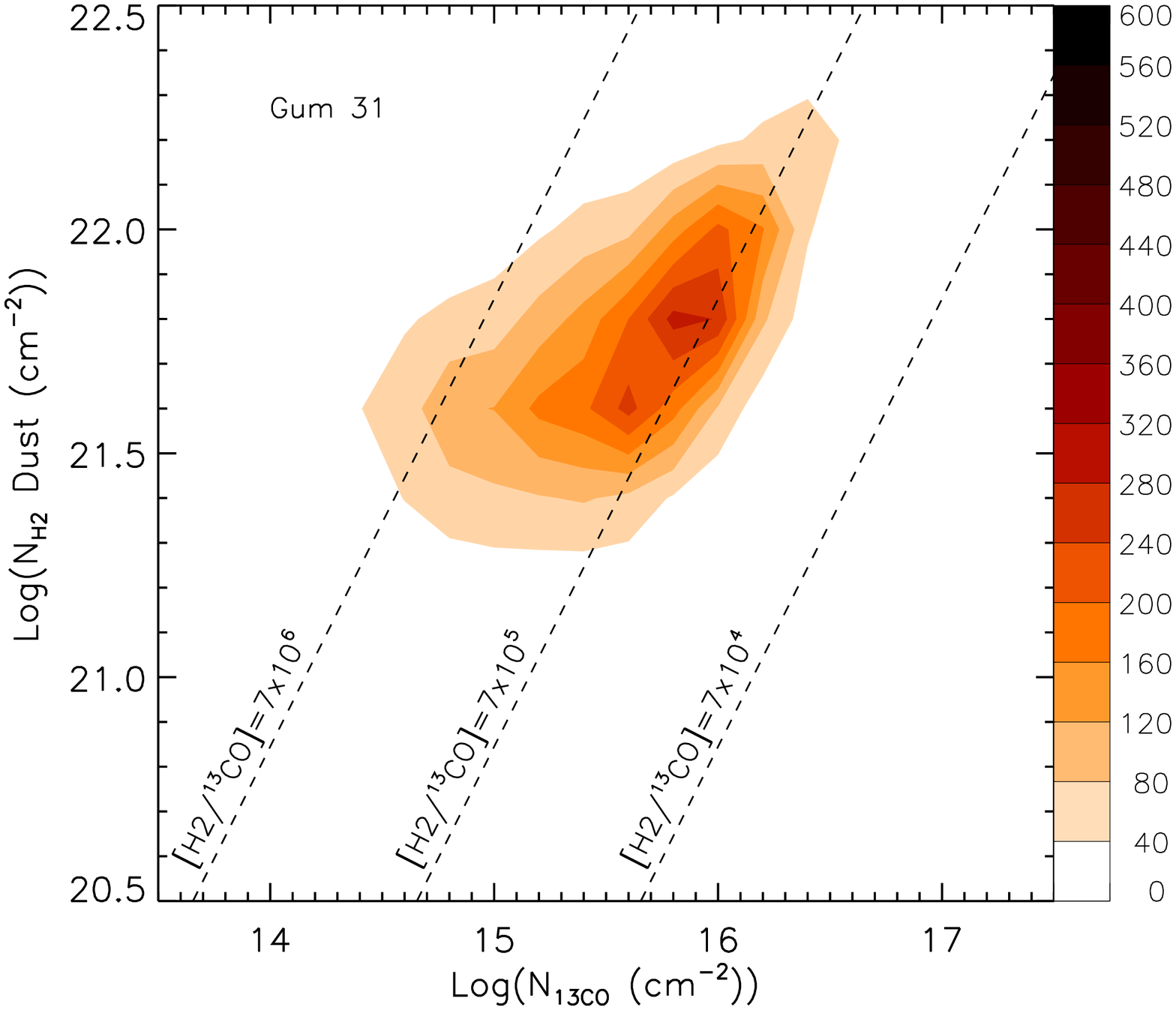,width=0.3\linewidth,angle=90}
\end{tabular}
\caption{Left column: Relation between the $\co$ integrated intensity and the gas column density derived from the dust SED fitting for each subregion.  Colour bars show the number of pixels.  The dashed black lines illustrate different values of the $X_\mathrm{CO}$ conversion factor.  Right column:  Relation between the $^{13}$CO column density and the gas column density derived from the dust SED fitting for each subregion.  Colour bars show the number of pixel.  The dashed black lines illustrate different values of the $[\mathrm{H}_2/\mathrm{^{13}CO}]$ abundance ratio.}
\label{12co_gas_comp}
\end{figure*}

\subsection{Variation of the conversion factors across the CNC-Gum 31 region}\label{factor-var}
In the previous section molecular gas column densities and masses were calculated using constant conversion factors for $\co$ and $\cother$, namely, the $X_\mathrm{CO}$ and the $[\mathrm{H}_2/\mathrm{^{13}CO}]$ factors.  In this section, we will investigate whether those factors change when individual regions are considered using the following approach.  Firstly, the gas column density map from the dust emission is set to the same grid of the CO maps.  Then, a pixel-by-pixel comparison is made between the $N_\mathrm{H_2}$(dust) and the $\co$ integrated intensity and the $N(^{13}\mathrm{CO})$ maps.  It is assumed that all the gas traced by the dust emission is in molecular form.  Although this approximation is certainly wrong for regions of low gas column density, the pixels included in this analysis are required to have significant $\co$ or $\cother$ emission.  Thus, we are selecting regions where the molecular gas is likely to be dominant.  In a companion paper, we investigate the fraction of atomic gas component in the CNC-Gum 31 region, and its importance in estimating the conversion factors (Rebolledo et.\ al in preparation). 

Figure \ref{12co_gas_comp} shows the pixel-by-pixel comparison between the integrated intensity maps from the $\co$ emission line and the column density from the dust emission for each region.  For the SP and SC, most of the pixels are consistent with $X_\mathrm{CO}$ values which are a factor of $\sim$ 3-4 larger than the value assumed in our previous calculation $X_\mathrm{CO}=2.0\times 10^{20}$.  For the NC and Gum 31 region, the majority of the pixels are consistent with our assumed $X_\mathrm{CO}$ factor.  This apparent change in the $X_\mathrm{CO}$ factor across the regions in the CNC-Gum 31 molecular complex can be interpreted as follows.  The SP and SC are heavily affected by the strong radiation of the massive star clusters in the CNC region.  The gas in this region could be predominantly in atomic rather than in the molecular phase.  If so, only a small fraction of the gas is molecular in the southern regions.  Additionally, as the CO molecule is less shielded against the radiation field than the H$_\mathrm{2}$ molecule, the CO molecule may not be a good tracer of the molecular gas in the southern regions.  For the NC, the gas appears to be dense enough to be predominantly in molecular phase, allowing the CO molecule to be effectively shielded despite the strong stellar feedback.  On the other hand, the much lower level of stellar feedback affecting the less dense gas present in the Gum 31 region is likely the predominant factor in setting the observed CO abundance.  Detailed observations of the atomic gas and the atomic carbon will provide a better picture of the different phases of the gas in this region of the complex.   

In Figure \ref{12co_gas_comp} we also show the pixel-by-pixel comparison between the $^{13}$CO column density and the column density from the dust emission for each region.  Although the number of pixels is smaller for the $^{13}$CO column density map, some differences are still evident.  The best match to our assumed value $[\mathrm{H}_2/\mathrm{^{13}CO}]=7\times 10^5$ is obtained in the Gum 31 region, but with a large scatter.  Most of the pixels in the SP and NC regions shows a $[\mathrm{H}_2/\mathrm{^{13}CO}]$ abundance ratio a factor of $\sim$ 2 larger than this value, but also a large scatter is observed.  For the SC, the small number of pixels yields a patchy pixel density map, but a similar behaviour is reproduced.

\section{Summary}\label{summary}
We have produced detailed maps of the molecular gas distribution in the Carina Nebula and the Gum 31 molecular complex.  With the Mopra telescope, we have observed $\co$ and $\cother$ over 8 deg$^2$ centered in $(l,b) \sim (287.5\degrees,-0.5\degrees)$, achieving a spatial resolution of $30\arcsec$.  The main results are summarized as follows:

\begin{enumerate}

\item The spatial distribution of the $\co$ and $\cother$ maps are very similar across the region covered by our observations, with the $\cother$ map tracing the regions of strongest emission of the $\co$ map.  Both maps revealed a complex molecular gas structure, with several regions of dense gas located on the inner regions of the SP, the centre of the NC and regions located at the edge of the shell-like structure surrounding the Gum 31 \hii\ region.  By combining position-velocity diagrams of the CO maps and a model of the spiral arms of the Milky Way, we have excluded any CO emission from regions not directly related to the CNC-Gum 31 molecular complex.

\item We estimated the molecular gas column density from $\co$ assuming a fixed $X_\mathrm{CO}$ factor typical of the Milky Way.  The molecular gas column density from the $\cother$ observations is estimated assuming LTE conditions over the CNC-Gum 31 region.  The column density distributions from both tracers cover a similar range of values, from $\sim 1.0\times 10^{19}$ cm$^{-2}$ to $\sim 5.0\times 10^{22}$ cm$^{-2}$.  Log-normal functions describe well the gas column density distributions for a limited range of column densities in both tracers, from our sensitivity limit to the high-column density regime starting at $\sim 5.0\times 10^{22}$ cm$^{-2}$.

\item Detailed maps of the dust temperature and the gas column density were produced by fitting a grey body function to the spectral energy distribution to each pixel of infrared images from Herschel.  The dust temperature map shows a clear spatial correlation between the warmer dust and the massive star clusters located on the centre of the CNC and the Gum 31 \hii\ region.  The regions of colder dust are coincident with regions of high column density, except for the eastern region of the NC which has been heavily heated by the Trumpler 16 and Trumpler 14 star clusters. 

\item The gas column density distribution calculated from the dust emission has values a factor of 2 larger than the maximum column densities traced by the CO maps.  A log-normal function fits the column density distribution for a limited range of values.  For column densities below this range, the sensitivity limit of the infrared maps reduces our ability to sample the low end of the distribution.  For values above this range, we observe a tail with respect to the log-normal function which has been identified as typical of regions with active star formation.

\item The mass traced by the $^{12}\mathrm{CO}$ map recovers 54\% of mass calculated from the dust emission, while the mass traced by $^{13}\mathrm{CO}$ map is 30\% of the mass traced by the dust map.  These fractions differ across the molecular complex.  These differences can be interpreted as the result of local variations of the massive star radiation field strength across the CNC-Gum 31 molecular complex and differences in the evolutionary stage of the NC and Gum 31 with respect to the southern regions.

\item The molecular mass gas fraction traced by $\co$ increases to 78\% when we calculate the gas masses only considering the pixels with detectable $\co$ emission.  The molecular gas mass traced by $\cother$ recovers 66\% of the total gas mass estimated from dust emission when we consider the mask defined by the $\cother$ map.  The southern regions still present smaller molecular mass gas fractions with respect to the northern regions when we only consider the pixels with detectable $\co$ emission.  In the case of the molecular mass gas fraction traced by $\cother$, this trend is lost when we only consider pixels with significant $N(^{13}\mathrm{CO})$. 

\item We find that assuming a fixed $X_\mathrm{CO}=2\times 10^{20}$ factor is a good approximation for the NC and Gum 31 region.  However, for the southern regions (SP and SC) an $X_\mathrm{CO}$ conversion factor $\sim 3-4$ times larger than the value originally assumed in this paper is a better fit.  This apparent variation of the $X_\mathrm{CO}$ factor may be related to local variations of the atomic-to-molecular gas phase transition in regions affected by the strong radiation of the massive stars, or possibly due to variations of the relative abundance of the CO molecule with respect to the H$_2$.

\item For the $[\mathrm{H}_2/\mathrm{^{13}CO}]$ abundance ratio, some regional variation across the CNC-Gum 31 complex has been detected.  The Gum 31 region presents the best match to our assumed abundance ratio of $7\times10^5$, but with a large scatter.  For the other regions, an abundance ratio $\sim$ 2 larger than this value would be needed to recover the column density traced by the dust emission.

\end{enumerate}

\section*{Acknowledgements}

The authors want to thank the anonymous referee for the excellent suggestions and comments that have significantly improved the presentation and the analysis presented in the paper.  The Mopra radio telescope is currently part of the Australia Telescope National Facility. Operation of the telescope is made possible by funding from the National Astronomical Observatory of Japan, the University of New South Wales, the University of Adelaide, and the Commonwealth of Australia through CSIRO. We acknowledge assistance with observations from members of the Mopra Southern Galactic Plane CO Survey Team. DR acknowledges support from the ARC Discovery Project Grant DP130100338.  DR acknowledges Tony Wong and Rui Xue for the IDL package to generate the integrated intensity and velocity maps of the CO intensity cubes.  DE and ES acknowledge support from the European Union FP7 program VIALACTEA (agreement no. 607380).




\bibliographystyle{mnras}
\bibliography{biblio} 


\bsp	
\label{lastpage}
\end{document}